\documentclass[pra,superscriptaddress,amsmath,amssymb,showkeys,longbibliography]{revtex4-1}

\pdfoutput=1 

\usepackage{amssymb}
\usepackage{amsmath}
\usepackage{graphicx} 
\usepackage{dcolumn}  
\usepackage{bm}       
\usepackage{float}

\usepackage{color}
\usepackage{soul}

\begin{document}


\title{Dynamical thermalization,  Rayleigh-Jeans  condensate,
vortexes and wave collapse in quantum chaos fibers and fluid of light}

\author{Leonardo Ermann}
\email{leonardoermann@cnea.gob.ar} 
\affiliation{%
  Departamento de Física Teórica, GIyA,
  Comisión Nacional de Energía Atómica.
  Av. del Libertador 8250, 1429 Buenos Aires, Argentina
}

\author{Alexei D. Chepelianskii}
\affiliation{%
  LPS, Université Paris-Sud, CNRS, UMR 8502, Orsay F-91405, France
}

\author{Dima L. Shepelyansky}
\email{dima@irsamc.ups-tlse.fr}
\affiliation{%
  Laboratoire de Physique Théorique,
  Université de Toulouse, CNRS, UPS, 31062 Toulouse, France
}

\date{September 4, 2025}


\begin{abstract}
We study analytically and numerically 
  the time evolution of a nonlinear field
  described by the nonlinear Schr\"odinger equation
  in a chaotic $D$-shape billiard. In absence of nonlinearity
  the system has standard properties of quantum chaos.
  This model describes a longitudinal light propagation
  in a multimode D-shape optical fiber and also those in a Kerr nonlinear medium
  of atomic vapor. We show that, above a certain chaos border of nonlinearity,
  chaos leads to dynamical thermalization with the Rayleigh-Jeans thermal distribution
  and the formation of the Rayleigh-Jeans condensate in a vicinity of the ground state
  accumulating in it about 80-90\% of total probability.
  Certain similarities of this phenomenon with the Fr\"ohlich condensate are discussed.
  Below the chaos border
  the dynamics is quasi-integrable corresponding to the Kolmogorov-Arnold-Moser integrability.
  We describe also the time evolution during the process of relaxation to the thermal state
  and the time dependence of quantum von Neumann and classical Boltzmann entropies
  during this process.
  At a  strong focusing nonlinearity
  we show that the wave collapse can take place even at sufficiently high positive energy
  being very different from the open space case. Finally for the defocusing case we establish
  the superfluid regime for vortex dynamics at strong nonlinearity.
  System parameters for optical fiber experimental studies of these effects are also discussed.
\end{abstract}

\maketitle

\section{Introduction}
\label{sec1}

In recent years a number of interesting physical phenomena
have been established for laser beam propagation in multimode optical fibers
(see e.g. overview \cite{aplbabin}). Among them there is a phenomenon
of self-cleaning of the laser beam when an injected laser power at high modes is transferred to
low-energy modes. It was realized that this effect appears due to
thermalization among linear fiber modes induced by nonlinear
four-wave interactions between modes
\cite{wabnitz,fiberrev2a,fiber1,fiber2,fiber3,fiber4,christo1,fiber5,fiberrev2b}.
Thus, even thermalization with negative temperatures has recently been
observed in such optical fibers \cite{fiber5}. It is argued that this thermalization
emerges as a result of purely Hamiltonian dynamics without any external
thermal bath.

The phenomenon of dynamical thermalization
attracts a deep interest of the scientific community,
starting from the famous Boltzmann-Loschmidt dispute
on thermalization emerging from the reversible dynamical equations of particle motion
\cite{boltzmann1,loschmidt,boltzmann2} (see also \cite{mayer}).
In modern times the interest in the problem of dynamical thermalization
was pushed forward by numerical experiments of Fermi, Pasta, Ulam (FPU) in 1955 \cite{fpu1955}
with the conclusion that ``The results show very little, if any, tendency toward
equipartition of energy between the degrees of freedom.'' \cite{fpu1955}.
The absence of thermalization in the FPU problem of linear oscillators
with nonlinear couplings was found to be linked to its
proximity to exactly integrable models with solitons
\cite{zabusky,greene,zakharovnse} and Toda lattice \cite{toda,benettin}.
Also, at weak nonlinearity there is no overlapping of main system resonances,
the system is below the chaos border,
and the dynamics remains quasi-integrable \cite{chirikovfpu1,chirikovfpu2,livi}.
Such a regime corresponds to a spirit of Kolmogorov-Arnold-Moser (KAM) integrability
which takes place at weak perturbation of integrable system
while chaos and thermalization may appear
only at relatively strong nonlinear perturbation
(see details in \cite{arnold,sinai,chirikov1979,lichtenberg}).
At present the FPU problem still remains an active area of research \cite{fpu50}.

\begin{figure}[H]
	\begin{center}
		\includegraphics[width=0.3\columnwidth]{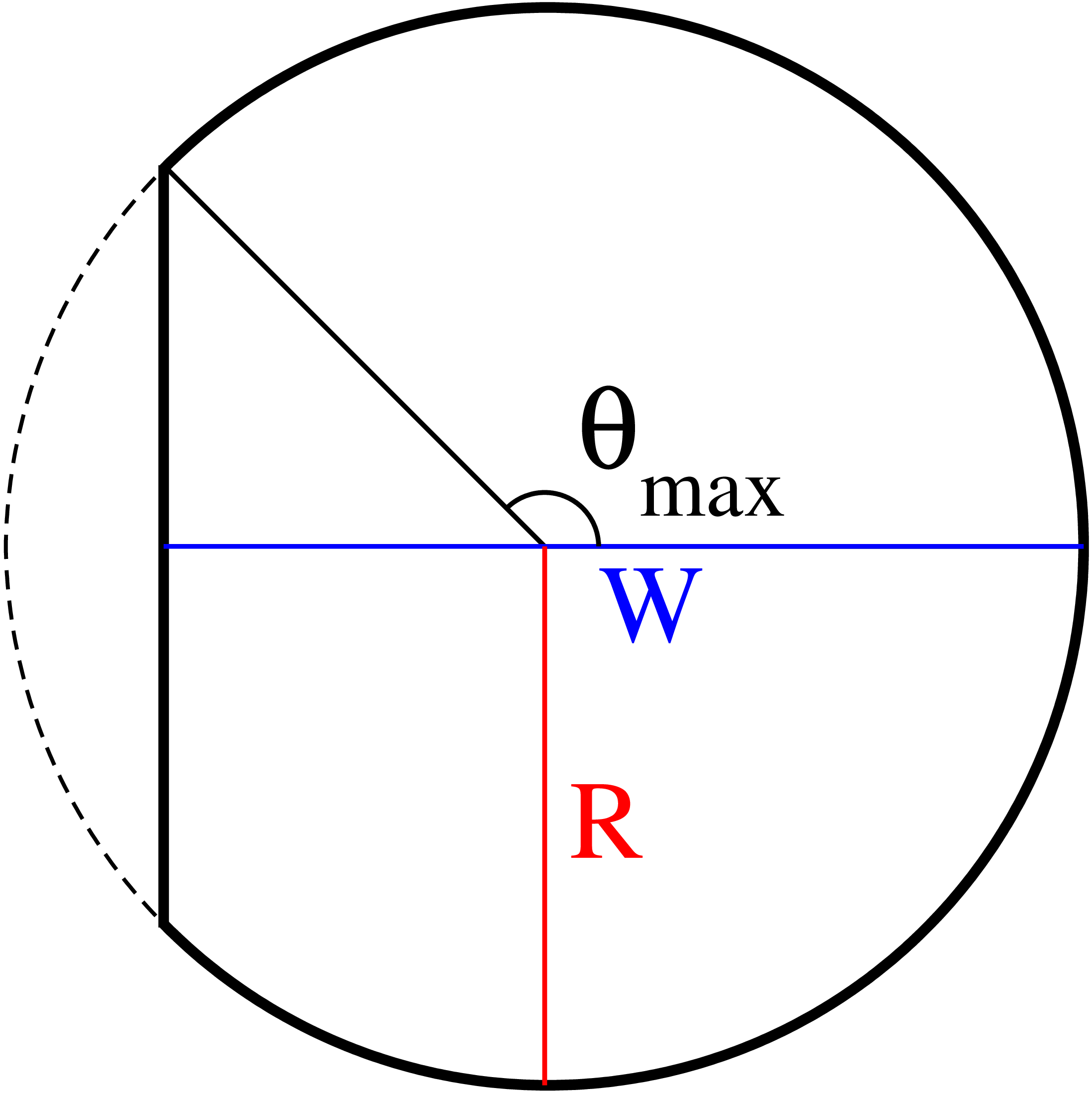}
	\end{center}
	\vglue -0.3cm
	\caption{\label{fig1} Geometry of the D-shape billiard with a straight cut of a circle
with length of a perpendicular to the cut being $W=wR$
where the diameter of the circle is $W=2R$,
here $\theta_{max}$ is given by the equation $\cos \theta_{max} = (R-W)/R$.
In our studies we use $R=1$, $w=(1+1/\sqrt{2})$.
	}
\end{figure}

In fact an example of FPU problem shows that for investigations of
dynamical thermalization in classical many-body  systems
it is important to choose a model corresponding to a generic situation
of oscillators  with nonlinear interactions.
Such a generic model was proposed in \cite{rmtprl} with unperturbed Hamiltonian of
linear oscillators described by a random matrix with additional
nonlinear perturbations. Thus the linear oscillators are described by
the Random Matrix Theory (RMT) invented by Wigner for
a description of complex quantum systems like
nuclei, atoms and molecules \cite{wigner}. At present RMT finds a variety
of applications in quantum systems, mesoscopic physics \cite{mehta,guhr}
and systems of quantum chaos \cite{bohigas,haake}. In \cite{rmtprl}
it was shown that a  nonlinear perturbation of RMT
leads to dynamical chaos and the Rayleigh-Jeans thermal distribution
over linear oscillator modes
if nonlinearity strength exceeds a certain chaos border.
In presence of additional dissipation and energy pumping 
this nonlinear random matrix  (NLIRM) model
describes also the spectra of Kolmogorov-Zakharov turbulence \cite{kzpre}.
A brief discussion of a model similar to NLIRM was also done
in \cite{prxshapiro}.

In practice it may be not so easy to reach an experimental
realization of such NLIRM model but
in  \cite{rmtprl} it was proposed that this system
can be approximately presented by a multimode optical fiber
with a fiber cross-section being a chaotic billiard,
e.g. D-shape billiard (circle with a cut. see Fig.~\ref{fig1}).
Indeed, it is known that classically chaotic billiards have many properties
being the same as for RMT \cite{bohigas,haake}. 
It was shown that the quantum chaos exists for
variety of cuts of a circle billiard \cite{reichl,miniatura,legrand}
even if for certain cut positions there can be present significant
integrable islands.
The advantage of D-shape billiard is
that it can be relatively simply realized in multimode fibers.
Indeed, observations of signatures of quantum chaos in D-shape billiard
had been reported in \cite{miniatura,legrand,stone}.
However, dynamical thermalization induced by nonlinearity
was not studied in these works.
We argue that  such quantum chaos fibers can be used for investigations
of dynamical thermalization for a generic case
of oscillators with nonlinear interactions.
In principle similar effects can be studied for lasing dynamics
in semiconductor chaotic microcavities realized in \cite{cao}
but in this work we restrict our studies to the case of
D-shape multimode fibers. 

It is natural to ask what is the situation
with dynamical thermalization observed with multimode fibers 
in \cite{fiber1,fiber2,fiber3,fiber4,fiber5}.
In fact in these circular shape fibers the  spectrum of linear modes
corresponds to a spectrum of two oscillators
with equal frequencies that leads to a significant
spectrum degeneracy. In such a case the KAM theory
for nonlinear perturbation of such a system is not
applicable \cite{arnold,sinai,chirikov1979,lichtenberg}.
Thus in \cite{chirikovyadfiz} it was shown that
for three oscillators with equal frequencies,
the measure of chaotic component remains
on a level of 50\% for arbitrary weak nonlinear
perturbation
(its strength only gives a reduction of the positive Lyapunov exponent
but not the measure of chaos).
Similar effects exist for other
nonlinear systems \cite{mulansky2}
and the FPU $\alpha$-model \cite{fpudls}.
Thus for  fibers, which have
spectrum of two oscillators with equal frequencies,
as those in  \cite{fiber1,fiber2,fiber3,fiber4,fiber5},
we expect that arbitrary small
four-wave nonlinear interaction,
that couples these modes,
should lead to a chaotic dynamics 
and eventual thermalization between degenerate modes.
However, since the Lyapunov exponent decreases
with the nonlinearity strength (as in  \cite{chirikovyadfiz,mulansky2,fpudls})
one would need to have longer and longer fiber
to reach such chaotization. It is possible
that certain non homogeneous perturbations along
a fiber act as a time-dependent noise
and facilitate development of chaos.
Due to these reasons we argue that
the experiments  \cite{fiber1,fiber2,fiber3,fiber4,fiber5}
are done in a specific regime of degenerate mode spectrum
when the KAM theory is not valid
and thus it is of fundamental interest
to study the generic case of D-shape fiber
when the KAM theory works and dynamical
thermalization appears only above a certain chaos
border. We should note that the mathematical  KAM theory results
for the NSE in billiards are very difficult to obtain
and the question about asymptotic time
behavior of nonlinear spreading 
on high energy modes remains open
for chaotic billiards and moderate nonlinearity
(see e.g. \cite{kuksin}).

With this aim we present here the numerical and analytical
study of dynamics of nonlinear waves in D-shape billiard
described by the Nonlinear Schr\"odinger Equation (NSE),
or the Gross-Pitaevskii equation (GPE) \cite{pitaevskii},
which  depicts the laser beam
propagation in multimode fibers \cite{aplbabin}.
We note that the onset of dynamical thermalization
for the NSE with defocusing nonlinearity $\beta>0$
in a chaotic Bunimovich billiard
was studied in \cite{eplbill}.
The results reported there show emergence of dynamical thermalization
above a chaos border when nonlinear perturbation exceeds
a typical energy level spacing in such a billiard.
However, in \cite{eplbill} only relatively short times
were reached in numerical simulations and
the Rayleigh-Jeans distribution was not reached.
Here we significantly
increased the time evolution scale
with emergence of the Rayleigh-Jeans thermalization in a D-shape fiber (billiard)
with defocusing nonlinearity.

Another important feature is  that  usually for fibers
a nonlinearity  is focusing $\beta <0$ and thus,
according to the so called Vlasov-Petrishchev-Talanov theorem \cite{talanov,kuznetsov},
a wave collapse can take place for strong nonlinearity
at finite times. Important implications
of such a collapse for Langmuir waves was discussed in \cite{zakharov}.
The collapse properties in multimode fibers
have been also studied in more recent publications (see e.g. \cite{turitsyn}).  
However, in \cite{talanov,kuznetsov,zakharov} a collapse was discussed
in a case of infinite system size.
For a D-shape fiber chaos and thermalization in a finite size billiard
may lead to an unusual interplay between collapse and dynamical thermalization
so that there is a fundamental interest to study the properties of NSE
in a D-shape focusing fiber. Here we present results of our studies of such a system
considering both  cases of negative (focusing) and positive (defocusing) signs of nonlinearity.
We show that that the dynamical thermalization with the Rayleigh-Jeans distribution
takes place at moderate strength of nonlinearity being above the chaos border.
Below this border the dynamics is quasi-integrable corresponding to the KAM theory.

It is interesting to note
that the GPE describes Bose-Einstein condensate (BEC) of cold boson atoms
in optical trap \cite{pitaevskii}. Thus there is a temptation to expect \cite{eplbill} that the
dynamical thermalization for GPE or NSE in a chaotic billiard would lead
to the thermal Bose-Einstein (BE) distribution over
eigenenergies $E_m$ of linear system \cite{landau}:
\begin{equation}
 \rho_m=\frac{1}{\exp[(E_m-\mu)/T]-1} \; ({\rm BE}) .
\label{eqbe}
\end{equation}
However, NSE describes classical
nonlinear fields without second quantization and thus in this case
we should have the classical Rayleigh-Jeans (RJ) thermal energy equipartition 
with chemical potential $\mu$ due to an additional conservation integral
of norm or number of particles, see e.g. \cite{landau,zakharovbook,nazarenko}:
\begin{equation} 
\rho_m = \frac{T}{E_m-\mu} \; ({\rm RJ}) .
\label{eqrj}
\end{equation}
In the distributions (\ref{eqbe}), (\ref{eqrj})
$T$ is a system temperature, $\mu$ is a chemical potential
and the total norm of probabilities $\rho_m$ on $N$ energy levels $E_m$
is preserved and normalized to unity (${\sum^N}_{m=1} \rho_m  =1$;
$1 \leq m \leq N$). We use here normalization of all $\rho_m$
values to be equal to unity so that $\rho_m$ can be considered
as a probability at a state $m$, but in a case of many particles $N_p$
the value $\rho_m$ gives simply a fraction of total number of particle $N_p$
in the state $m$.
In optical fibers such classical thermal distribution  (\ref{eqrj}) is usually 
called as Rayleigh-Jeans (RJ) distribution (see e.g. \cite{fiber1}).
Indeed, in \cite{rmtprl} for the NLIRM and related models
it was shown that above the chaos border the dynamical chaos
at moderate nonlinearity leads to the RJ thermal distribution
and not to the BE one. For dynamical systems like NSE
we assume that the nonlinear term  is relatively weak or moderate and
it leads to dynamical chaos for nonlinearity above the chaos border
producing thermal distribution (\ref{eqrj})
over linear eigenenergies (or mode eigenfrequencies)
that are not significantly affected by weak nonlinearity.

In fact the RJ distribution is a limiting case of the BE one
in the limit of high temperature significantly exceeding typical energies of
the linear system ($T \gg E_m$). Using this RJ description of BE distribution
at high temperature Fr\"ohlich showed in 1968 that for the RJ distribution
there is a significant condensation of probability at the lowest system energy level
that may be close to hundred percent
of total probability norm or total number of bosons \cite{frohlich1,frohlich2}.
The model proposed by Fr\"ohlich assumes a presence of external pumping of the system
in presence of dissipative process leading to a certain steady-state.
The total number of phonons in the system depends on a pumping strength
leading to condensation in the ground state above
a certain critical strength. Thus this model does not directly
correspond to the RJ thermal distribution studied here,
but there are still certain similarities due to
the fact that the steady-state in  \cite{frohlich1,frohlich2}
is described by the RJ distribution with a certain
prefactor depending on pumping and dissipation.

On the basis of this result Fr\"ohlich made a conjecture
that such a condensate, appearing at high temperatures, concentrates energy
in a single mode and thus it
can play an important role in coherent properties of cell membranes
and biomolecules. This conjecture
attracted a significant interest of physicists and biologists
with a large number of theoretical and experimental
studies of this phenomenon called  Fr\"ohlich condensate,
see e.g. \cite{austin,reimers,scully,li,wang,kats}.

In this work we show that in the RJ thermal distribution (\ref{eqrj})
there is also condensation at the ground state, which we call
Rayleigh-Jeans (RJ) condensate.
This RJ condensate captures almost all norm, or number of particles, 
it  emerges at low temperatures ($T \ll E_m$) and high number of
levels in the system. Since RJ condensate appears only at low temperatures
it describes the thermalization of classical fields
since for quantum systems the thermalization is described by
the BE distribution (\ref{eqbe}) which at low temperatures
is very different from RJ one (\ref{eqrj}).
Thus the RJ condensate exists only at low temperatures
($T \ll E_m$) at norm conservation and absence
of external pumping. Hence it is very different
from the Fr\"ohlich condensate which
exists at high temperatures ($T \gg E_m$)
in presence of external pumping.

It is established experimentally and theoretically
that the self-cleaning effect \cite{aplbabin,wabnitz,fiberrev2a,fiber1}
is an example of the RJ condensate appearing
at the lowest frequency fiber mode.
Here we show that this happens due to Hamiltonian
chaos induced by nonlinear interaction of waves in multimode
quantum chaos fibers.
The emergence of such condensation
was also reported for numerical simulations
of  NSE in two and three dimensional 
finite systems with quadratic spectrum of waves
\cite{picozzi1,picozzi2,picozzi3,picozzi2024}.
In these works the emergence of RJ distribution
was attributed to the Kolmogorov-Zakharov spectra
of turbulence \cite{zakharovbook,nazarenko}.
However, as argued above, we attribute the appearance
of RJ thermal distribution in these numerical simulations
to the degeneracy of linear eigenmode frequencies
due to which the KAM theory
cannot be applied to these systems
and thus chaos appears at very weak nonlinearity strength.
For the case of focusing fibers
we discuss an interesting interplay
between the wave collapse and chaos.
We also consider the problem of ultraviolet catastrophe
at high energy modes in a chaotic regime.

Finally, we note that a laser beam  propagation in a nonlinear
Kerr media with  atomic vapor \cite{pavloff}
is also described by the NSE which we study here.
The important advantage of this physical system
is that the sign of NSE nonlinearity
can be easily changed by a frequency detuning of a laser beam.
In \cite{pavloff} it was experimentally and numerically demonstrated a formation of
vortexes of such light fluid described by a dissipative  NSE 
within a circular integrable billiard cross-section.
Here we show that in the case of chaotic D-shape cross-section
there are also long living vortexes described by the unitary NSE evolution.
The NSE dynamics of laser beam propagation in fiber or other nonlinear media
is also considered as a fluid of light \cite{picozzi2024,pavloff}.
In fact the NSE and GPE  are closely related to the Ginzburg-Landau equation
used for the description of superconductivity, superfluidity and quantum turbulence
as it is described in the reviews \cite{ginzburg,aranson,tsubota,qturbu}.

The article is composed as follows: Section 2 presents the model
and description of quantum chaos properties in absence of nonlinearity,
numerical methods of integration of NSE are given in Section 3,
results for  simple models with RJ condensate are displayed in Section 4,
the results for dynamical thermalization for defocusing fiber NSE
are described in Section 5,
properties of the wave collapse for focusing NSE are presented in
Section 6, the dynamical thermalization for focusing fiber NSE
is described in Section 7, 
properties of vortexes in defocusing fiber NSE
are displayed in Section 8,
NSE long time dynamics is discussed in Section 9,
experimental parameters for a quantum chaos fiber are displayed in Section 10,
discussion and conclusion are given in Section 11. Appendix presents some additional results.

\section{Model description}
\label{sec2}

The D-shape fiber model is described by the NSE with the Dirichlet boundary conditions:
\begin{equation}
 i\hbar{\partial\psi(\vec{r},t)\over\partial t}= -{\hbar^2\over 2m}
 \Delta \psi(\vec{r},t)+\beta \vert\psi(\vec{r},t)\vert^2\psi(\vec{r},t)
\label{eqnse}
\end{equation}
where  we consider $\hbar=1$, $m=0.5$, $\vec{r}$ is a vector in $(x,y)$ plane.
The form of D-shape fiber billiard  is shown in Fig.~\ref{fig1}.
The time variable $t$ corresponds to $z$ direction of a beam propagation along fiber
(see e.g. Eq.(1) in \cite{aplbabin}), $\psi$ represents a field inside fiber and $\beta$
is a field strength coefficient of nonlinearity. This equation has two integrals of motion being
the norm $\int \vert\psi(\vec{r},t)\vert^2 dx dy =1 $
and energy $\;\;\;\;\;$ $E=<\psi(t)|\hat{H_0}|\psi(t)> +\beta\vert \psi(t) \vert^4/2 $
where $\hat H_0$ is the operator of linear Hamiltonian at $\beta=0$.

\begin{figure}[H]
	\begin{center}
		\includegraphics[width=0.7\columnwidth]{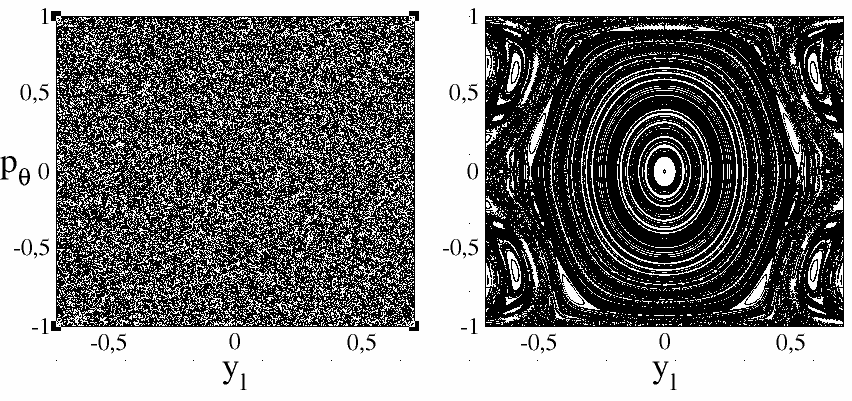}
	\end{center}
	\vglue -0.3cm
	\caption{\label{fig2}
Poincar\'e section $(y_l,p_\theta)$ is shown for
1 trajectory for time up to $t= 10^5$ with
the hard chaos case $(w = 1 + 1/\sqrt{2})$ (left panel), and
for 361 trajectories of 1000 points for each trajectory for the case
of divided phase space  with large integrable islands  $(w = 1 -
1/\sqrt{2})$
(right panel) with large integrable component.
Here, $(y_l, p_{\theta})$ are conjugate variables,
where $y_l$ is the position at which the collision
with the vertical line $x = -1/\sqrt{2}$ occurs.
The term $p_\theta = \sin\theta$ is the tangential
component of the particle's momentum, with $\theta$
being the angle of incidence in the collision.
	}
\end{figure}

Here we study the D-shape billiard with $w=1+1/\sqrt{2}$ shown in Fig.\ref{fig1}.
In this case the Poincar\'e section \cite{lichtenberg} of classical ray dynamics is shown in
the left panel of Fig.~\ref{fig2}. The phase space is fully chaotic even if
very small islands of regular motion are not excluded.
For other values of $w$ it is possible to have
large islands with regular dynamics as shown in the right panel of Fig.~\ref{fig2}.
Our results show that in the range of $1.7 \leq w \leq 1.95$ the dynamics is fully
chaotic similar to the case of $w=1+1/\sqrt{2}$.

Millions of eigenvalues $E_m$ and eigenvectors of linear Hamiltonian $\hat{H_0}$
can be found numerically with the efficient methods
developed in the field of quantum chaos (see e.g. \cite{vergini}). 
At $\beta =0$ we  consider the eigenstates  $\psi_m$
with even-odd parity in $y$ ($\psi(x,y,t) = \pm \psi(x,-y,t)$).
Usually multimode optical fiber may have up to about  thousand of modes
that determined our consideration with up to 1000 states.

\begin{figure}[H]
	\begin{center}
		\includegraphics[width=0.95\columnwidth]{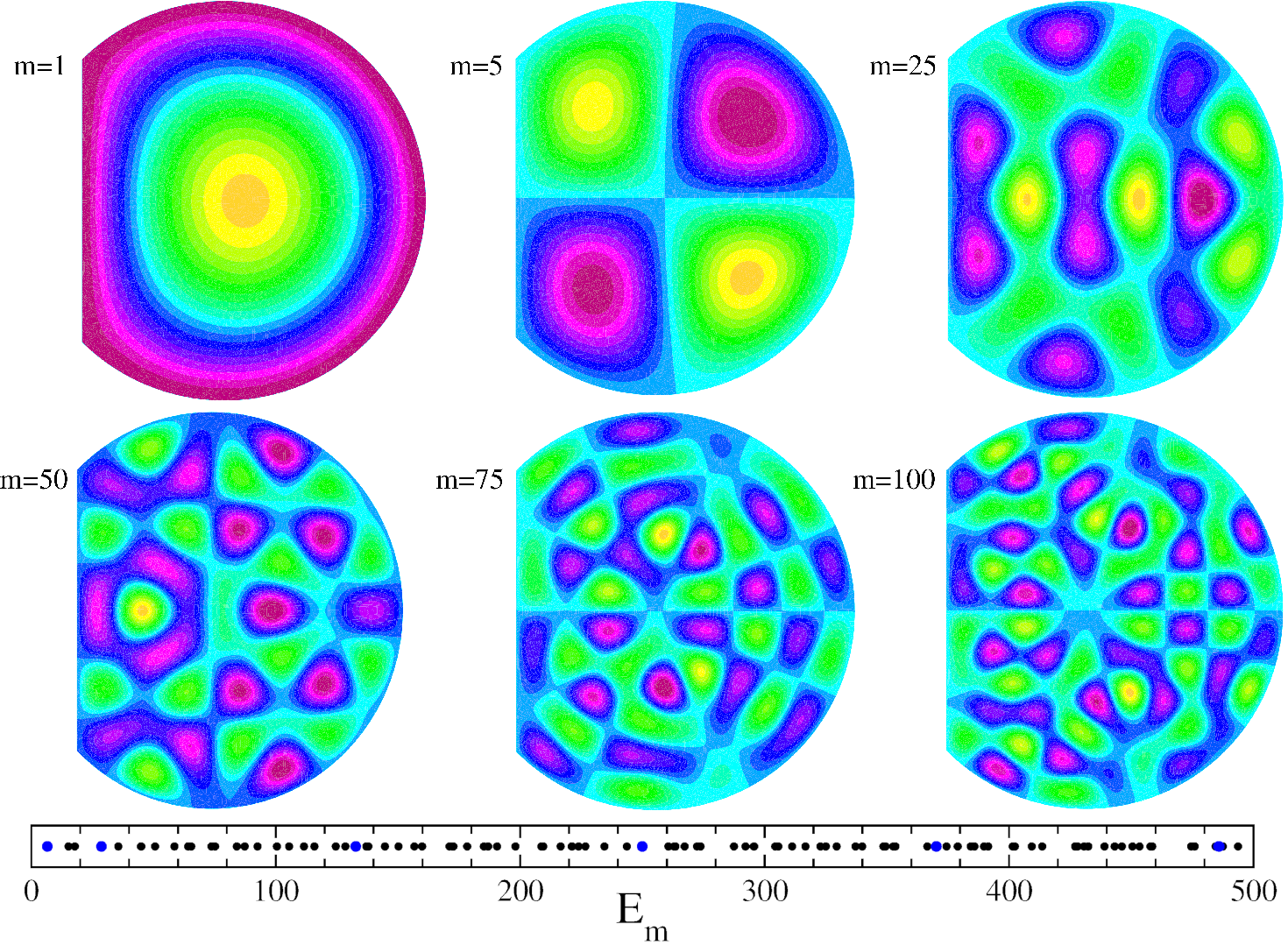}
        \end{center}
	\vglue -0.3cm
	\caption{\label{fig3} Bottom panel shows first energy values $E_m$
          for both parities with respect to the y-axis (odd and even);
          six top panels show the amplitude of real eigenfunctions
          $\psi_m(x,y)$ for $m = 1,5, 25, 50, 75, 100$ (marked by blue points
          of bottom energy panel).
          The color scale of wave functions $\psi_m(x,y)$ a
          includes negative values for $m>1$ and ranges from minimum (red) to maximum (yellow),
          passing through blue, light blue, and green.
          For $\psi_1(x,y)$ at $m=1$ red corresponds to zero for the ground state ($m=1$)
          while light blue represents zero for the other states.
	}
\end{figure}

\begin{figure}[H]
	\begin{center}
		\includegraphics[width=0.7\columnwidth]{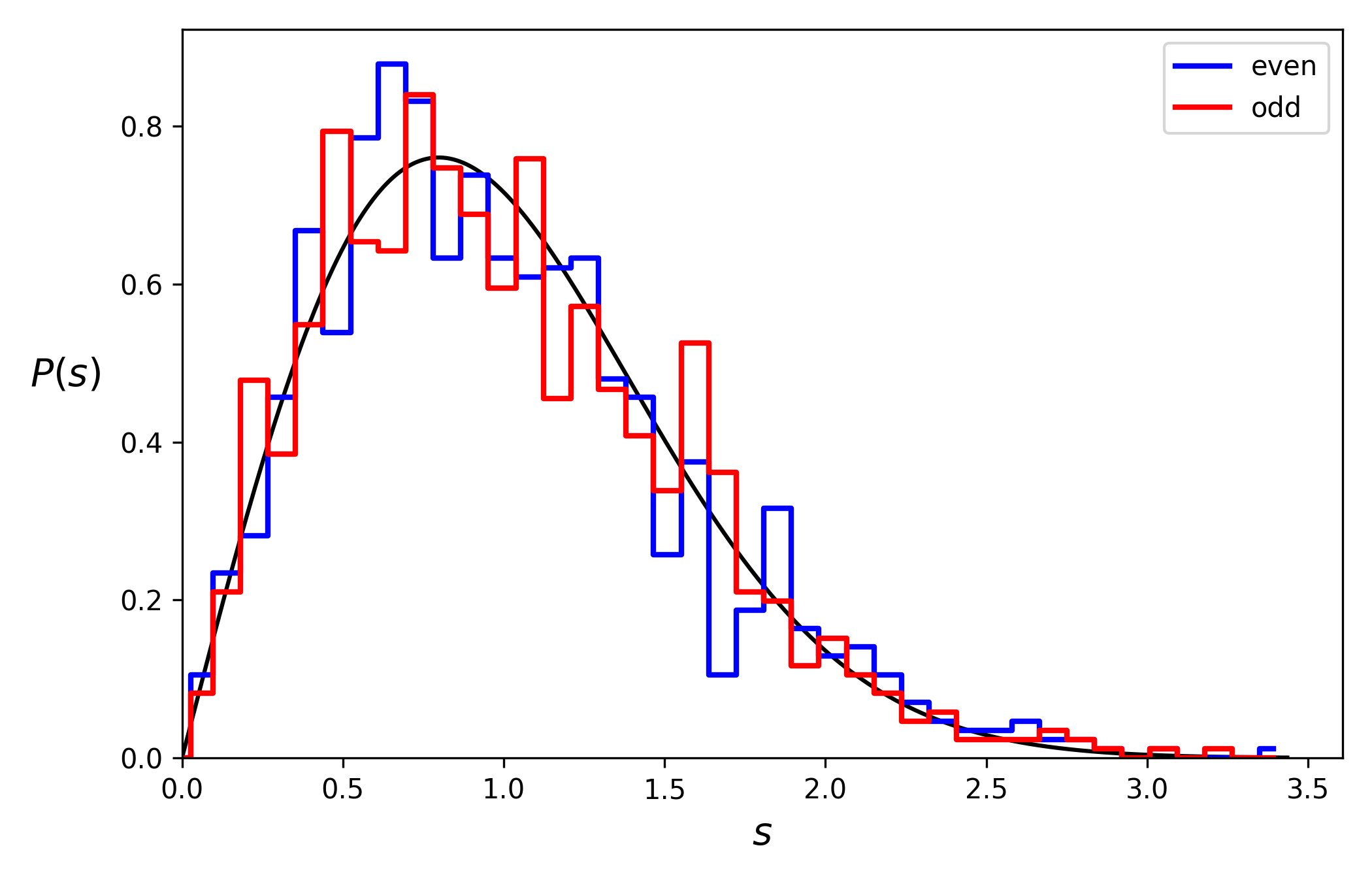}
	\end{center}
	\vglue -0.3cm
	\caption{\label{fig4} Histogram for the level spacing statistical 
          distribution $P(s)$ is shown by blue and red bins for even and odd parity respectively.
          This distribution is 
    calculated for both even and odd parity eigenstates. The theoretical Wigner 
    surmise  $P(s) = (\pi s/2) \exp(-\pi s^2/4)$
    is shown by a full curve. Here, 
    $s$ is  the energy level 
    spacing rescaled by the average level spacing for fixed parity states. 
    The histogram is obtained with the lowest 
    2000 energy levels.
	}
\end{figure}

Examples of several eigenstates $\psi_m(x,y)$ are shown in Fig.~\ref{fig3}
with eigeneneries $E_m$ of first 100 states.
A chaotic structure of wave function eigenstates at high energies is evident.
According to the Weyl theorem  the average energy level spacing
between adjacent levels $\Delta$ is independent of energy $E$
being $\Delta \approx 4\pi /A \approx 4.4$ 
where $A$ is the billiard area \cite{haake} ($A = 3\pi/4 + 1/2 \approx 2.856$).
The level spacing statistical distribution $P(s)$ for the lowest 2000
energies $E_m$ for even and odd parity eigenstates is shown in Fig.~\ref{fig4}.
It is in a good agreement with the Wigner result from RMT
and other results for quantum chaos billiards \cite{bohigas,haake}.

\section{Numerical integration of NSE} 
\label{sec3}

For numerical integration of NSE (\ref{eqnse}) we use two different methods.
The first one uses the same strategy as in \cite{eplbill}.
The time evolution of (\ref{eqnse}) is integrated
by small time steps of Trotter decomposition of linear and nonlinear terms
with a step size going down to  $\Delta t =4 \times 10^{-5}$.
There  are up to $N_l=1085$ linear eigenmodes $\phi_m$ 
for linear part of time propagation;
then the wave function is transferred in coordinate space $(x,y)$
with $N_p \approx 3N_l$ points inside the billiard.
The change of basis from coordinates to energies (and vice versa) is given by a unitary matrix
in double precision.
At any step the wave function is expanded in the basis of linear
modes $\phi_m$ so that $\psi(x,y,t)= \sum_m C_m(t) \phi_m(x,y)$.
A special aliasing procedure \cite{kolmoturb}
is used with an efficient suppression of nonlinear numerical instability at high modes. 
This integration scheme exactly conserves the probability norm
providing the total energy conservation with an accuracy better than $2\%$
at large values $\beta=10$ and better than $1\%$  at lower $\beta$ values.
But this approach allows to obtain numerical results
only on relatively short times with $t \leq 40$.

Due to the numerical restrictions of the first approach
we perform  all presented numerical simulations with the second approach
using the public codes of FREEFEM package \cite{freefem}.
The number of net nodes of FREEFEM used in simulations is usually up to
$N_F = 136241$ with the integration time step $\Delta t = dt =10^{-4}$
for the case of vortex dynamics at such high $\beta$ values as $\beta=45$.
These parameters depend on nonlinearity and initial state
usually composed from linear eigenmodes $\psi_m$.
For typical cases with $\beta =  10$ we use
$\Delta = 5 \times 10^{-4}$ and $N_F = 34501$ for maximal time studied $t=1000$.
For such typical cases the norm is conserved with accuracy of  approximately 3\% and 
total energy conservation is at 6\% at $t=1000$
($\beta = 10$, initial state at $m=11$).
Another typical example of computation
is $\beta=-10$, with initial $m=11$, it has a CPU run time of 2 weeks
on a workstation with 100 processors
with $N_F = 136241, \Delta t =10^{-4}$
and reached time $t=800$
with accuracy of norm and energy relative conservation of
0.05\%   and 1\%  respectively.
In certain cases with collapse we used
even higher $N_F$ and smaller $\Delta t$ values (see below).

We checked that the obtained results
are not sensitive to a variation of $N_F$ and $\Delta t$.
At short times $t \leq 40$ both methods
give close results but such long times
at $t=1000$ can be reached only with the FREEFEM codes
with which all presented NSE results are obtained.

\section{Model systems with RJ condensate} 
\label{sec4}

\subsection{Simple estimates}

To understand the properties of RJ condensate 
we consider a simple model with
$N$ equidistant energy levels $E_m=  m \Delta$ with
level index $m=1, 2, \cdots, N$ and  $\Delta$ neing a level spacing
which we suppose to be unity.
The probability norm or number of particles (bosons)
is conserved so that ${\sum^N}_{m=1} \rho_m = 1$
where $\rho_m $ is a probability at level $m$.
The system ground state has energy $E_g=E_1=1$.
We assume that this system of linear modes $m$
is coupled to an external thermostat or
that an additional  weak nonlinear interaction
of modes (e.g. as in (\ref{eqnse})) leads to dynamical thermalization.

Since we have classical modes or fields
and two integrals of motion,
being total energy $\epsilon_0$, initially injected in the system,
and the total norm being unity,
the thermal distribution is the RJ one given by Eq.(\ref{eqrj})
with temperature $T$ and chemical potential $\mu$ as it was shown in \cite{rmtprl}
for the NLIRM system.
These two parameters are determined by two integrals of norm and total energy:
\begin{equation} 
  \sum_m \rho_m = T \sum 1/(E_m - \mu) =1 \; ; \;\;
  \sum_m E_m \rho_m = T \sum E_m/(E_m - \mu) = \epsilon_0 \;
  ; \;\; \epsilon_0 = NT +\mu \; ,
\label{eqrj1}
\end{equation}
where the last equality directly follows from (\ref{eqrj}).
If we have a case of a system coupled to a thermostat
with temperature $T$ then the chemical potential is
determined by the first equality of norm being unity
and then the total energy $\epsilon_0$ is given
by the last equality. In the case of dynamical thermalization
we have initial system energy $\epsilon_0$
being the total energy and since energy is the integral of motion
then the first and second equalities in (\ref{eqrj1})
determine the system temperature $T$ 
and chemical potential $\mu$.

A case when energy levels grow approximately linearly with
the level number is rather generic and standard.
Due to that we call this simple model with
equidistant levels as
the RJ Standard (RJS) model.

As for the case of BEC \cite{landau,pitaevskii}
at low temperature we have chemical potential approaching to
ground state energy $E_g=E_1$ from below ($\mu \leq E_1$)
leading to a concentration of probability at the 
ground state. Since $\mu$ is very close to $E_1$
we obtain from (\ref{eqrj1}) the approximate  condensation temperature
$T_c$ at which the fraction of condensate at the round state $E_g$
is very close to unity:
\begin{equation} 
  T_c \approx (\epsilon_0 - E_g)/N \; .
\label{eqrj2}
\end{equation}
Then the fraction at the excited states at $2 \le m \le N$ is $W_{ex}$
and the condensate fraction is $W_c$:
\begin{equation} 
  W_{ex} \approx T {\sum^N}_{m=2} 1/(E_m - E_g) \approx (T/\Delta)\ln N
  \approx (\epsilon_0 \ln N)/N \Delta\; ;
  \; W_c \approx 1 - W_{ex} \; ; \; \Delta =1 \; . 
\label{eqrj3}
\end{equation}
Thus, in the RJS model for a fixed initial or total energy $\epsilon_0$
the fraction of non condensate drops to zero as $W_{ex} \propto (\ln N)/N$
with growing number of system levels $N$.
When the condensate fraction is close to unity $W_c \rightarrow 1$ and
$W_{ex} \ll 1$ we see that this situation
takes place at very low temperatures $T \approx (\epsilon_0 - E_g)/N \ll E_m$
where $E_m \sim N\Delta /2$ is a typical system energy of excited levels.
This shows that the RJ condensate is very different
from the  Fr\"ohlich condensate \cite{frohlich1,frohlich2}
which  exists  at high temperatures $T \sim E_m \sim N\Delta/2$.
We have $W_c \approx W_{ex} \approx 1/2$ at $\epsilon_0 \approx N\Delta /(2 \ln N)$.
If an initially injected energy is $\epsilon_0 \sim B \Delta$
with a constant $B \sim 1$ (e.g. a few units of $\Delta$) then
with increasing $N$ almost all probability is located in the ground state
while probability at $m \ge 2$ drops as $1/N$. Thus for such an initial energy $\epsilon_0$
there is no ultraviolet catastrophe since the total probability on
levels $m \ge 2$ drops as $(\ln N)/N$.

\begin{figure}[H]
	\begin{center}
		\includegraphics[width=0.7\columnwidth]{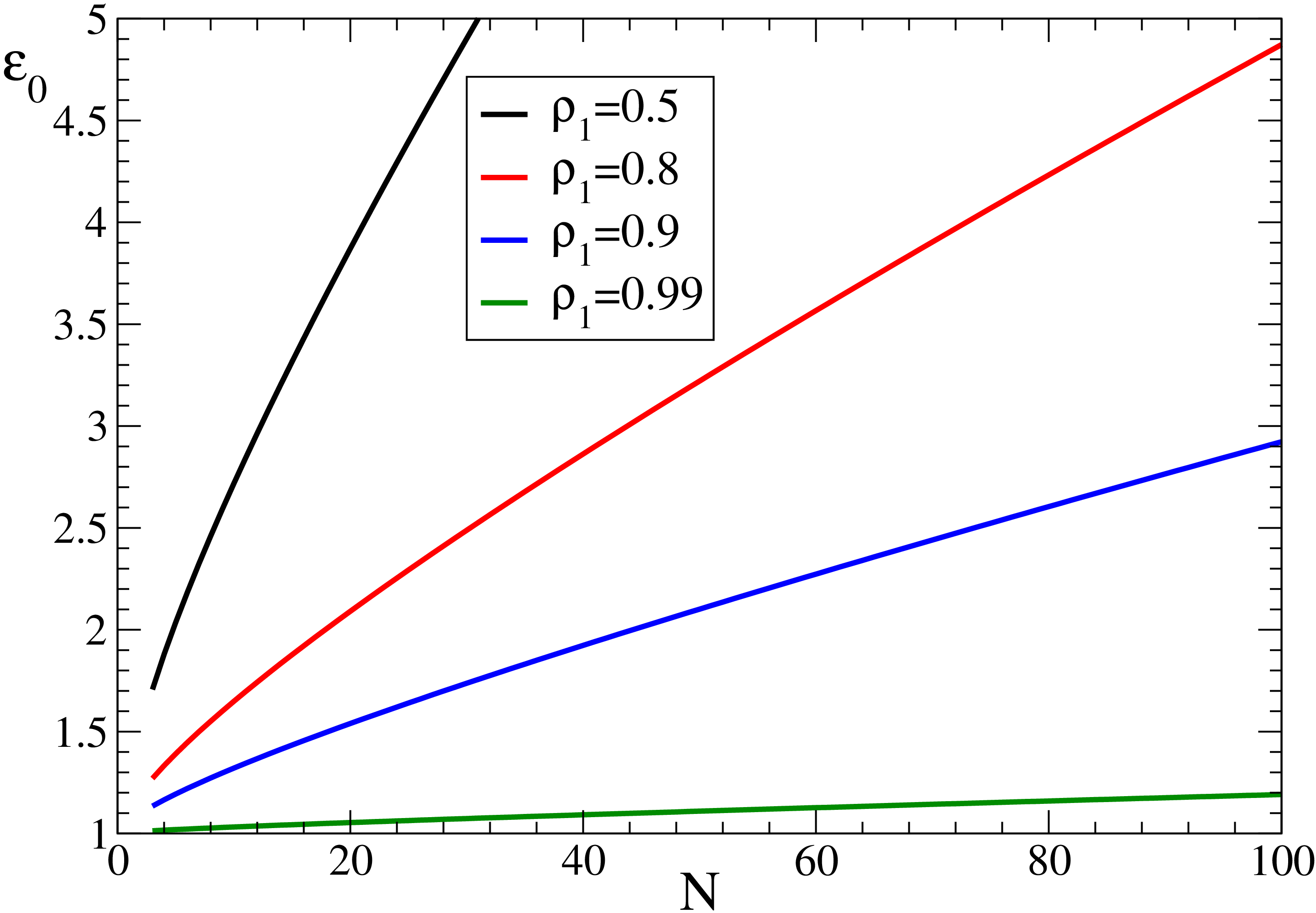}
	\end{center}
	\vglue -0.3cm
	\caption{\label{fig5} Dependence of initially energy $\epsilon_0$ on
          total number of levels in the RJS model with $E_m = m \Delta$,
          $m=1, 2. \cdots N$, $\Delta =1$ for fixed
          probability of condensate at the ground state $m=1$
          with $\rho_1 = 0.5,0.8; 0.9; 0.99$.
	}
\end{figure}

\subsection{Numerical results}

Here we present the numerical results that confirm
the estimates of previous subsection. We assume that
an initial energy $\epsilon_0$ is injected in the above
RJS model with eigenenergies $E_m =m \Delta$ of $N$ modes.
The two equalities for norm and total energy in (\ref{eqrj1})
determine temperature $T$ and chemical potential $\mu$
of the RJ thermal distribution (\ref{eqrj}).
The solution always exists, at fixed $\epsilon_0$
and $N \gg 1$ we have $\mu \rightarrow E_g$.

In Figure~\ref{fig5} we depict the curves of fixed condensate probability $\rho_1$ at
the ground state $m=1$ at different values of initial energy $\epsilon_0$
and total number of states $N$. The results show that very high values of $\rho_1 =0.9; 0.99$
can be obtained for high values of $N$ and low values of $\epsilon_0$
that corresponds to very low temperature $T \approx \epsilon_0 /N \ll \epsilon_0$.

\begin{figure}[H]
	\begin{center}
           \includegraphics[width=0.8\columnwidth]{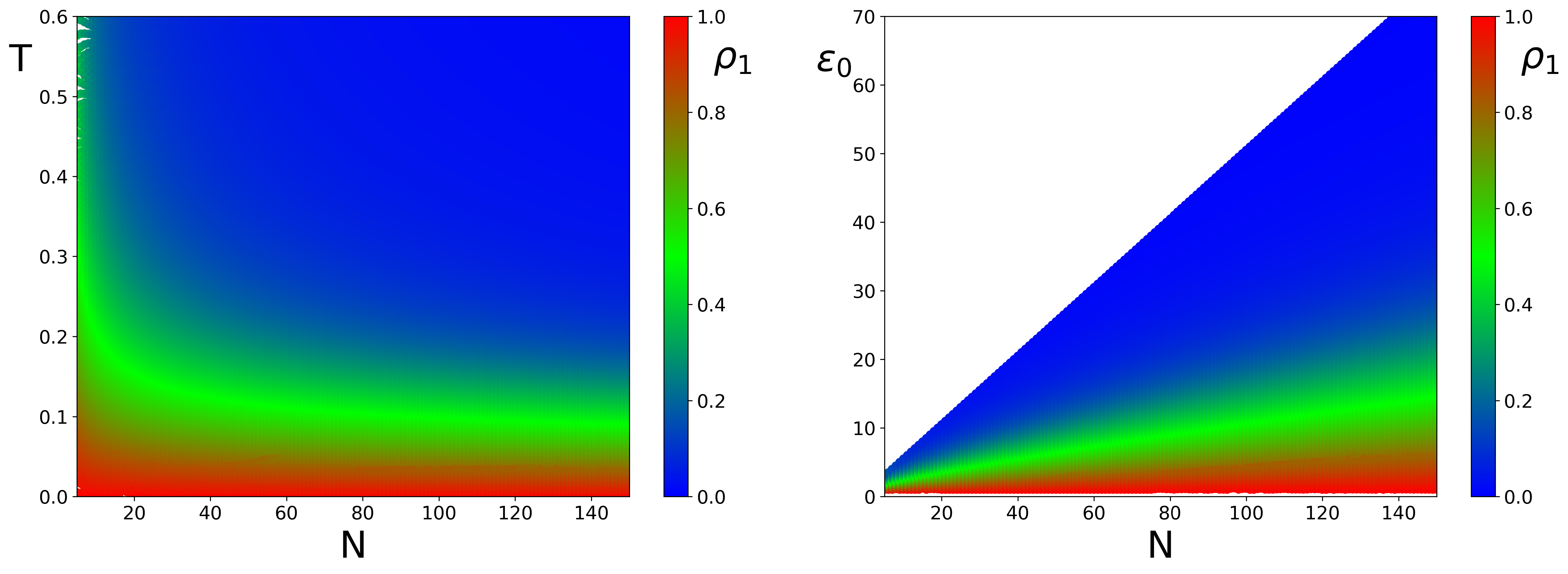}
	\end{center}
	\vglue -0.3cm
	\caption{\label{fig6} Color map of condensate probability $\rho_1$
          at corresponding system temperature $T$ and number of system modes $N$ (left panel);
          right panel shows the same $\rho_1$ as in left panel
          but in the plane $(N, \epsilon_0)$;
          $\Delta =1$ obtained in the frame of RJ thermal distribution (\ref{eqrj}).
	}
\end{figure}

Indeed,  the results of Figure~\ref{fig6} in the left panel
show that high probability of condensate $\rho_1$
can be obtained at high $N$ only at very low temperature $T < \Delta$
in agreement with above simple estimates (\ref{eqrj2}),(\ref{eqrj3}).
The right panel shows the same $\rho_1$ but in the plane $(N, \epsilon_0)$;
it also shows that the condensate exists
only at low temperature $T \approx \epsilon_0/N < \Delta$.

\begin{figure}[H]
	\begin{center}
		\includegraphics[width=0.7\columnwidth]{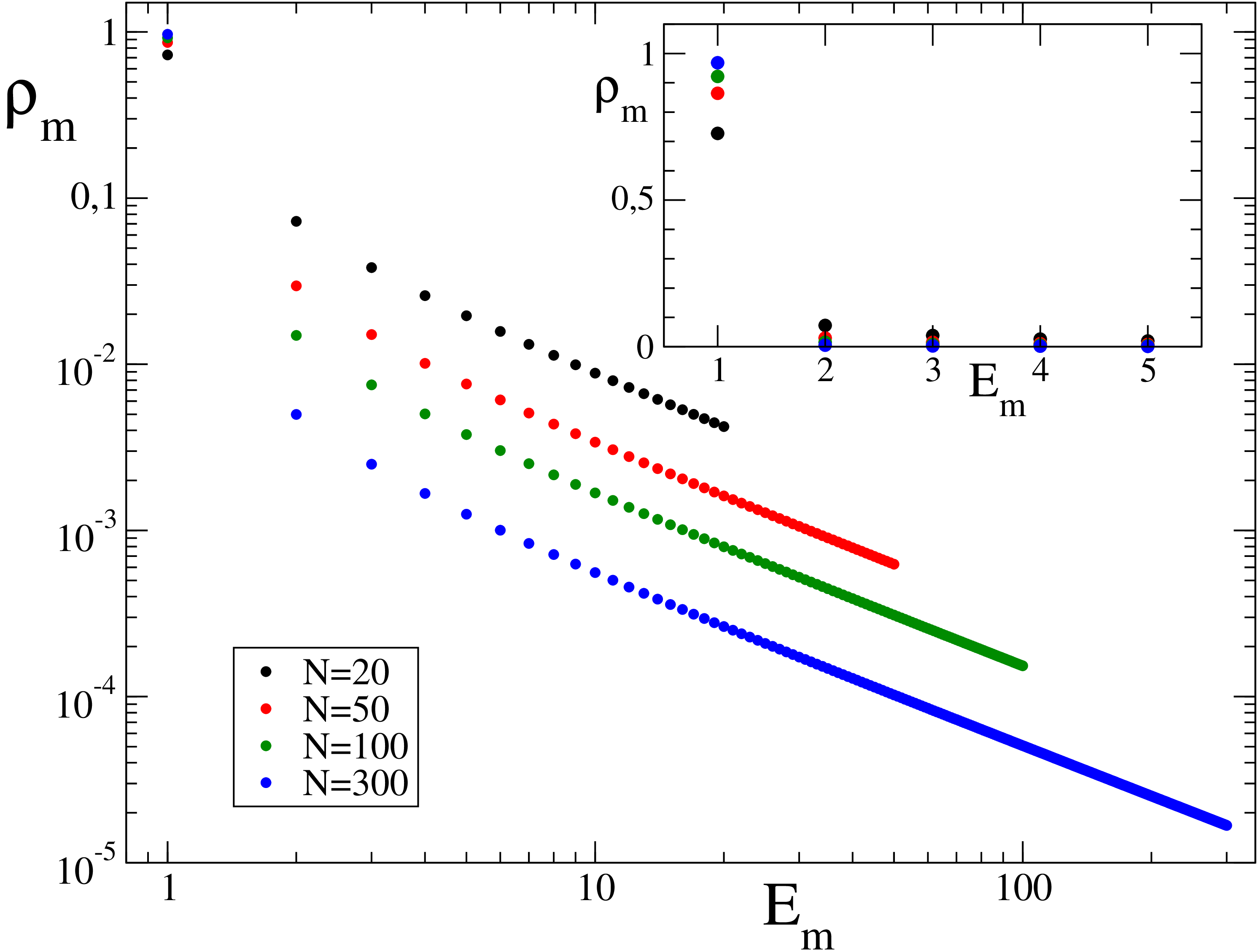}
	\end{center}
	\vglue -0.3cm
	\caption{\label{fig7} Thermal RJ distribution $\rho_m$ (\ref{eqrj})
          for various  mode energies $E_m$ at $N=20, 50, 100, 300$
          and initial total energy $\epsilon_0 \ 2.5$; $\Delta=1$;
          data is presented in log-log scale.
	}
\end{figure}

\begin{figure}[H]
	\begin{center}
		\includegraphics[width=0.7\columnwidth]{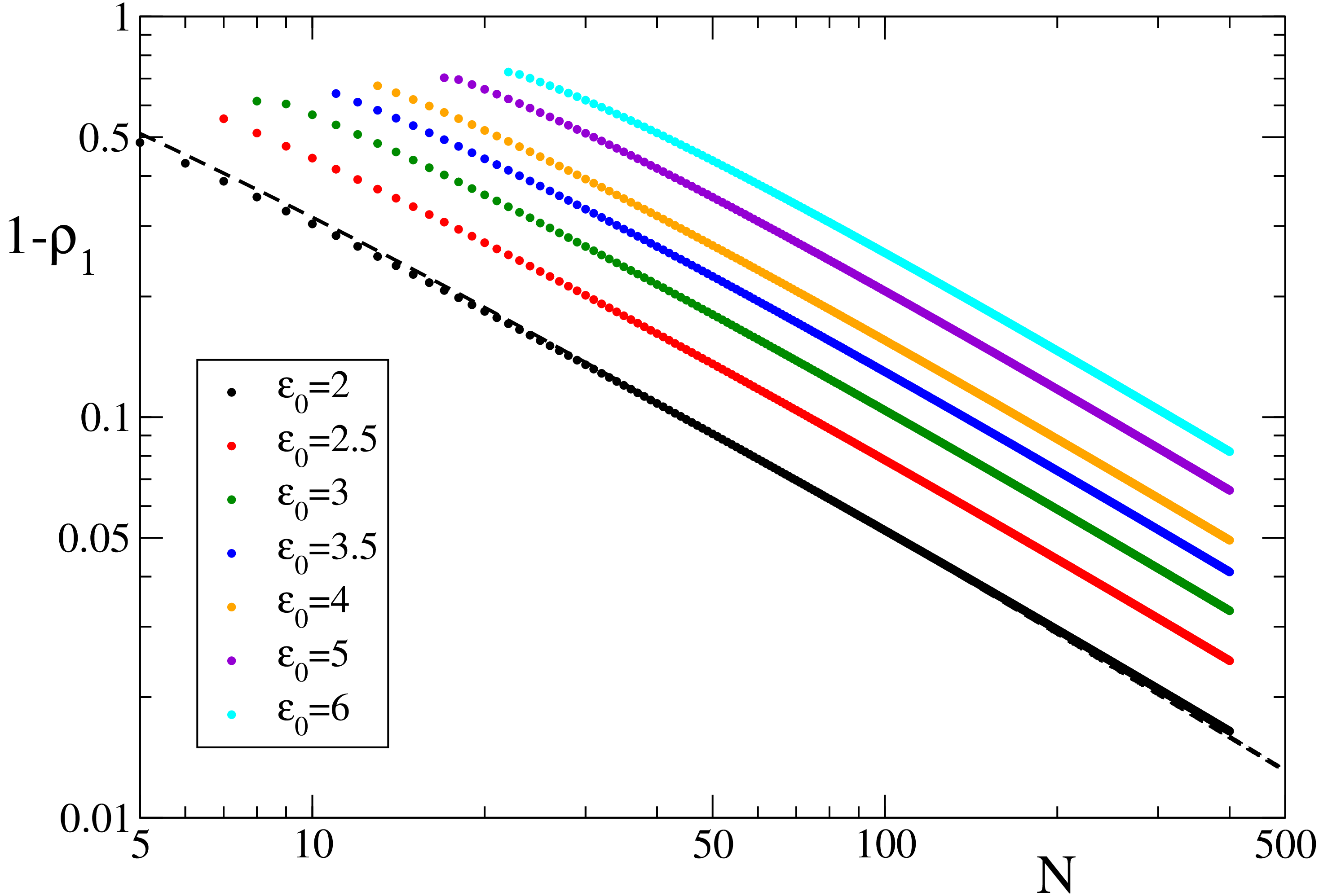}
	\end{center}
	\vglue -0.3cm
	\caption{\label{fig8} Dependence of the non-condensate probability, $W_{ex} = 1 - \rho_1$,
          on the total number of modes, $N$,
          at a fixed total energy of $\epsilon_0 = 2;2.5;3;3.5;4;5;6$.
          Here, $\rho_1$ represents the ground state population.
          The dashed curve shows the analytical theory given by the average Gumbel distribution
          with $\rho_1=\epsilon_0(\log{N}+\gamma)/N$,  for $\epsilon_0=2$
          where $\gamma$ is the Euler-Mascheroni constant (see  text).
	}
\end{figure}

The probability thermal distributions $\rho_m$ are shown
in Figure~\ref{fig7} at various total number of modes $N$
and fixed total energy $\epsilon_0 =2.5$. In agreement with
estimates (\ref{eqrj2}),(\ref{eqrj3}), the probability at $m \ge 2$ drops with increase of $N$
while condensate probability $\rho_1$ approaches unity.
The dependence of total non condensate probability $W_{ex}=1-\rho_1$
at $m \ge 2$ is shown in Figure~\ref{fig8} for several values of $\epsilon_0$.
It decreases as $W_{ex} \propto (\ln N)/N$ is agreement
with estimate (\ref{eqrj3}) and analytical result of next subsection.

We also should point that for the RJS model we consider the RJ condensate at low
positive temperatures corresponding to the case when initial energy $\epsilon_0 < N \Delta/2$.
However, there is a symmetric situation
for negative temperature being close to zero from below $T<0$
when due to symmetry (see \cite{rmtprl}) the RJ condensate appears on the top
level $m=N$ (negative temperatures appear when $\epsilon_0 > N\Delta/2$).
This RJ condensate at negative temperature appears in systems with bounded
energy spectrum $E_m$ of linear eigenmodes. For our NSE billiard (\ref{eqnse})
the spectrum $E_m$ is unbounded and due to that we do not
discuss here RJ condensation at negative temperatures (we remind that 
RJ distribution at negative temperatures are realized in fiber experiments \cite{fiber5}).

We note that a high fraction  for RJ condensate,
close to hundred's percent,  has been obtained in numerical
studies of nonlinear oscillator models
NLIRM and NLIRM plus extra diagonal  in \cite{rmtprl} (see there Fig.3 top left panel,
Fig.S4 left panels and Fig.S14 top left, right panels).
However, the dependence of this  condensate fraction
on initial energy $\epsilon_0$ and total number of oscillators $N$ was not analyzed there.

\begin{figure}[H]
	\begin{center}
		\includegraphics[width=0.85\columnwidth]{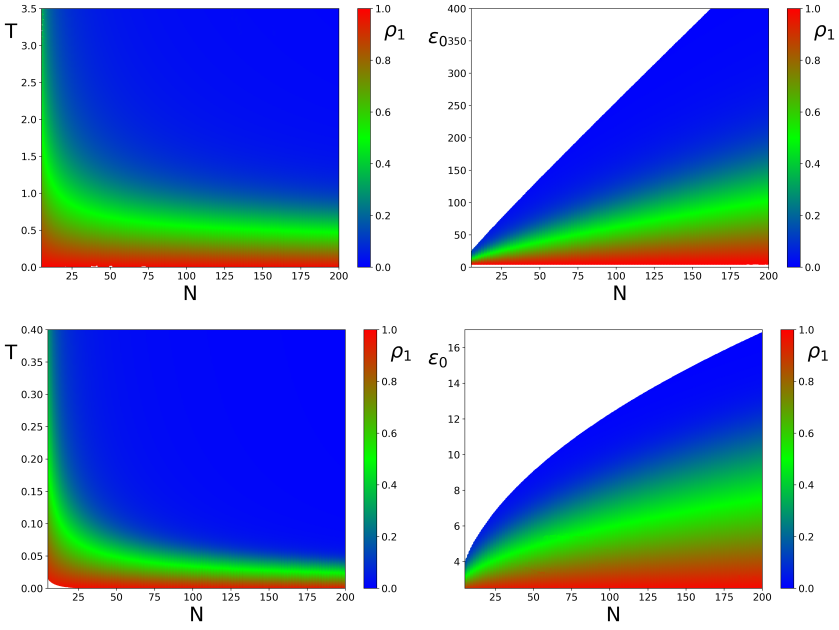}
    \end{center}
	\vglue -0.3cm
	\caption{Color plot of condensate probability $\rho_1$.
The left panels show plots of $T$ vs. $N$, 
while the right panels show the distributions of $\epsilon_0$ vs. $N$. 
The top panels display the case of the D-shape billiard, with energies $E_m$
of the D-shape fiber described in Section \ref{sec2}.
The bottom panels show the corresponding plots for the Sinai oscillator,
with energies $E_m$ from \cite{sinaiosc}.
	}
	\label{fig9} 
\end{figure}

Above we considered the case of RJS model with mode energies
$E_m = m \Delta$. But the same approach of RJ thermal distribution (\ref{eqrj})
can be used for other systems with other spectrum. Thus in Figure~\ref{fig9}, top panels,
we show the color plot of condensate probability in
the planes  $(N, T)$ and $(N, \epsilon_0)$ for the spectrum $E_m$ of linear modes
in the D-shape fiber  of Figure~\ref{fig1}
(a part of spectrum is shown in Figure~\ref{fig3}).
Here we use the modes of odd and even parity since they
become both excited due exponential instability induced
by nonlinearity in NSE. In the bottom panels of Figure~\ref{fig9}
  we show the condensate probability for the NSE in the Sinai oscillator
  studied in \cite{sinaiosc}.

For comparison between BE and RJ distributions we show in Appendix Figure~\ref{figA1}
the BE condensate probability $\rho_1$ in the RJS model of equidistant levels
at various system parameters, this Figure~\ref{figA1} is analogous to
the RJ case shown in Figure~\ref{fig6}. For the BEC case $\rho_1$ is practically
independent of number of levels due to an exponential decay
of probability at high levels with $E_m > T$ that allows to
have relatively high BEC probability $\rho_1 \approx 0.5$ even at $T > \Delta$
in contrast to the RJ case in Figure~\ref{fig6}.
We note that relations between BE and RJ condensates are discussed in the recent work
\cite{picozzi2024}.

\subsection{Classical Boltzmann entropy $S_B$ and quantum von Neumann entropy $S_q$}

The classical Boltzmann entropy $S_B= \ln \Gamma$ is determined by a phase volume $\Gamma$ of
microscopic states at a given energy \cite{boltzmann1,mayer,landau}.
For multimode fiber with a finite number of modes,
as e.g. for the above RJS model, it was shown
\cite{picozzi2,picozzi3,picozzi2024,christo2021} that
\begin{equation} 
S_B = \sum_m \ln (\rho_m) + C \;\; ,
\label{eqentB}
\end{equation}
where $C$ is some constant that
can make entropy to be positive, 
but we take $C$ to be zero.
We note that the total number of particles, or norm,
and the total energy $E=\epsilon_0$ are preserved in our system
and thus these two integrals determine the functions
$T(E)$ and $\mu(E)$ entering in RJ distribution (\ref{eqrj}).
From the above expression for $S_B$ and RJ thermal distribution (\ref{eqrj})
it follows that $S_B$ satisfies the standard thermodynamic relation \cite{landau}
${dS_B}/dE = 1/T$ where $E$ is the total system energy on which depends
temperature $T(E)$ and chemical potential $\mu(E)$ from the expression for $\rho_m$ (\ref{eqrj}).
Indeed, the expression for ${dS_B}/dE$ directly follows from (\ref{eqrj}), (\ref{eqrj1})
that gives: $dS_B/dE = \sum_m (d \rho_m/dE)/\rho_m = (N dT/dE)/T + (d \mu/dE)/T = 1/T$.
We note that the relation $dS_B/dE =1/T$ works for any
sequence of level  energies $E_m$. 

The expression (\ref{eqentB}) for $S_B$ can be used
during the time evolution 
of the initial distribution $\rho_m(t)$ over modes $m$ at $t=0$
and  higher $t$, 
so that we have time dependence $S_B(t)= \sum_m \ln\ (\rho_m(t))$.
In \cite{picozzi2,picozzi3} it was shown that for
a fiber with a finite number of modes $N$
nonlinearity leads to a monotonic growth of
$S_B(t)$ with time in agreement with
the Boltzmann H-theorem \cite{boltzmann1,mayer,landau}. 

\begin{figure}[H]
	\begin{center}
		\includegraphics[width=0.85\columnwidth]{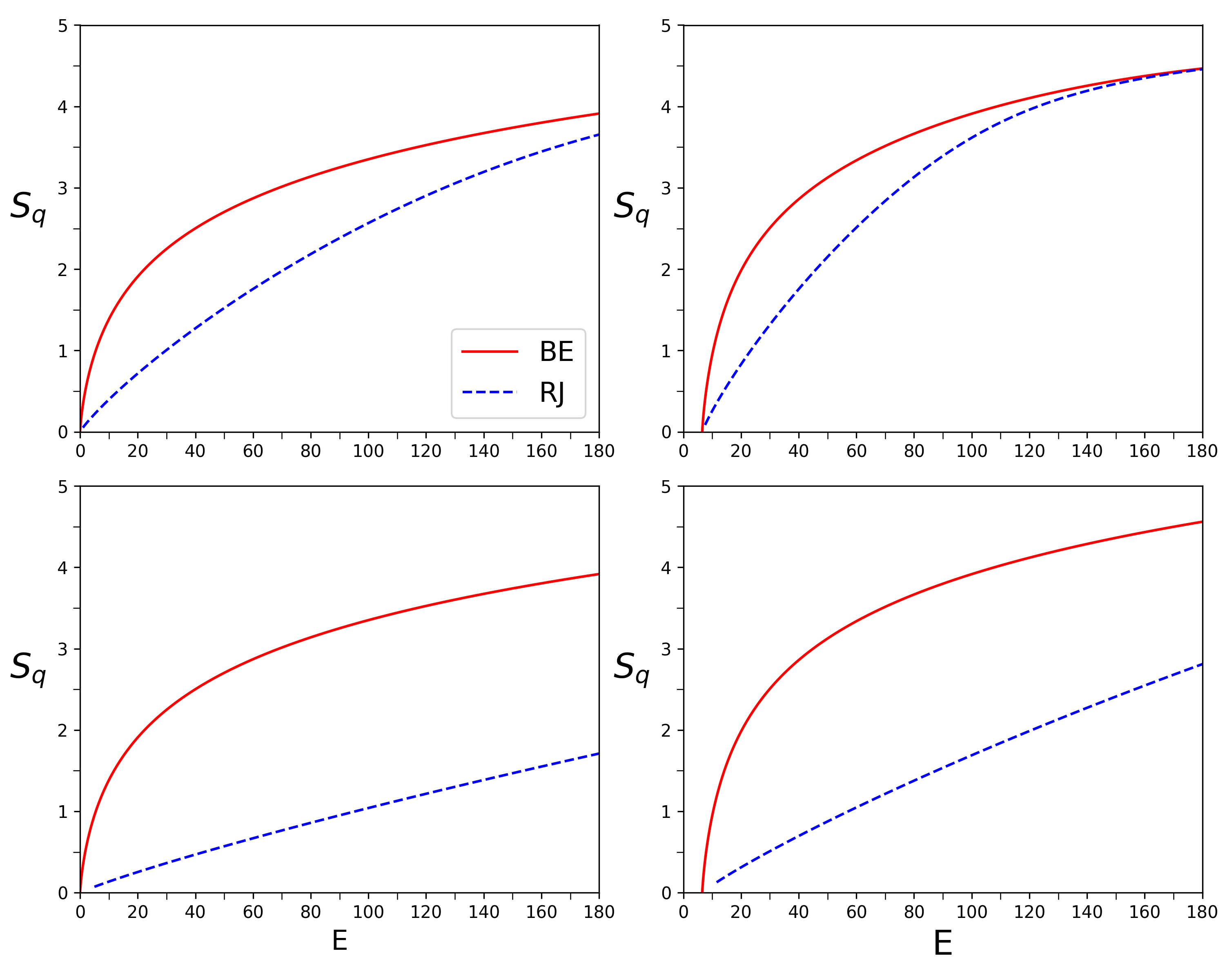}
    \end{center}
	\vglue -0.3cm
	\caption{\label{fig10}  Quantum von Neumann entropy $S_q(E)$ as a function of
    energy ($E$) for Bose-Einstein (BE) distributions 
    (red solid line, Eq. \ref{eqbe}) 
    and Rayleigh-Jeans (RJ) distributions 
    (blue dashed line, Eq. \ref{eqrj}). 
    The left panels show the RJS model 
    with equidistant energy levels, 
    while the right panels are for energy states 
    given by the $D$-billiard. 
    The top panels show the case of $N = 100$ states, 
    and the bottom panels for $N = 500$ states.
    For the D-shape billiard the lowest energies
    are $E_1 =6.55, E_2=15.18, E_3=17.8, E_4=28.59, E_5 = 28.65$.
    	}
\end{figure}

At $\beta =0$ we have a quantum system (\ref{eqnse})
and it is tenting to assume that the small or moderate nonlinear $\beta$-term
acts as a some kind of dynamical nonlinear thermostat \cite{eplbill,sinaiosc}.
In such a case the distribution $\rho_m$
of probability over billiard eigenmodes
would be described by the BE thermal distribution (\ref{eqbe})
with the system quantum  Neumann entropy \cite{neumann} (see also \cite{mayer,landau,petz}):
\begin{equation} 
S_q = - \sum_m \rho_m\ln (\rho_m) \;\; .
\label{eqentq}
\end{equation}
However, in the system (\ref{eqnse}) there is no second quantization
and in fact we have a nonlinear dynamics of classical field where the thermal distribution
of $\rho_m$ is given by the RJ relation (\ref{eqrj})
as it was shown for nonlinear perturbation of RMT \cite{rmtprl}.
Thus in our classical system (\ref{eqnse})
with the RJ distribution (\ref{eqrj}) we have the Boltzmann entropy $S_B$
that satisfies the thermal relation $dS_B/DE=1/T$
while for the quantum entropy $S_q$ (\ref{eqentq}) and RJ distribution (\ref{eqrj})
we have $dS_q/dE \neq 1/T$.

However, it is still useful to study the properties of quantum entropy $S_q$
since it gives the effective number of populated eigenmodes
$\Delta m_{eff} = \exp(S_q)$. The time dependence $S_q(t)$ is given
by the relation (\ref{eqentq}) where one should take
$\rho_m(t)$ at a given moment of time so that
$S_q(t)=-\sum_m \rho_m(t) \ln (\rho_m(t))$ .

In the works \cite{rmtprl,eplbill,sinaiosc} the dependence of
quantum entropy on system energy $S_q(E)$
was used to obtain a distinction between
BE (\ref{eqbe}) and RJ (\ref{eqrj}) distributions.
Thus in Figure~\ref{fig10} we present these two theoretical dependencies for the cases
of RJS model of equidistant levels (left panels)
and the system with energies of D-shape billiard.
The theory shows that there is a significant difference between two
dependencies that grows with the increase of total number $N$ of levels in the system
that boosts the condensate fraction (see Figure~\ref{fig8}).
At low energies $E$ the RJ entropy is significantly smaller
compared to the BE case due to RJ condensation.

The results obtained from the NSE dynamics (\ref{eqnse}) determine
the dependences $S_q(E,t)$ and $S_B(E,t)$ which are discussed in the next Section~\ref{sec5}.

\subsection{Analytical results}

To describe analytically the population of eigenstates in the thermodynamic
limit of a large number of modes $N$, we assume a microcanonical ensemble
for a simplified model with evenly spaced eigenenergies
$\epsilon_n = n-1$. In this case the total energy is given by:
\begin{equation}
\epsilon = \sum_{n=1}^{N} (n-1) \rho_n 
\end{equation}
while the total population normalized to $\sum_{n=1}^{N} \rho_n = p$.

The probability distribution of the populations
$\rho_n$ can be obtained from the partition functions:
\begin{equation}
  s_N(\epsilon, p) = N! \int_{\rho_i>0} d\rho_1 ... d\rho_{N}
  \delta( \epsilon - \sum_{n=1}^{N} (n-1) \rho_n ) \delta(p - \sum_{n=1}^{N} \rho_n)
\end{equation}
with the short notation $s_N(\epsilon) =   s_N(\epsilon, p=1)$. 

For example the probability distribution of
the ground state population $p_1^{(N)}(\rho_1)$ is given by:
\begin{equation}
  p^{(N)}_1(\rho_1) = N (1 - \rho_1)^{N-3} \frac{s_{N-1}( (\epsilon + \rho_1 - 1)/(1 - \rho_1) ) }
  { s_N(\epsilon) } \label{eq:p0exact}
\end{equation}

The following analysis has close similarities with the approach from \cite{Evans}.
The Laplace transform of $s_N(\epsilon)$ can be easily computed:
\begin{eqnarray}
  {\hat s}_N(\beta, \mu)& = & 
\int_0^{\infty} d\epsilon e^{-\beta \epsilon}  
\int_0^{\infty} dp e^{-\mu p} s_N(\epsilon, p) \\
    &=& N! \prod_{n=0}^{N-1} \frac{1}{\beta n + \mu}\\
\end{eqnarray}
we now need to inverse the Laplace transform to recover $s_N(\epsilon, p)$.

The inverse Laplace transform over $\mu$, is known:
\begin{equation}
  {\cal L}_\mu^{-1} \left( N! \prod_{n=0}^{N-1} \frac{1}{\beta n + \mu} \right)
  = N \left( \frac{1 - e^{-p \beta}}{\beta} \right)^{N-1}
  \label{Lmu}
\end{equation}

For a fixed $N$ we can find the exact inverse Laplace transform over
$\beta$ expanding the power in Eq.~(\ref{Lmu}), this gives the explicit formula:
\begin{equation}
  s_N(\epsilon) = \frac{N}{(N-2)!} \sum_{n=0}^{N-1} C^{n}_{N-1} (-1)^n (\epsilon - n)^{N-2}
  \theta(\epsilon - n) \label{eq:snexact}
\end{equation}
but which is not convenient to study the $N \gg 1$ limit.

\begin{figure}[H]
	\begin{center}
		\includegraphics[width=0.85\columnwidth]{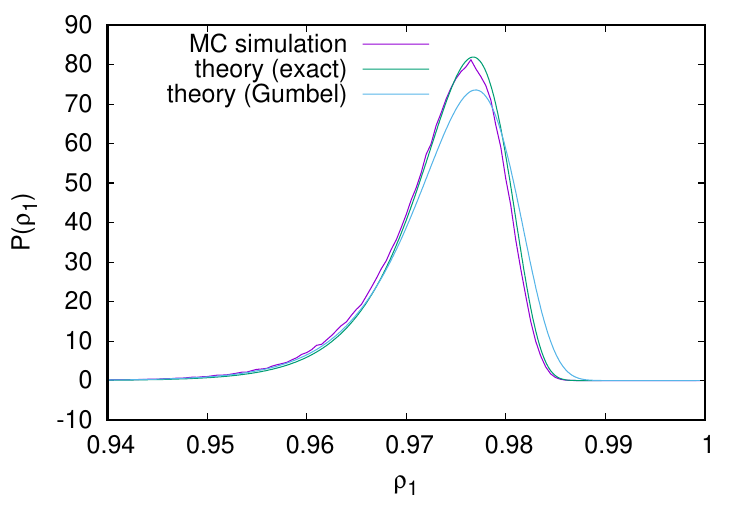}
    \end{center}
	\vglue -0.3cm
	\caption{\label{fig11} 
          Probability distribution $P(\rho_1)$ for fluctuations of RJ condensate population
          at the ground state $\rho_1$ for the RJS model of $N$ equidistant energy levels
          in RJ thermal distribution (\ref{eqrj})  with  energy and
          norm conservation, comparing Monte Carlo simulations,
          exact analytical expression Eq.~(\ref{eq:snexact}) and
          asymptotic Gumbel distribution Eq.~(\ref{eq:p0gumbel}). }
\end{figure}

Instead, in this case we perform a saddle point approximation on the inverse Laplace transform integral:
\begin{equation}
  s_N(\epsilon, p) = \frac{N}{2 \pi i} \int d\beta \exp\left( \epsilon \beta + (N-1) \log
  \frac{1 - e^{-p \beta}}{\beta}  \right)
\end{equation}
We now fix $p=1$, in this case the saddle point approximation gives the following equation for $\beta$:
\begin{equation}
\frac{N-1}{e^{\beta}-1} - \frac{N - 1}{\beta} + \epsilon = 0
\end{equation}
Using the asymptotic expansion near the saddle point in the $\epsilon \ll n$ limit,
we can find an analytic expression for $p_1(x_1)$:
\begin{equation}
  p_1(\rho_1) = \frac{N}{\epsilon} f(y) \;,\; y = \frac{N}{\epsilon}
  \left(1 - \frac{\epsilon \log N}{N} - \rho_1\right) \;,\;   f(y) = e^{-y - e^{-y}}
\label{eq:p0gumbel}
\end{equation}
where $f(.)$ is the Gumbel distribution. 
The Gumbel distribution often appears in the description of rare events.
It seems here it occurs because we are looking at the intersection of two microcanonical ensembles
at fixed energy and at fixed population which is a rare event. 
This asymptotic agrees very well with both Monte Carlo simulations and
the exact Eq.~(\ref{eq:snexact}) (see Fig.~\ref{fig11}).

The mean of the Gumbel distribution is given by the Euler-Mascheroni constant $\gamma = 0.57721$ 
This gives the following asymptotic expression for the population of
the mean ground state population $\langle \rho_1 \rangle$: 
\begin{equation}
\langle \rho_1 \rangle = 1 - \epsilon \frac{\log N + \gamma}{N}
\end{equation}
which coincides from the behavior expected from the Rayleigh-Jeans distribution.
Thus the condensate fraction $\langle \rho_1 \rangle$ tends to 100\% in the $N \gg 1$ limit.
In the context of the time evolution of non-linear Schr\"odinger equation,
this would correspond to a slow depletion eigenmodes at intermediate energies,
in favor of a small and continuously decreasing population of modes of higher energy. 

It is interesting to know what can limit the condensation in the large $N$ limit
in this statistical framework. Including a non-inform eigenenergy energy distribution
as well as a weak nonlinearity in the expression of the total energy $\epsilon$
as function of the eigenlevel populations $\rho_n$ does not seem to stop condensation
as long as there is a gap between the ground state and higher excited states.
A possible constraint that can limit the condensation fraction is the requirement
that the population $\rho_{i}$ do not enter the stability island around
the functional ground-state of the nonlinear Schr\"odinger equation.
Since we do not know the equation for the boundary of this stability island
we cannot perform realistic simulations, but we checked that simple approximation
for the shape of the stability island indeed limit the condensation.
For example, if the stability island imposes $\sum_{i=2}^{N} \rho_{i} > \rho^*$,
then the condensate fraction will converge to $1-\rho^*$,
with $\rho_{2} \rightarrow \rho^*$ in the limit $N \gg 1$. 

\subsection{RJ condensate and Fr\"ohlich condensate}

  To summarize, for  all systems  discussed  above we see that
  the RJ condensate fraction can be high
  only at low temperature $T \approx \epsilon_0 /N $ being comparable
  with a typical energy level spacing (see Figures~\ref{fig6},~\ref{fig9}).
  In contrast to this condition $T \ll \Delta \sim E_m$
  the BE thermal distribution (\ref{eqbe})
  can be approximated by the RJ distribution (\ref{eqrj})
  only at high temperature $T \gg E_m$.
  Thus it is not possible to start to approximate
  the BE distribution (\ref{eqbe}) of a quantum system at high temperature,
  obtain the RJ distribution (\ref{eqrj}) and then
  to have RJ condensate since it appears only at small
  temperature $T \sim T_c \sim \Delta$ when the approximation
  of BE by RJ distribution is not valid.
  In fact such a logical scheme was used by Fr\"ohlich in
  \cite{frohlich1,frohlich2}. Thus Fr\"ohlich
  in fact follows this RJ approximation for BE at high
  temperatures and then finds that RJ distribution
  has the condensate phase
  (that formally is not logically correct
  since RJ condensate exists only at low temperature
  while transition from BE to RJ distribution
  is valid only at high temperature).
  However, at the same time  Fr\"ohlich
  assumes that there is external energy pumping of the system
  which is absorbed by certain dissipative processes
  in the system so that the conditions of possible realization
  of  Fr\"ohlich condensate are significantly complicated
  by the system couplings with external sources
  while the RJ condensate appears in
  a completely isolated system.
  This system is a Hamiltonian conservative system
  described by classical nonlinear fields
  so that from the begging the thermal distribution
  of this classical system is given by the RJ distribution (\ref{eqrj})
  without any links with the quantum BE distribution (\ref{eqbe}).

\section{Dynamical thermalization for defocusing fiber NSE}
\label{sec5}

\subsection{Thermalization features}

Here we present the numerical results for the time evolution
in the defocusing NSE (\ref{eqnse}) for  typical values of moderate
and weak nonlinearity $\beta =10, 1, 0.5$. Due to chaos and exponential
instability of motion the numerical integration corrections in NSE
lead to symmetry parity breaking and for an initial odd state
with $\psi(x.y) = - \psi(x,-y)$ the even component
of positive parity $\psi(x,y)=\psi(x,-y)$ appears with time.
Due to this we state with an initial state
which contains from the beginning at $t=0$ both parities
using usually $\psi(t=0) =(\psi_m + \psi_{m+1})/\sqrt{2}$.

\begin{figure}[H]
	\begin{center}
		\includegraphics[width=0.85\columnwidth]{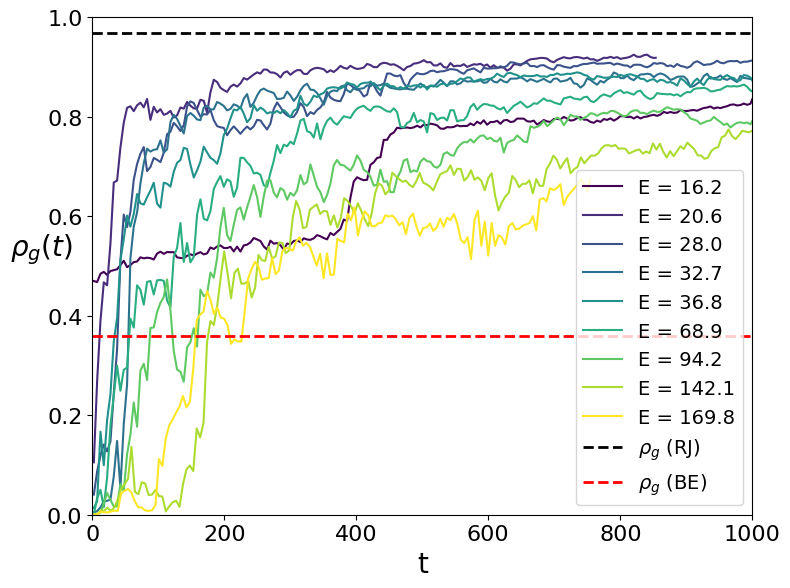}
    \end{center}
	\vglue -0.3cm
	\caption{\label{fig12} Time dependence of the ground state probability $\rho_g(t)$ shown up
          to $t = 1000$ for various initial states. The color of each curve corresponds to
          the initial total state energy, as indicated in the legend;
          the initial states are $(\psi_m + \psi_{m+1})/\sqrt{2}$ for
          $m=1; 2; 3; 4; 5; 11; 16; 26; 31$
          (from minimal to maximal total energies $E$ given in the figure panel);
          the plot shows a temporal average over time interval $\delta t = 1$; here $\beta=10$.
          RJ and BE ansatz are shown for $\rho_g$ by the  black and red dashed lines respectively
          for $E=20$ and $N=500$; the values of parameters obtained for both ansatz
          are $T=0.0269$, $\mu=6.52$ for RJ and $T=16.27$, $\mu=-15.1$ for BE.     
  }
\end{figure}

The time evolution of the projection probability to the linear
ground state $\rho_g(t) = \rho_1(t) = |<\psi(x,y,t)|\psi_1(x,y)>|^2$ is shown
for various initial states  in Figure~\ref{fig12} at $\beta=10$.
The initial state with a minimal total energy $E=16.2$
is $\psi(t=0) = (\psi_1 + \psi_2)/\sqrt{2}$ so that $\rho_g(t=0)=0.5$.
For all other initial states we have $\rho_g(t=0)=0$.
The results of Figure~\ref{fig12} show
that for all initial states the 
probability $\rho_g$ approaches to its limiting
value being approximately in the range $0.75 \leq \rho_g \leq 0.9$.
Thus  the probability fraction of RJ condensate at the linear
ground state is very high but still it does not reach
so high values as in the case of RJS model
(see Figure~\ref{fig8} with the maximal value $\rho_g=\rho_1 \approx 0.98$).
We discuss the reason of this limiting value of
$\rho_g$ at the end of this Section. 
For the states with higher initial energy, e.g. $E=142.1; 169.8$
the values $\rho_g$ are a bit smaller compared to those of lower energy,
e.g. $E=20.6$. We attribute this to the fact that it takes more time
to reach steady-state RJ thermal distribution (\ref{eqrj})
from high energy states
with a transfer of probability from high energy
to the ground state. In Figure~\ref{fig12} we also
show the expected values of $\rho_g$ according to the BE ansatz  (\ref{eqbe})
and to the RJ ansatz RJ (\ref{eqrj}) assuming the total energy is
$E=20$ (and total number of states $N=500$ for the RJ case).
We have for the BE ansatz $\rho_g \approx 0.38$ while for RJ ansatz
$\rho_g \approx 0.98$ being very close to unity thus  showing a drastic
difference between BE and RJ thermalization.

In global the results of Figure~\ref{fig12}
demonstrate a good agreement with the RJ thermal condensate theory.

\begin{figure}[H]
	\begin{center}
		\includegraphics[width=0.85\columnwidth]{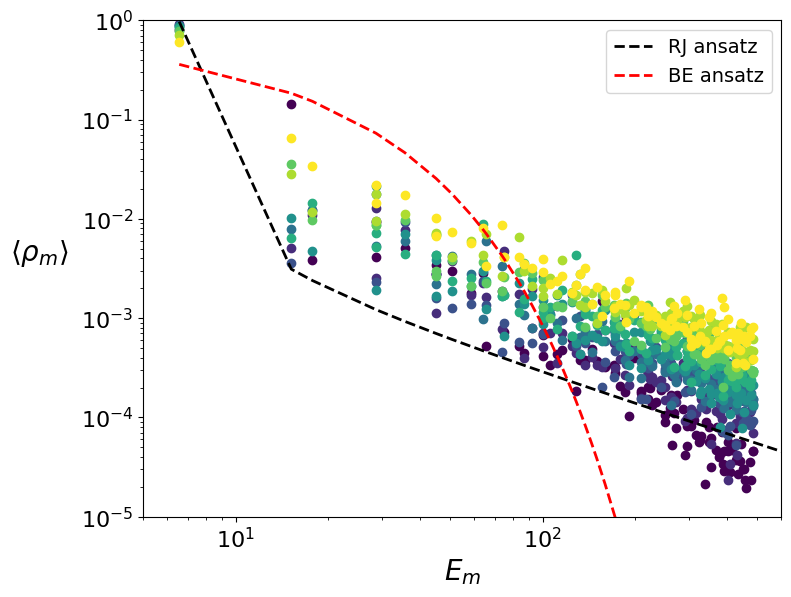}
    \end{center}
	\vglue -0.3cm
	\caption{\label{fig13}
                    Probability at eigenstates $\psi_m$ ($\langle \rho_m \rangle$)
          as a function of linear eigenstate energy $E_m$ for different initial states
          and averaged over the time interval $t = 500$ to $t \leq 1000$;
          other parameters are as in Figure~\ref{fig12}.
          Colors correspond to the same initial states as in \ref{fig12};
          RJ and BE ansatz parameters are shown by dashed curves described
          in Figure~\ref{fig12}.
}
\end{figure}

In Figure~\ref{fig13} we show the probability distribution
$<\rho_m(t)>$ over linear eigenmodes $\psi_m$
averaged over a long time interval $500 \leq t \leq 1000$
for the same initial states as in Figure~\ref{fig12} at $\beta=10$.
For all initial states we have a high fraction of RJ condensate
at the linear ground state $\psi_g=\psi_1$. The probabilities $\rho_m$
at higher modes with $m>1$ approximately follow the RJ thermal distribution (\ref{eqrj})
which is characterized by a decay $\rho_m \sim 1/(E_m -\mu)$
shown by a dashed black curve for initial energy $E=20$.
The BE ansatz at the same energy $E=20$ is also shown in Figure~\ref{fig12}
by the dashed red curve and it is obviously very different from
the presented numerical results. Even if there are visible fluctuations of $\rho_m$
values, which we attribute to not very long times reached in heavy numerical simulations,
on average we show that the dynamical thermalization and chaos
in the defocusing fiber NSE (\ref{eqnse})
leads to the RJ thermal distribution (\ref{eqrj}) providing
a good description of numerical results including the formation of RJ condensate.

\begin{figure}[H]
	\begin{center}
		\includegraphics[width=0.85\columnwidth]{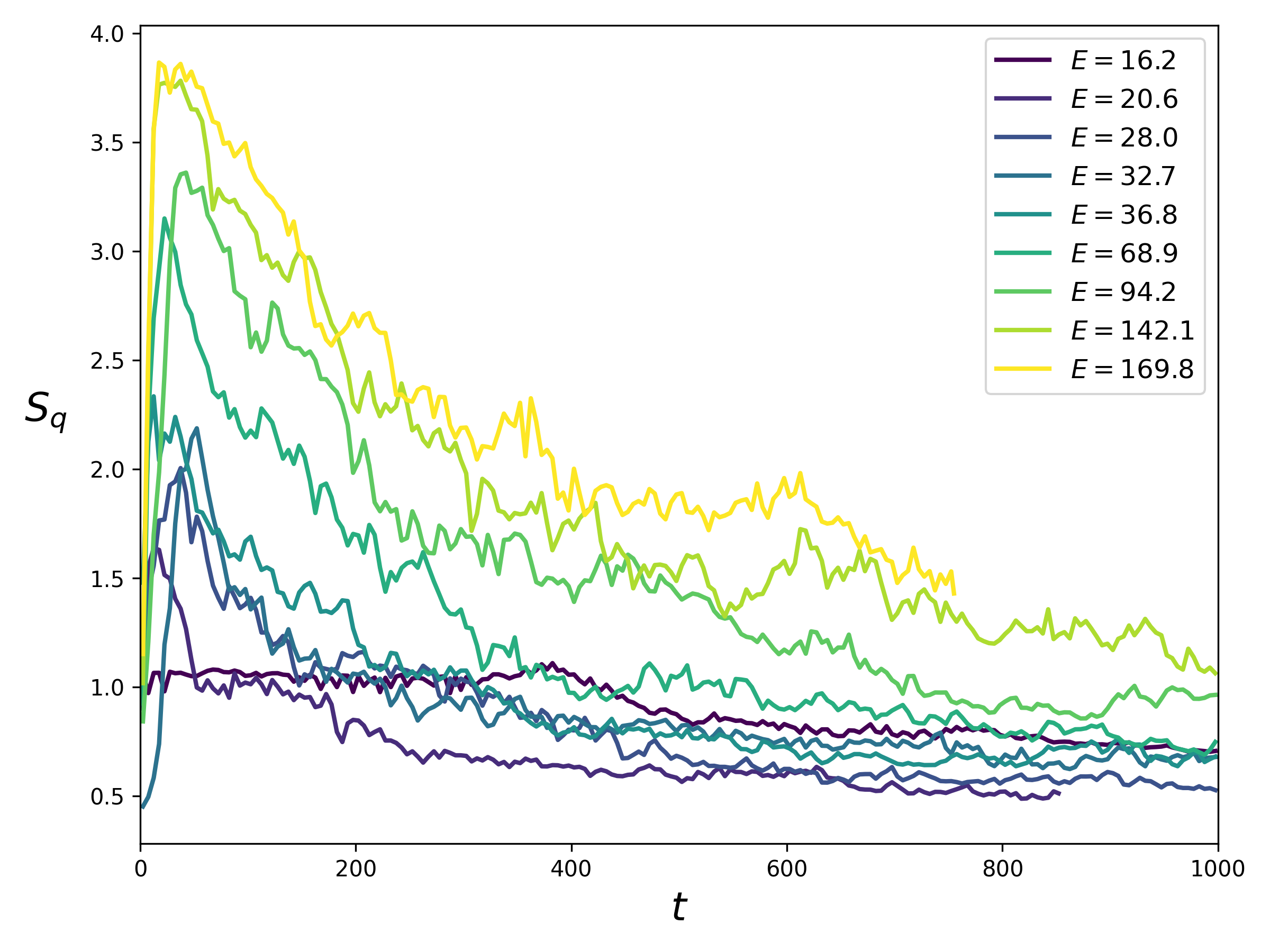}
    \end{center}
	\vglue -0.3cm
	\caption{\label{fig14} Dependence of quantum von Neumann entropy
          $S_q(t)= - \sum_m \rho_m(t) \ln \rho_,(t)$ on time $t$
          for $\beta=10$ and other parameters as in Figure~\ref{fig12};
          values of $S_q$ are averaged in the time window $\delta t =5$
          to reduce the fluctuations; short time behavior of
          $S_q(t)$ is shown in Appendix Figure~\ref{figA2}.
}
\end{figure}

The Boltzmann H-theorem of 1872 \cite{boltzmann1}, see also e.g. \cite{mayer,landau},
states that during a thermalization process
in a classical system entropy $S_B(t)$ cannot decrease with time $t$,
thus $S_B(t)$ can only  grow with time or stay constant at
its maximal value in a thermal equilibrium.
For the quantum von Neumann entropy $S_q(t)$
we have a different situation
in our system (\ref{eqnse}) with the dynamical thermalization and RJ condensation.
Thus the results presented in   Figure~\ref{fig14}, obtained at
the same parameters as in Figures~\ref{fig12},~\ref{fig13},
show that the quantum von Neuman entropy $S_q(t) = - \sum_m \rho_m(t) \ln \rho_m(t)$
initially grows with time, reaches a maximal value
and then decreases up to a certain smaller value
(being up to a factor 3 smaller than at maximum) 
corresponding to the RJ thermal distribution (\ref{eqrj})
with RJ condensate.

Our qualitative explanation
of this surprising behavior of $S_q(t)$ is the following:
initially $S_q(t\approx 0) \approx \ln 2 $ since initially
two linear eigenmodes are populated, then
at higher times transitions to other
linear eigenmodes are produced by nonlinear $\beta$ term
and $S_q$ is increased reaching its maximal value,
but then a slow thermalization
due to chaos leads to the RJ thermal distribution (\ref{eqrj})
where the fraction of RJ condensate is close to unity
(see Figure~\ref{fig12}), since RJ condensate fraction
is so high the entropy $S_q$ of steady-state has a relatively small
value being significantly smaller than its maximal value.

The above time dependence of $S_q(t)$ is not in a contradiction
with the Boltzmann H-theorem formulated for the Boltzmann entropy
$S_B(t) = \sum_m \ln \rho_m(t)$. However, the time dependence of the Boltzmann
entropy $S_B(t)$  shown in Appendix Figure~\ref{figA3}
is also non-monotonic in contradiction with the H-theorem.
The maximal value of $S_B$ is about 50\% higher than its value at high times
where the thermalization takes place.
Such a behavior is found bith for $\beta > 0$  and $\beta <0$
(see Appendic Figure~\ref{figA3}).
At the same time we should point on the problem of
the definition of $S_B$ in our system (\ref{eqnse})
with infinite number of linear eigenmodes $1 \leq m \leq \infty $.
Indeed, it is clear that probabilities $\rho_m$ at high modes should be
small that leads to a formal divergence for $S_B$ value
given by the above expression. Due to this reason we present
$S_B(t)$ time dependence in Appendix Figure~\ref{figA3}
where $S_B$ is computed by a summation over
a finite interval of eigenmodes with $1 \leq m \le m_{max}$.
The change of $m_{max}$ does not eliminate the non-monotonic
time dependence of $S_B(t)$. Strongly negative values of $S_B$
can be in principle compensated by
an arbitrary constant $C$ in the definition of $S_B$ in \ref{eqentB}).
There is also the problem of $S_B$
definition at small initial times
$t \ll 1$ but it can be avoided by taking average over
a time interval that in any case should be used
to suppress fluctuations.

We note that the numerical simulations of defocusing NSE presented in \cite{picozzi3} 
show a monotonic increase of $S_B(t)$. However, the numerical simulations
there were done with a finite number of about 40 modes
present in a fiber. Our numerical simulations here
with FREEFEM codes have much larger number of eigenmodes
that may be at the origin if non-monotonic dependence
of $S_B(t)$ found here in Appendix Figure~\ref{figA3}.
Formally, it is possible to say that the H-theorem
is formulated
for a gas of many particles and there is no
guaranty that it should be valid for the NSE case
of our system (\ref{eqnse}).
At the same time we should note that
our system (\ref{eqnse}) in the thermalized regime
is always located in the RJ condensate phase
since number of levels is formally unlimited here.
It is also possible that we do not reach complete thermalization
on very high eigenmodes $m$ that gives a significant
contribution to $S_B$ that may be also responsible for non-monotonic
time dependence of $S_B(t)$. We restrict ourselves by
pointing on the non-monotonic time dependence of
Boltzmann entropy $S_B(t)$ and leaving the resolution
of this problem to future studies.

\begin{figure}[H]
	\begin{center}
		\includegraphics[width=0.85\columnwidth]{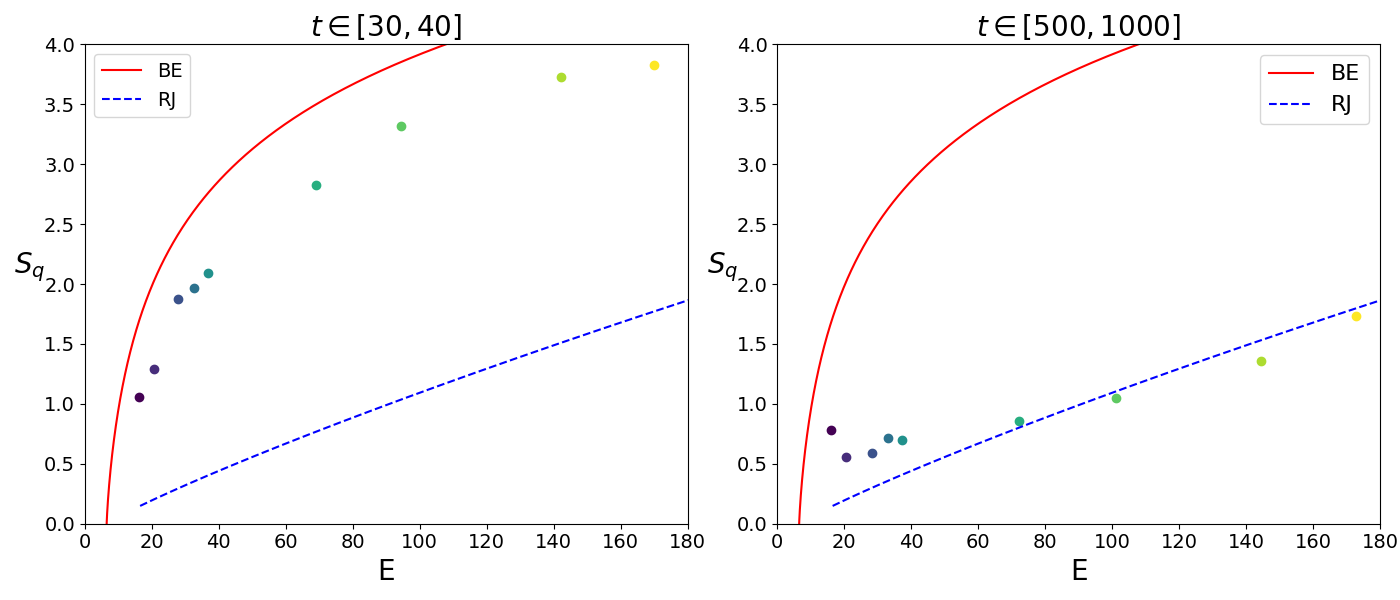}
    \end{center}
	\vglue -0.3cm
	\caption{\label{fig15} Dependence of quantum von Neumann entropy $S_q$ on energy $E$
          is shown by points 
for the initial states of Figure~\ref{fig12} and Figure~\ref{fig13}.
The left panel displays the time average for times between 
30 and 50, while the right panel shows the average for times
between 500 and 1000; here as in Figure~\ref{fig12}
$E$ is the total energy.
In both panels, the curves are shown for the Bose-Einstein (BE) 
and Rayleigh-Jeans (RJ) distributions for the billiard 
with $N=1000$.
}
\end{figure}

In Figure~\ref{fig15} we compare the dependence of quantum von Neumann entropy $S_q$ values, averaged over
a certain time interval at low and high times, on initial energy of linear modes
$E_m$. On short times $30 \leq t \leq 40$ this dependence $S_q(E_m)$ is close to those 
of one given by the BE thermal distribution (\ref{eqbe})
while at longer times $500 \leq t \leq 1000$ it is more close to those of  RJ
thermal distribution (\ref{eqrj}). The reason of this is related to
the unusual entropy time dependence discussed for Figure~\ref{fig14}
where $S$ grows at initial times and later decays to a smaller value
induced by the RJ thermal condensate.
On the basis of obtained results
we argue that the result of \cite{eplbill,sinaiosc} that
the dependence $S_q(E)$ is close to those of the BE distribution (\ref{eqbe})
instead of RJ one (\ref{eqrj})
is related to significantly shorter times reached there.
Here  the FREEFEM codes allow us to reach times being by a factor 30 longer
where the effects of dynamical RJ thermalization and
RJ condensate start to play the dominant role.

\subsection{Stability island near the ground state linear eigenmode}

The obtained results exposed in Figures~\ref{fig12}.~\ref{fig13} 
show that the fraction of RJ condensate
does not exceed 80\%-90\% even if for the RJS model
(see Figure~\ref{fig8}) this fraction goes to unity
with the increase of total number of linear system modes $N$.
Of course, such an effect can be related
to a finite number of linear eigenmodes
$N$ effectively populated during the NSE (\ref{eqnse}) evolution.
However, we argue that there is another reason behind the
bounded fraction of RJ condensate.
In fact this reason is related to existence
of a stationary self-consistent solution
of the NSE (\ref{eqnse}).
To find this solution $\psi_{s}(x,y)$
we put the time derivative in (\ref{eqnse}) equal to zero.
Then for the remained stationary Hamiltonian
$H_{sf}(x,y) =- \Delta + \beta |\psi_{s}(x,y)|^2$ 
we find the self-consistent solution $\psi_{s}$.
This is done by the recursive iterations
$\psi_g(k) = \alpha \psi_g(k-1) + (1-\alpha) \psi_g(k-2)$
where $\psi_g(k-1)$ is the ground state
of the Hamiltonian $H_{sf}(k-1) = -\Delta + \beta |\psi_g(k-2)|^2$.
With the parameter $\alpha$ being close to unity,
e.g. $\alpha=0.9$, this iterative process rapidly converges
to the self-consistent stationary ground state $\psi_{s}(x,y)$
of the NSE (\ref{eqnse}). The same procedure allows to find
other self-consistent eigenstates $\psi_{s2}, \psi_{s3} ...$
for higher modes.

\begin{figure}[H]
	\begin{center}
		\includegraphics[width=0.85\columnwidth]{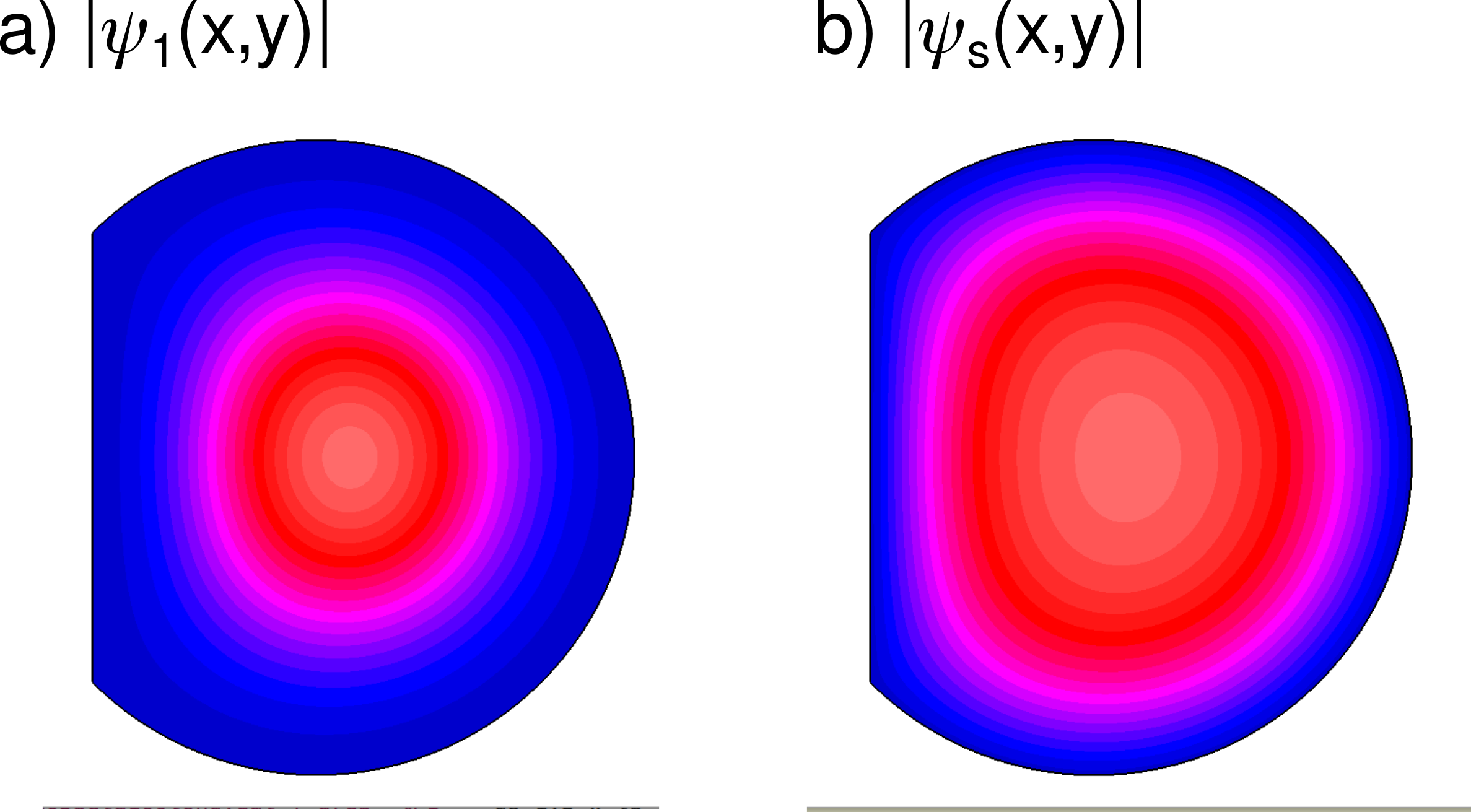}
    \end{center}
	\vglue -0.3cm
	\caption{\label{fig16} 
          Left panel: amplitude of wave function of the ground state
          $|\psi_1(x,y)|$ at $\beta=0$; right panel: amplitude of the self-consistent
          solution $|\psi_s(x,y)|$ of the NSE (\ref{eqnse}) at $\beta = 10$;
          overlap probability between these states is $0.991$.
          Amplitude is changing from zero (blue color) to high value (red color)
          and maximal value (light red in the center); wave functions are normalized to unity.
}        
\end{figure}

The comparison
of this stationary self-consistent state $\psi_s(x,y)$ at $\beta=10$
with the linear ground state $\psi_1(x,y)$ at $\beta=0$
is shown in Figure~\ref{fig16}. The projected probability
of $\psi_s$ on $\psi_1$ is 0.991 so that
both states have similar shape but $\psi_s$ has also about 0.009
probability at other high energy linear eigenstates $\psi_m$.

\begin{figure}[H]
	\begin{center}
		\includegraphics[width=0.85\columnwidth]{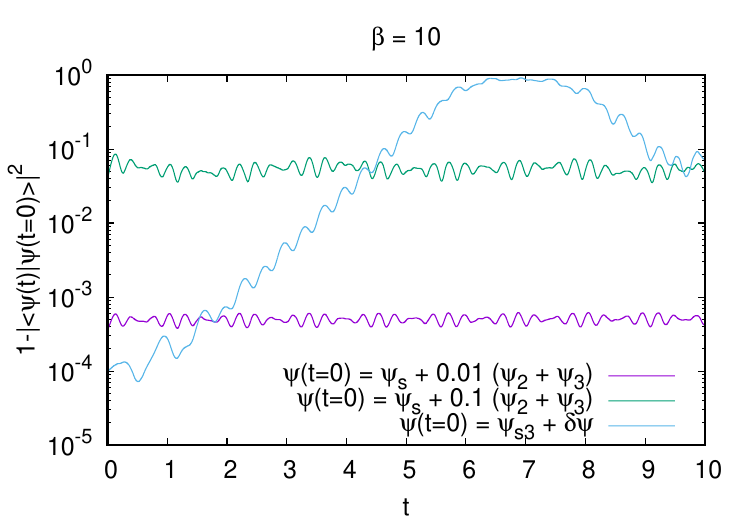}
    \end{center}
	\vglue -0.3cm
	\caption{\label{fig17} 
          Stability of self-consistent states obtained near the ground state
          and first excited state: the error of initial deviation
          remains approximately constant in time $t$ showing that
          the state $\psi_s$ is taken near an integrable vicinity of the ground state
          while in a vicinity of the first excited state $\psi_{s2}$ the error
          grows exponentially with time showing that the corresponding self-consistent
          state is unstable.
}        
\end{figure}

Even if we find the stationary self-consistent eigenstates
at finite $\beta=10$ the main question is if these states
are stable or not. To determine their stability
we follow the time evolution of NSE (\ref{eqnse})
for an initial state being a slightly
perturbed self-consistent state.  The time evolution
of the deviation error, defined as
$1- |<\psi(t)|\psi(t=0)>|^2$, is shown
in Figure~\ref{fig17}. The results show
that the self-consistent ground state $\psi_s=\psi_{s1}$
is stable in respect to perturbations
while the excited self-consistent state $\psi_{s3}$
is exponentially unstable and
in this case the deviation is growing exponentially with time.
These results show that the self-consistent ground state $\psi_s$
is stable, e.g. at $\beta=10$,
and hence there is a certain stability region around this
state. In contrast, the higher energy self-consistent stationary states
are unstable.
The  initial states
started at high linear modes $\psi_m$ at $m>1$
are located in a chaotic component and thus they cannot populate
the integrable stability region located around
the stable self-consistent ground state $\psi_s$.
We argue that this is the main reason
due to which the fraction of RJ condensate
remains bounded to 80\%-90\% (see Figures~\ref{fig12},~\ref{fig13}).

\subsection{Quasi-integrable KAM regime}

\begin{figure}[H]
	\begin{center}
		\includegraphics[width=0.85\columnwidth]{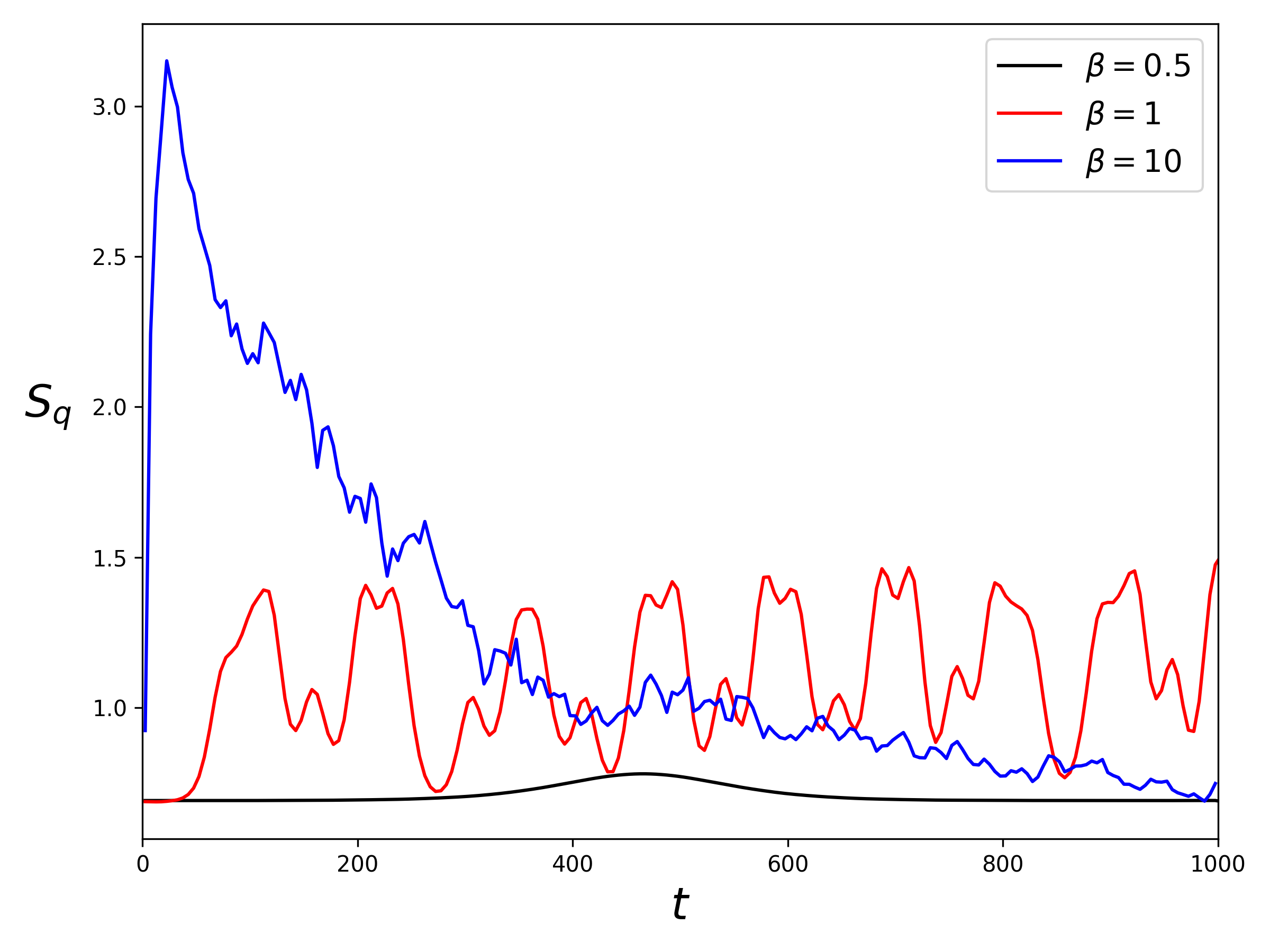}
    \end{center}
	\vglue -0.3cm
	\caption{\label{fig18} Time dependence of the quantum von Neumann entropy $S_q(t)$ 
          for the initial state $\psi(t=0) =(\psi_m + \psi_{m+1})/\sqrt{2}$ at $m=11$
          and $\beta = 0.5; 1; 10$ (values of $S$ are averaged over a time window $\delta t =5$).
}        
\end{figure}

According to the KAM theorem \cite{arnold,sinai,chirikov1979,lichtenberg,kuksin}
we expect that at small nonlinearity $\beta \ll \beta_{ch}$
the NSE dynamics is integrable in the major part of system phase space,
even if certain small chaotic regions can remain, e.g. around thin separatrix layers
with the Arnold diffusion. Indeed, the results of Figure~\ref{fig18} show that at  small $\beta$ the
entropy  $S(t)$ is  not growing with time showing only regular oscillations at $\beta=1$
or almost constant small value $S <1$ at $\beta=0.5$. Even if $S$ values at $\beta=1; 0.5$
are comparable with $S$ value at long time $t=1000$ for $\beta=10$
the situation is qualitatively different: 
at $\beta =10$ the dynamics is chaotic leading to RJ thermalization
with a high fraction of RJ condensate (that gives small $S$ values at $t=1000$)
while
at $\beta \le 1$ the probabilities $\rho_m(t)$ are 
simply regularly oscillating near there initial values
as it is shown in Appendix Figure~~\ref{figA4}.

We use here he term quasi-integrable KAM regime
to point that even in the limit of very small nonlinearity
there are small regions of chaotic component, usually around
separatrix curves (see e.g. \cite{chirikov1979,lichtenberg}).

It is not easy to determine the chaos border for the NSE (\ref{eqnse})
in a chaotic billiard. Similar to the arguments \cite{dls1993,eplbill}
we estimate that the developed chaos appears in the system
when the nonlinear energy spread $\delta E \sim \beta|\psi|^2 \sim \beta/A$
becomes larger than the level spacing $\Delta \approx 4\pi /A  $
that leads to the chaos border:

\begin{equation}
 \beta > \beta_{ch} \sim 4\pi \;\; .
\label{eqchbor}
\end{equation}

This is an approximate estimate while the obtained results
described above  show that
developed chaos and dynamical thermalization
appear already at $\beta =10$. It is not so easy to determine the exact
position of the chaos border due to complexity of various regimes
present in the fiber NSE (\ref{eqnse}) and thus this question requires further studies.

\section{Wave collapse for focusing fiber NSE}
\label{sec6}

\begin{figure}[H]
	\begin{center}
		\includegraphics[width=0.85\columnwidth]{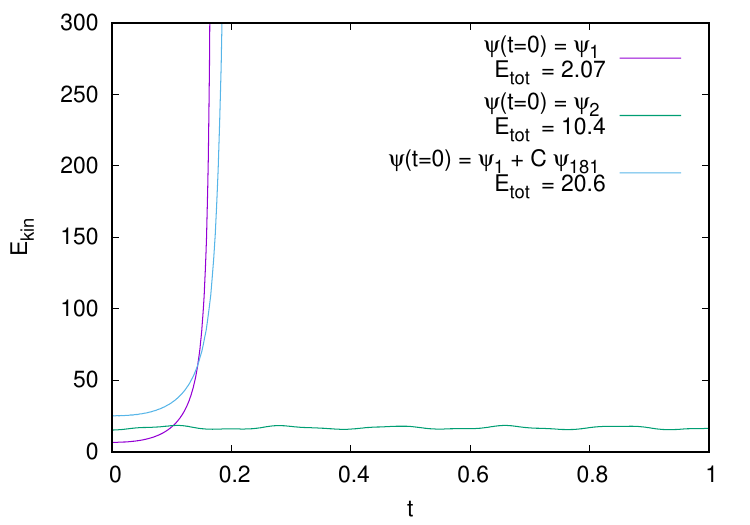}
    \end{center}
	\vglue -0.3cm
	\caption{\label{fig19} 
          Dependence of kinetic energy $E_{kin} = -<\psi|\Delta| \psi>$
          on time $t$ in the NSE (\ref{eqnse}) at $\beta=-12$;  the initial state at $t=0$
          is the linear ground state
          $\psi_{m=1}$ with the  initial total energy  $E=2.076$ (magenta curve),
          here the collapse takes place at $t \approx 0.16$ with the divergence of $E_{kin}$;
          the initial state  is given by a linear combination
          of ground state $m=1$ and excited state at $m=181$
          with $\psi(t=0) = (\psi_1 + C \psi_{181})/\sqrt{1+C^2}$,
          $C=0.15$, $E_{tot}=20.66$ and collapse happens at $t \approx 0.18$;
           the initial state is taken as  the first linear excited state $\psi_2$ at
          the initial total energy $E=10.4$ (green curve) for which
          there is no collapse and only oscillations of $E_{kin}$ with time $t$.
           Here FREEFEM size is $N_F = 416299$,
          integration time step $\Delta t = 10^{-5}$, the relative accuracy of
          norm conservation is typically $10^{-6}$, with an increase up to maximal $10^{-5}$
          at specific time moments, and  a relative accuracy of total energy conservation
          is typically $10^{-5}$ with maximum being $6 \times 10^{-5}$.
}        
\end{figure}

It is known that for the focusing NSE on an unrestricted 2D plane
the wave collapse takes place in a finite time
if initial total energy of the system is negative \cite{talanov,kuznetsov}.
Here we show that the collapse in a finite D-shape billiard
can take place at moderate negative nonlinearity
$\beta <0$ and certain positive energy.
A typical example is shown in Figure~\ref{fig19} at $\beta = -12$
where the initial state at $t=0$ is taken as the ground state of linear system.
Here the collapse happens at rather short time $t_{col} \approx 0.16$
even if the initial total energy at this $\beta=-12$ is positive
being $E=2.076$. At $t_{col}$ the kinetic energy of the system
$E_{kin} =<\psi(t)| -\Delta|\psi(t))>$ diverges growing to infinity.
Indeed, approaching the collapse time  $t_{col}$ the wave function is
localized in smaller and smaller area $A_{col}$ in $(x,y)$ plane
and due to norm conservation $|\psi(x.y.t)|^2 \sim 1/A_{col}$ is growing
so that negative nonlinear energy is also increasing
as $\beta |\psi(x.y.t)|^2 \sim \beta/A_{col} < 0 $.
Thus due to the conservation of total energy
we have divergence of kinetic energy $E_{kin} \sim |\beta|/A_{col}$.
Special checks ensure that these results
are independent of the numerical integration
parameters $N_F$ and $\Delta t$.
For the initial state taken at
the next energy level with $\psi(t=0) = \psi_2(x,y)$ at $m=2$ 
with the total energy $E=10.4$
there is no collapse and kinetic energy
$E_{kin}$ simply oscillates
around its initial value.
A similar collapse is also present at $\beta =-10$.

\begin{figure}[H]
	\begin{center}
		\includegraphics[width=0.85\columnwidth]{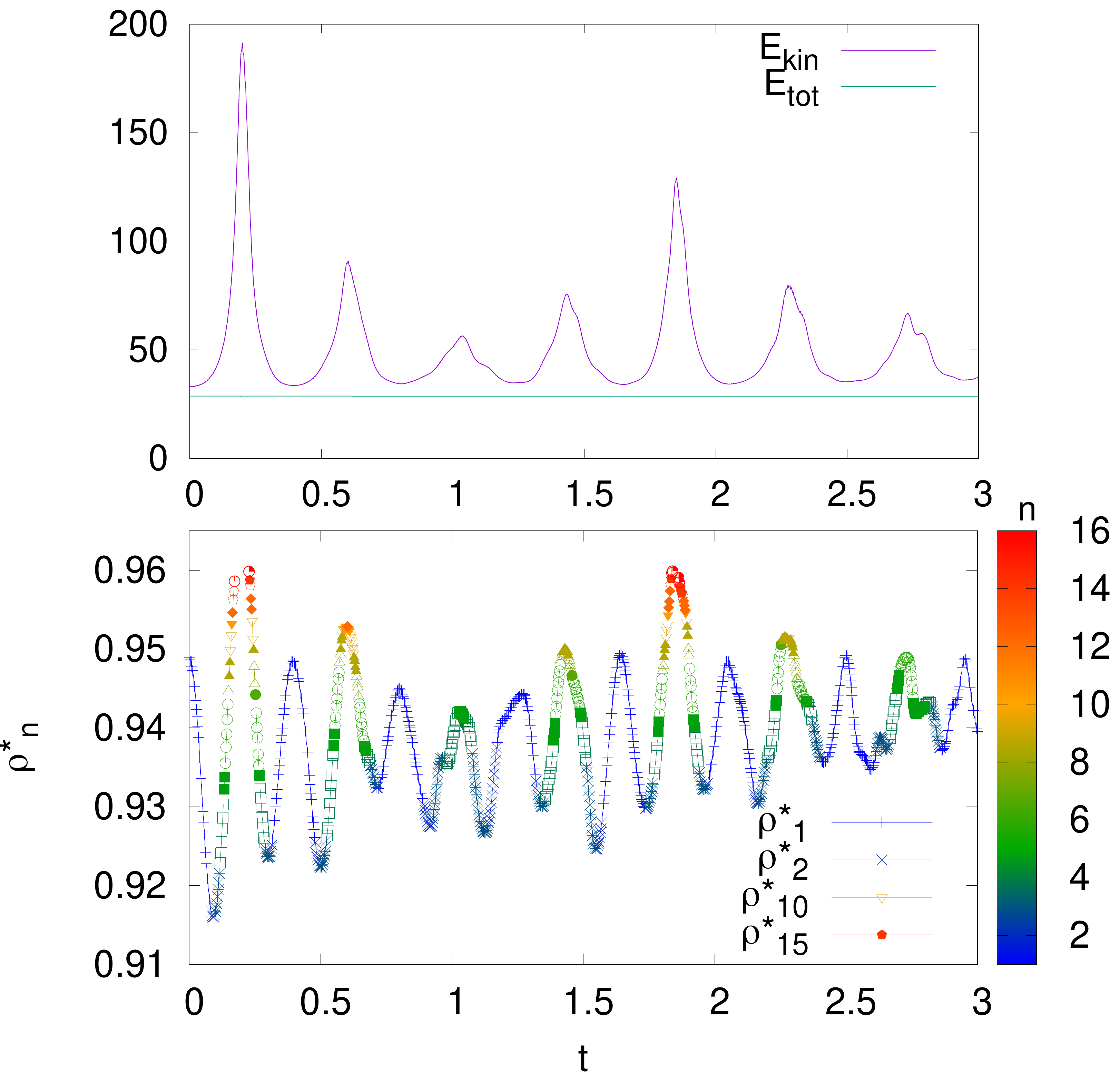}
    \end{center}
	\vglue -0.3cm
	\caption{\label{fig20} Top panel: oscillating dependence of   kinetic energy $E_{kin}$
          on time $t$ (total energy $E_{tot}$ is constant)
          for $\beta = -12$ and the initial state is given by a linear combination
          of ground state $m=1$ and excited state at $m=181$
          with $\psi(t=0) = (\psi_1 + C \psi_{181})/\sqrt{1+C^2}$
          and $C=0.18$, $E_{tot}=28.58$, $\beta=-12$: no collapse
          (compare with collapse at $C=0.15$ in Figure~\ref{fig19}).
          Bottom panel:  for the case $C=0.18$ in the top panel
          we show the time dependence of the main part of probability in
          the eigenstates $n$ of instantenious potential
          $V_{inst} = -\beta |\psi(x,y,t)|^2$ given by  $\rho^{*}_n$;
          there are diabatic transitions between $n$-levels, shown by color bar,
          see text for details. Numerical integration parameters are as in Figure~\ref{fig19}.
}        
\end{figure}

We saw above that for $\beta >0$, e.g. $\beta=10$,
due to dynamical thermalization
with RJ thermal distribution (\ref{eqrj}),
a significant
fraction of total probability  is accumulated 
in the RJ condensate at the ground state $\psi_1(x,y)$ with $m=1$.
This probability $\rho_g=P_0$ can be around 80\%-90\%
(see Figures~\ref{fig12},~\ref{fig13}).
For negative values of $\beta$ we expect that the
dynamical thermalization
leads also to emergence of RJ condensate
with a high condensate probability fraction.
Thus it is important to
see if the collapse can take place for an initial state
with a high probability at the linear ground state 
and a relatively small probability at high energy state.
To study such a situation we
choose the initial state
being $\psi(x,y,t=0) = (\psi_1 + C\psi_{181})/\sqrt{1+C^2}$
with $C \sim 0.18$. Thus at $t=0$
there are about 96.8\% of probability in the ground state $m=1$
and 3.2\% at the excited state $m=181$.
For $C=0.18$ the  results show  that there are time
oscillations of $E_{kin}$ 
with a significant increase of  $E_{kin}$
but the collapse is avoided
due to presence of high energy component
of wave function (see top panel of Figure~\ref{fig20}).
However, for $C=0.15$ still the collapse takes place at
$t_{col} \approx 0.18$ even if the initial
total energy is rather high $E_{tot}=20.66$
(see  Figure~\ref{fig19}).

In the pre-collapse regime at $C=0.18$
there is an interesting behavior of wave function
$\psi(x,y,t)$ that can be seen by its projection probability ${\rho^*}_n$ on the
eigenbasis $1 \leq n < \infty$ of the Schr\"odinger equation (\ref{eqnse})
in which the nonlinear term is considered
as the instantaneous effective attractive potential
$U_{eff} = \beta |\psi(x,y,t)|^2 \leq 0$.
The dependence of ${\rho^*}_n$ on time $t$ is shown in the bottom
panel of Figure~\ref{fig20}. At small times $t < 0.1$
there is about ${\rho^*}_1 = 0.95$ to $0.915$ probability located at the
ground state $n=1$ of the instantaneous potential $U_{eff}(x,y,t)$.
But $U_{eff}$ is changing with time that leads to
a transition of almost all fraction
of  ${\rho^*}_1$ to other eigenstate at $n=2$ at higher moment of time $t$.
Then with growth of $t$ there are transitions to high
and higher states $n$ of instantaneous potential $U_{eff}$.
The maximal $n$ values (up to $n=16$) are reached at the minimal size of wave function
corresponding to the maximal values of $E_{kin}$ in the top panel.
Then, when the packet squeezing passed, its maximal value
and the packet size starts to increase the population ${\rho^*}_n$
starts again go to minimal $n$ values
returning back to almost all probability at the ground state $n=1$ of $U_{eff}$
when the packet returns of its maximal site
(minimal of $E_{kin}$ in the top panel).
It is interesting to note that the transitions between states $n$ happens
in adiabatic manner with almost all probability transferred
from one $n$ value to another one.
These results show that in the pre-collapse regime there is
rather nontrivial time evolution of
wave function with its oscillating size.
It would be interesting to study this behavior in more detail
but this goes outside of scopes of this work.

Thus we show that the wave collapse in a focusing NSE inside
a chaotic billiard can take place even at rather highly
positive values of total system energy
when a probability fraction on high energies is relatively small.
Thus there should be a certain boundary in the space of total energy
and probability on high levels
($C^2$ in Figures~\ref{fig19},~\ref{fig20}).
In global this boundary determines the regime of
collapse stability in respect of
perturbations produced by high energy waves.
Since the collapse remains stable only at rather
small probability at high energies we expect
that in the regime of dynamical thermalization
and appearance of RJ condensate
there will be no wave collapse
at moderate negative $\beta$ values.
The results of dynamical thermalization at negative $\beta$
are presented in the next Section.

\section{Dynamical thermalization in focusing fiber NSE}
\label{sec7}

\begin{figure}[H]
	\begin{center}
		\includegraphics[width=0.85\columnwidth]{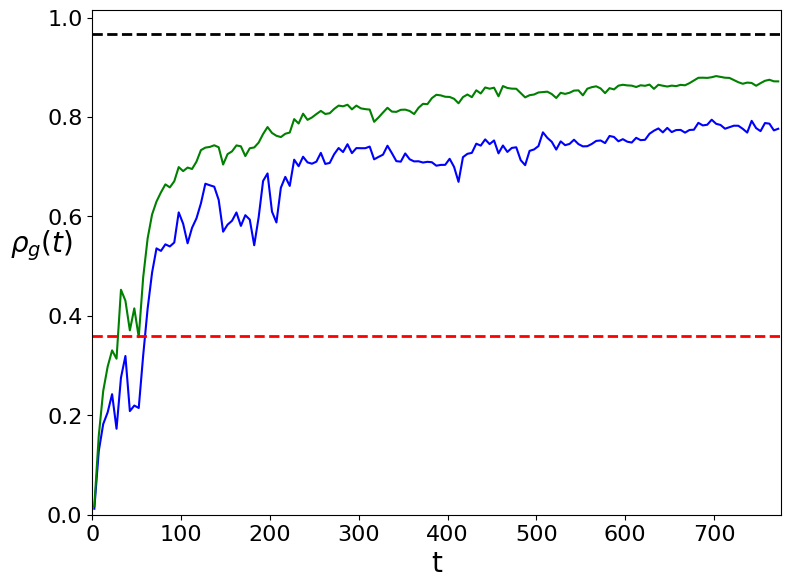}
    \end{center}
	\vglue -0.3cm
	\caption{\label{fig21} Dependence of ground state probability $\rho_g(t)$ on time $t$
            as in Figure~\ref{fig12} but for $\beta=-10$ and only one
            initial state  $\psi(t=0) = (\psi_m + \psi_{m+1})/\sqrt{2}$ at $m=11$;
            the projection probability on the linear ground state $\psi_1$ is shown by blue curve
            and those on the  ground state of instantaneous potential $U_{eff} = \beta |\psi(t)|^2$
            is shown by green curve; total energy is $E=60.36$;
            the dashed lines show  BE ansatz (red) and RJ ansatz (black)
            with the same values of $E$, $T$, $\mu$, $N$ as in Figure~\ref{fig12}.
}        
\end{figure}

For negative nonlinearity $\beta < 0$ in (\ref{eqnse})
  the dynamical thermalization has the features similar to
  the case $\beta > 0$ with certain specific aspects related to
  the wave collapse discussed above. Thus in Figure~\ref{fig21}
  we show the time dependence of projection probability $\rho_g$
  on the linear ground state $\psi_1$ for
  a typical case with $\beta = -10$ and initial
  state $\psi(t=0) = (\psi_m + \psi_{m+1})/\sqrt{2}$ at $m=11$.
  At long times $t$ it approaches to $\rho_g \approx 0.8$
  showing the formation of RJ condensate.
  This behavior is similar to those at $\beta =10$
  in Figure~\ref{fig12} with the same initial state.
  In Figure~\ref{fig21} we add also the projection probability ${\rho^*}_g$
  on the ground state of the instantaneous effective
  potential $U_{eff} = \beta |\psi(x.y.t)|^2 < 0$
  as it was done in Figure~\ref{fig20} bottom panel
  for the case of wave collapse. At long times when the steady-state
  is established we obtain ${\rho^*}_g \approx 0.9 $ being higher than $\rho_g \approx 0.8$.
  Thus a higher fraction of RJ condensate is captured
  by attractive instantaneous potential $U_{eff}$. However,
  these fractions of RJ condensate are still not sufficiently high
  and there is no wave collapse. We also find the similar $\rho_g(t)$ 
  time dependence for other initial states $\psi(t=0) =  (\psi_m + \psi_{m+1})/\sqrt{2}$
  at $m=5$ and $m=7$ (see Appendix Figure~\ref{figA5}).

\begin{figure}[H]
	\begin{center}
		\includegraphics[width=0.85\columnwidth]{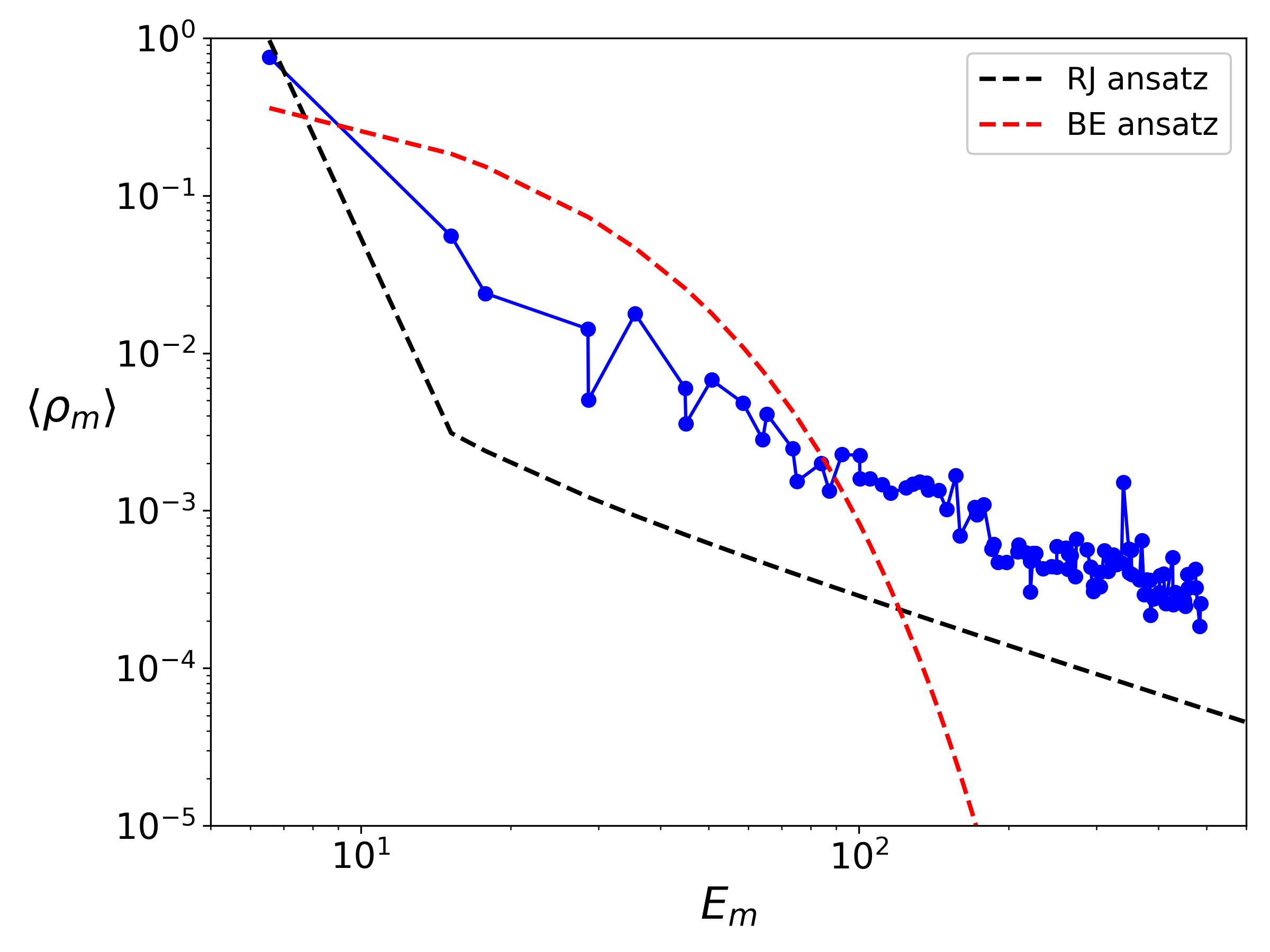}
    \end{center}
	\vglue -0.3cm
	\caption{\label{fig22} Projected probabilities $<\rho_m>$ 
          in the linear eigenstates $\psi_m$ and averaged in the time interval $400 \leq t \leq 800$
          are shown in dependence on linear eigenenergies $E_m$ at $\beta=-10$
          and initial state as in Figure~\ref{fig21};
          dashed curves show the BE and RJ ansatz with same parameters as in Figure~\ref{fig12}.
}        
\end{figure}

In Figure~\ref{fig22} we show the probabilities $\rho_m$
averaged in the time interval $400 \leq t \leq 800$ 
with dashed curves showing the RJ ansatz (\ref{eqrj}) and
BE ansatz (\ref{eqbe}) corresponding
to energy $E$ of the initial state (for RJ case we assume
the total number of levels $N=500$). As for the case of Figure~\ref{fig13}
at $\beta =10$ we find that the RJ ansatz provides a good
description of the steady-state distribution
even if the RJ fraction $\rho_g $ is higher
comparing to the numerically obtained value $\rho_g \approx 0.8$
(due to that the RJ curve is located below the numerical $\rho_m$
values but the energy dependence $\rho_m$ on $E_m$ remains the same).
Also the results show that the BE ansatz is not working.
The similar results at $\beta = -10$ are shown in Appendix Figure~\ref{figA6}
for  other initial states $\psi(t=0) =  (\psi_m + \psi_{m+1})/\sqrt{2}$
  at $m=5$ and $m=7$.

\begin{figure}[H]
	\begin{center}
		\includegraphics[width=0.8\columnwidth]{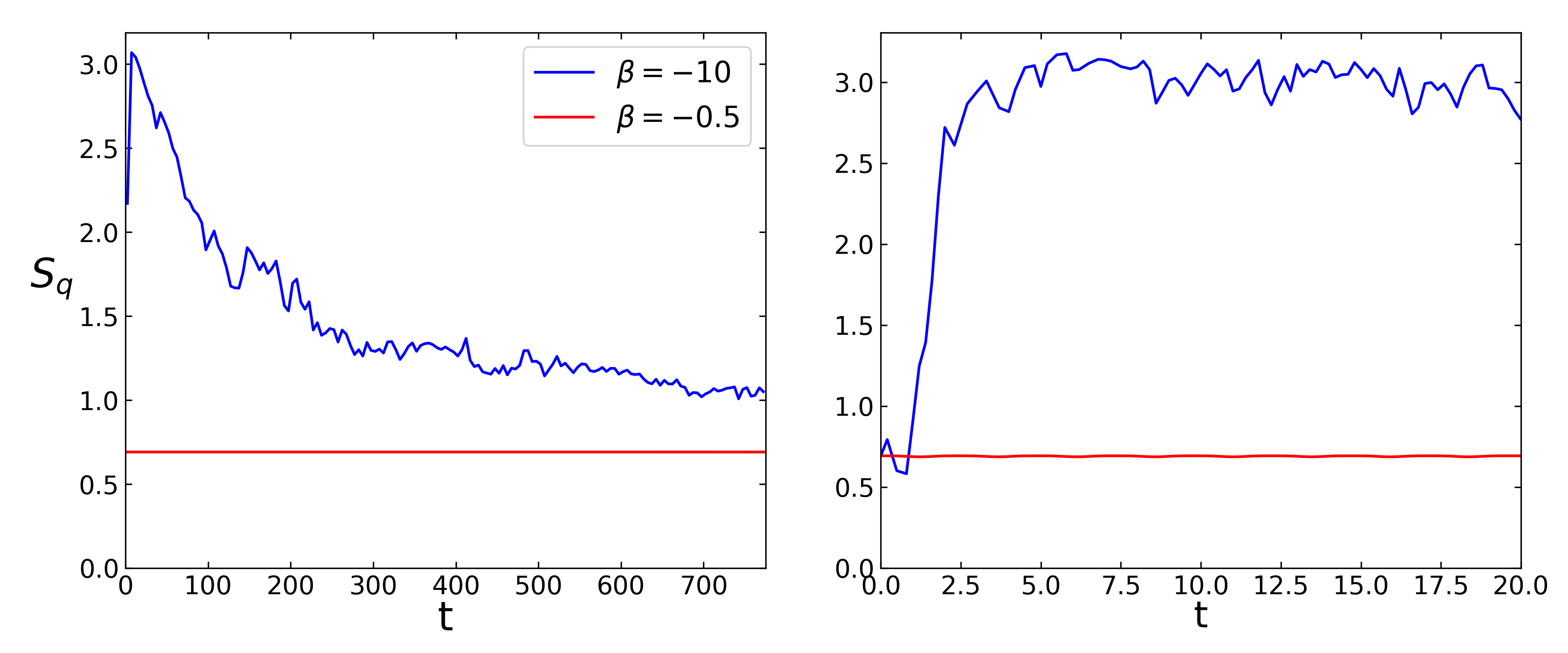}
    \end{center}
	\vglue -0.3cm
	\caption{\label{fig23} Dependence of the quantum von Neumann entropy
          $S_q(t) = - \sum_m \rho_m(t) \ln \rho_m(t)$
          on time $t$ for $\beta = -10$ (blue curve ) and $\beta = -0.5$ (red curve)
          with initial state  $\psi(t=0)=(\psi_m + \psi_{m+1})/\sqrt{2}$ at $m=11$;
          left/right panels show long/short time behavior. The time dependence of classical Boltzamann
          entropy $S_B(t) = \sum_m \ln \rho_m(t)$ is shown in Appendix Figure~\ref{figA3}
          at the same system parameters.
}        
\end{figure}

The time dependence of quantum von Neumann entropy $S_q(t)$  at $\beta=-10; -0.5$
and initial state $\psi(t=0) = (\psi_m + \psi_{m+1})/\sqrt{2}$ at $m=11$
is shown in Figure~\ref{fig23}.
This dependence is similar to those in Figure~\ref{fig14}
at $\beta = 10$. At small $\beta = -0.5$ the entropy $S_q$ remains practically unchanged
corresponding to the KAM regime. 
In this quasi-integrable KAM regime at $\beta = -0.5$
only regular probability oscillations in time
(see Appendix Figure~\ref{figA7}).

Thus we conclude that the process of dynamical thermalization
is very similar for positive and negative values of nonlinearity $\beta$
in (\ref{eqnse}). For significantly negative $\beta$ value the collapse
can take place even at positive total energies
but then the initial state should have probability at the linear ground state
being very close to unity.

\section{Vortexes in defocusing fiber NSE}
\label{sec8}

It is well known that the defocusing NSE in 2D unrestricted plane have
vortexes at strong nonlinearity $\beta$ \cite{tsubota,qturbu}.
The properties of vortexes have been studied theoretically,
numerically and experimentally with superfluidity in liquid helium \cite{tsubota,qturbu}
and with light fluid in nonlinear media of atomic vapor \cite{pavloff}.
It is known that an effective size of vortex scales
in dimensionless units as $1/\sqrt{\beta}$   \cite{tsubota,qturbu}.
A good agreement  between numerical NSE modeling and experimental results
has been demonstrated in \cite{pavloff}.

\begin{figure}[H]
	\begin{center}
		\includegraphics[width=0.85\columnwidth]{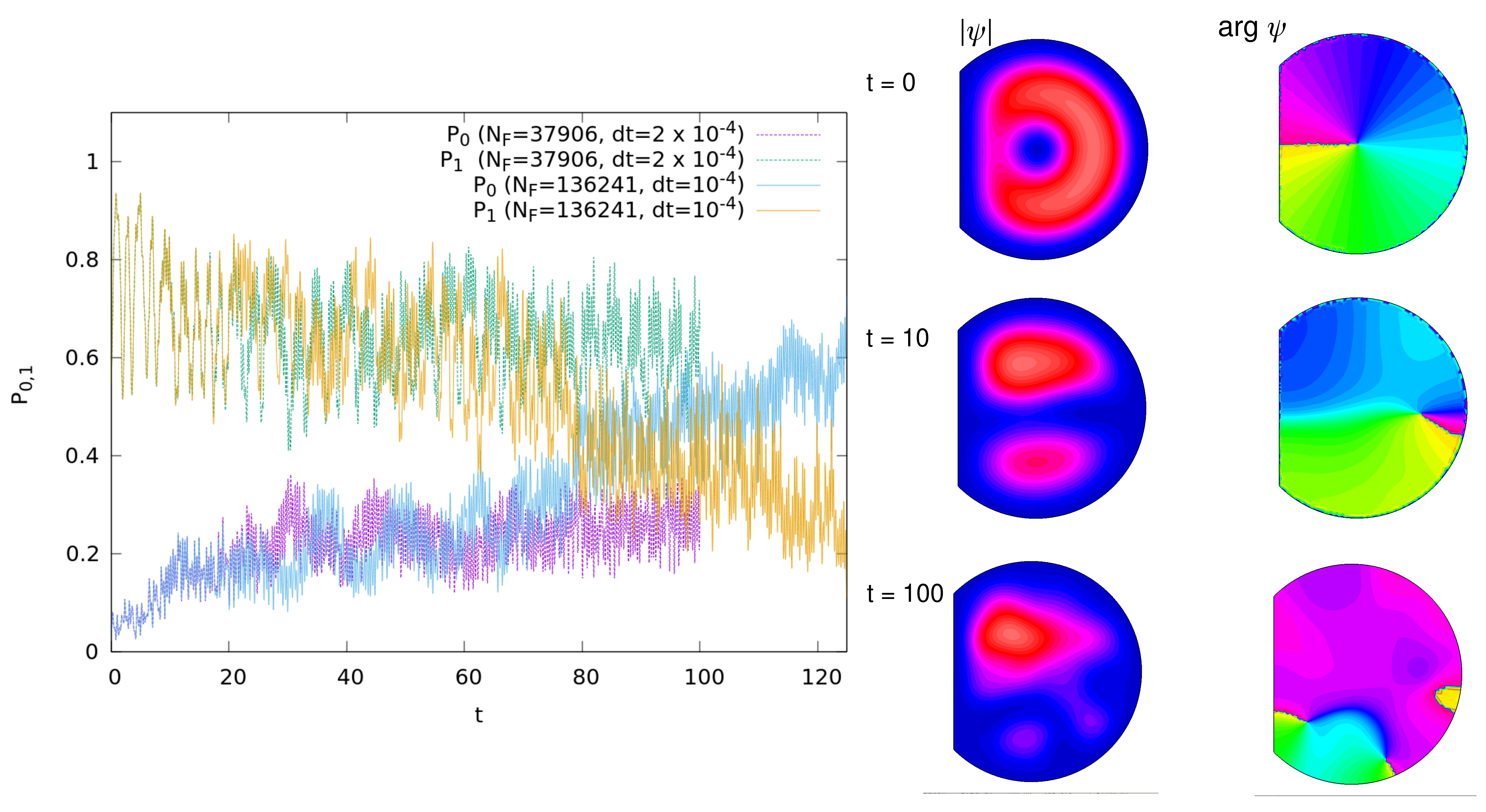}
    \end{center}
	\vglue -0.3cm
	\caption{\label{fig24}
          Left panel: probability in the linear ground state $P_0$
          and first excited state $P_1$ (which is relatively close to vortex)
          as function of time $t$ for different integration steps $\Delta t= dt = 2\times 10^{-4}$ 
          and $10^{-4}$ with the number of intergration net nodes $N_F=37906$
          and $N_F=136241$ respectively at $\beta =10$. Right panel: color panels show
          the wave function amplitude $|\psi|$ and its phase
          $\arg \psi$ at different moments of time; color scale as in Fig.~\ref{fig16}.
}        
\end{figure}

Here we study the properties of vortex in a chaotic D-shape billiard in
the frame of defocusing NSE (\ref{eqnse}).  In the results presented above
the initial state $\psi(x,y,t=0)$ was always chosen as a certain combination
of linear modes. However, to study the vortex dynamics we need to start
with initial state having initial vorticity. Thus
we start with an initial state shown in Figure~\ref{fig24},
it is close to a vortex field in an open 2D plane
with $\psi(r,\theta) \propto r \exp(i\theta)$ \cite{tsubota,qturbu}.
This state has a significant 
projection probability $P_1$ on the first excited linear mode 
$\psi_2(x,y)$ at $m=2$. Thus the high value of   probability $P_1$
indicates that the vortex is well survived in time even if
the classical dynamics of rays is chaotic in the billiard at $\beta=0$.
In contrast, if the projection probability $P_0$ on
th ground state linear mode $\psi_1(x,y)$ at $m=1$ becomes high
and $P_1$ becomes low
then this indicates that the vortex is destroyed by
quantum chaos and nonlinearity.

The time evolution of projection probabilities $P_0, P_1$
for vortex time dynamics is shown in Figure~\ref{fig24} (left panel)
for moderate nonlinearity $\beta =10$.
For a small integration time step $\Delta t = dt = 10^{-4}$ and high number
of FREEFEM net nodes $N_F = 136241$ we have high numerical conservation
of norm and total energy integral. For time $t \ge 80$ the results 
show that $P_0$ probability starts to grow significantly
while $P_1$ probability starts to drop rapidly that tells about the
complete destruction of vortex at times $t > 80$. This is
confirmed by color vortex images in the right panel of Figure~\ref{fig24}:
vortex structure is well visible at $t=10$ both for
wave function amplitude $|\psi)x,y,t)|$
and its phase  $\arg \psi(x,y,t)$ showing clear phase rotation.
In contrast at $t=100$ both $|\psi(x,y,t)|$ and  $\arg \psi(x,y,t)$
show disappearance of vortex. For times $t >80$ the vortex
is destroyed and the dynamical RJ thermalization takes place
for nonlinear field $\psi(x,y,t)$. Indeed,
the results for $P_0$ show its significant
growth so that its value becomes close to
the probability fraction of RJ condensate 
and $P_1 \approx 0.8 - 0.9$ (see Figure~\ref{fig12}).
At the same time the projection probability $P_1$ 
drops significantly similar to the cases shown in Figure~\ref{fig13}.
Thus the dynamical RJ thermalization works also 
for the initial vortex states at moderate nonlinearity $\beta=10$.

It is interesting to note in the left panel of Figure~\ref{fig24} that
at smaller number of nodes $N_F=37906$
the evolution of vortex is more stable with higher projection probability
$P_1$ at $t \approx 100$ comparing to the case
with significantly higher integration
parameters $dt=10^{-4}$ and $N_F = 136241$.
We interpret this by assuming that the lower value
$N_F=37906$ leads to a certain effective
roughness of induced effective potential
that reduces velocity of vortex center
decreasing the quasi-particle excitation 
thus increasing vortex life-time.

\begin{figure}[H]
	\begin{center}
		\includegraphics[width=0.85\columnwidth]{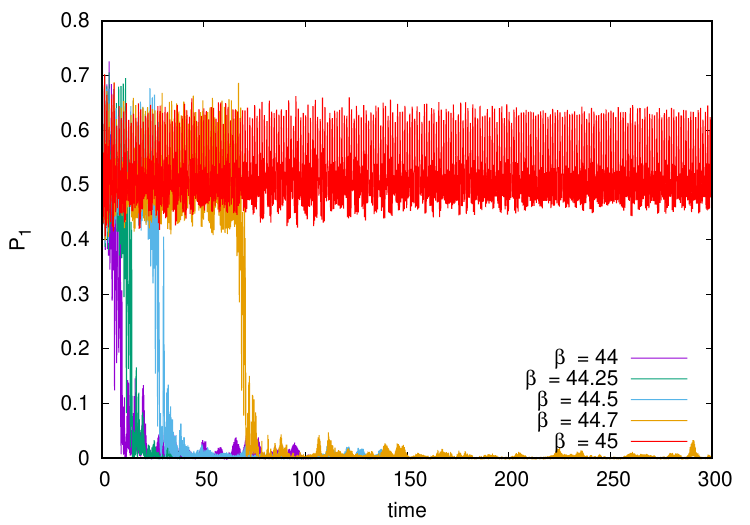}
    \end{center}
	\vglue -0.3cm
	\caption{\label{fig25}
          Probability in the first linear excited state $P_1$ (close to vortex)
          as function of time $t$ for different $\beta$ values; there is a sharp
          disapperance of vortex for $\beta < 45$; for $\beta = 45$
          the same behavior continues
          up to $t=600$ and may be continued beyond this maximal computation time.
}        
\end{figure}

\begin{figure}[H]
	\begin{center}
		\includegraphics[width=0.85\columnwidth]{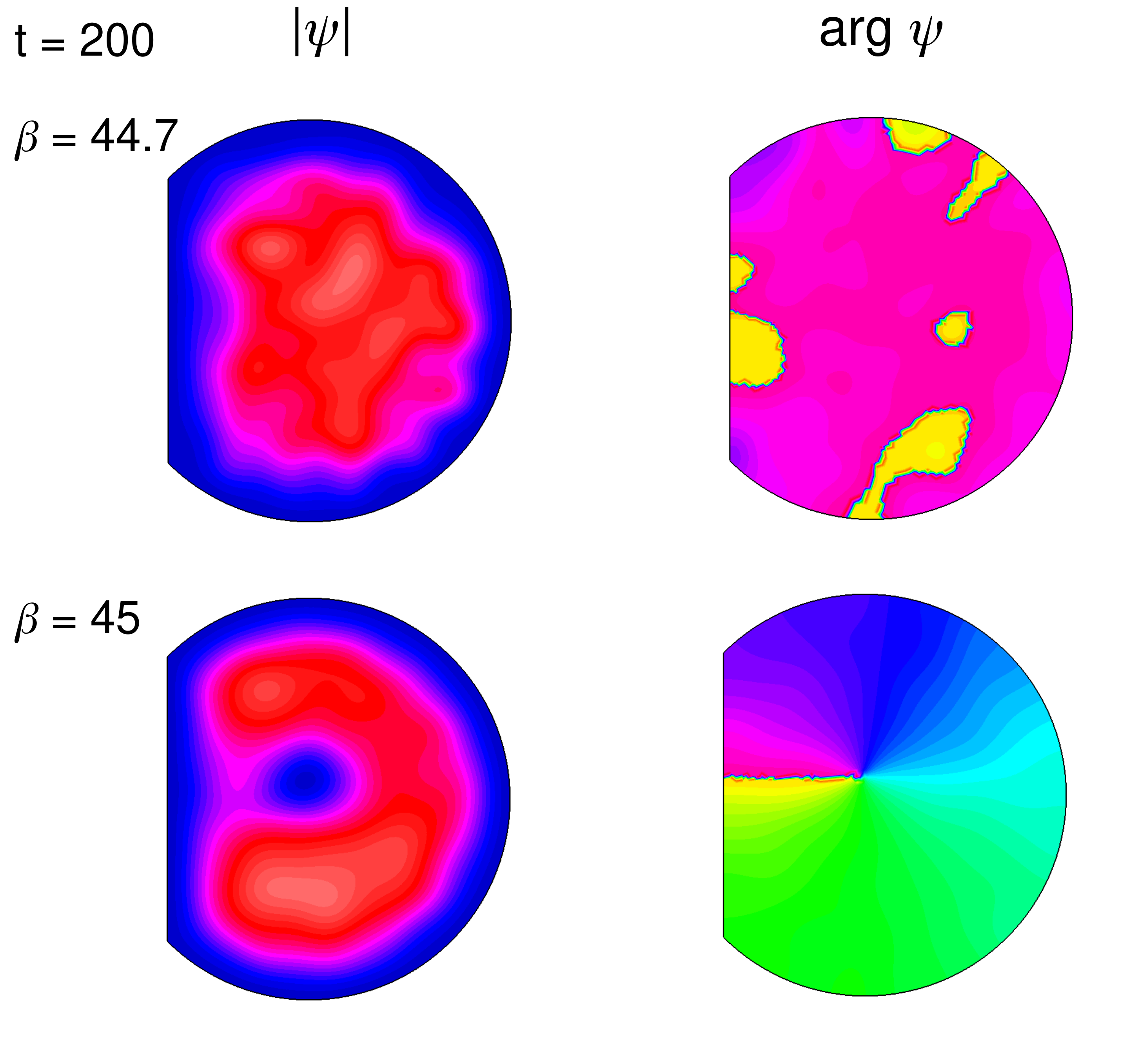}
    \end{center}
	\vglue -0.3cm
	\caption{\label{fig26}
          Nonlinear vortex fields $\psi$ (amplitude and phase)
          are shown at time $t=200$ for $\beta=44.7$
          (vortex is destroyed) and $\beta =45$ (vortex survive).
          The video of time evolution of $|\psi|$ is shown   for
          $\beta =44$ (left image) and $\beta =45$ (right image)
          up to $t \approx 30$   in the Supplementary Material file videovortex.mp4.
}        
\end{figure}

The dynamics of vortexes at a moderate nonlinearity $\beta \sim \beta_{ch}$
tends to approach the  dynamical thermalization
even if the votrex life time $t_v$ can be relatively high.
At high $\beta$ values the vortex size decreases as $1/\sqrt{\beta}$
and its interactions with billiard border becomes weaker
with the growth of vortex life-time $t_v$ as it is show in Figure~\ref{fig25}
where the projection probability $P_1$ remains high for longer and longer times.
However, we find here a striking effect when
this vortex life-time $t_v$ changes  abruptly 
from $t_v \approx 70$ for $\beta=44.7$
to $t_v > 600$ for $\beta = 45$.
This sharp transition in $t_v$
is also well seen in Figure~\ref{fig26} where the snapshots
of nonlinear field $\psi(x,y,t)$ are shown at $t=200$:
for $\beta=44.7$ the vortex is completely destroyed
while for $\beta=45$ it is well preserved.
The vortex time evolution is also shown
in Supplementary Material (SupMat) video (file videovortex.mp4)
for vortex time evolution at $\beta=44$
(left panel, vortex is destroyed at maximal shown $t \approx 30$)
and at $\beta =45$ (right panel, vortex is well
survived at $t > 30$). This video shows that
at $\beta=44$ the center of vortex moves more rapidly
compared to those at $\beta=45$. Thus we make a
superfluid transition conjecture:
at $\beta < \beta_{sf} \approx 45$
the velocity of vortex center is higher than the critical
superfluid velocity of light fluid
and it starts to radiate quasi-particles
breaking superfluidity, while
for  $\beta > \beta_{sf} \approx 45$
this velocity is below critical
superfluid velocity and the vortex life-time
is increased enormously or may become even
infinite. The verification of this
superfluid transition conjecture requires
further studies which are beyond the scope of this work.

\section{NSE dynamics at long times}
\label{sec9}

The question about asymptotic properties
of NSE dynamics in the D-shape chaotic billiard at long time scales is rather complex.
The NSE (\ref{eqnse}) can be rewritten in the
basis of linear eigenmodes $\psi_m(x,y)$  of the billiard (at $\beta=0$)
where it reads:
\begin{equation}
i {{\partial C_{m}}}/{\partial {t}}
= E_{m} C_{m}
+ \beta \sum_{{m_1}{{m_2}}{{m_3}}}
V_{{{m}}{{m_1}}{{m_2}}{{m_3}}}
C_{{m_1}}C^*_{{m_2}}C_{{m_3}} .
\label{eqnse2}
\end{equation}
Here $C_m(t)$ are wave function amplitudes 
in the eigenbasis $\psi_m$ so that
$\psi(x,y,t) = \sum_m C_m(t) \psi_m(x,y)$
and interaction matrix elements
induced by nonlinearity are
\begin{equation}
V_{{m}{m_1}{m_2}{m_3}} =
\int_{x,y} {\psi^*}_m \psi_{m_1} {\psi^*}_{m_2} \psi_{m_3} dx dy \;.
\end{equation}

This type of equation appears also in the models
on nonlinear perturbation 
of quantum chaos \cite{dls1993},
Anderson localization 
(DANSE model for Discrete Anderson Nonlinear Schr\"odinger Equation)
\cite{dls2008,garcia,flach,flach2}
and Random Matrix Theory (RMT) \cite{rmtprl} (NLIRM model).
In these models the norm and total energy are exactly conserved
as for NSE model (\ref{eqnse}).
For DANSE type models \cite{dls1993,dls2008,garcia,flach,flach2}
the linear eigenstates are exponentially localized
due to quantum interference and Anderson localization
so that the coupling matrix elements are $V_{{{m}}{{m_1}}{{m_2}}{{m_3}}} \sim 1/\xi^{3/2}$
for states $\psi_m, ...\psi_{n_3}$ located on a distance of localization length $\xi$
from each other on a lattice,
while outside of this range the values of $V$ drop exponentially.
Here $\xi$ is a number of lattice states in a localization length
 and for a $d$-dimensional lattice with disorder we have $\xi \sim \ell^d$ \cite{garcia}.

For such type DANSE models it was shown that for a nonlinearity
above a certain chaos border $\beta > \beta_{c}$ there is
a slow subdiffusive spreading of probability over the lattice sites $n$
with the width growing as $(\Delta n)^2 \sim t^\nu$
with the subdiffusive exponent $\nu \approx 0.3 - 0.4$ \cite{dls1993,dls2008,garcia,flach,flach2}.
This growth continues during all computational times up to extremely
long times $t \sim 10^{10}$.

It is interesting to note
that for the DANSE models considered in \cite{dls1993,dls2008,garcia,flach,flach2}
the spectrum of linear modes is bounded in a certain finite energy band.
However, the DANSE type model studied in \cite{garcia2} has
a Stark lattice with on-site energies  growing linearly with site $n$
index $E_n \sim f n$  and still the unlimited
spreading was present up to maximal times $t =10^8$
with an exponent $\nu \approx 0.24 - 0.30$
(in a chaotic regime with
$ \beta > \beta_c , f<f_c$).
Due to energy growth with site number
the model \cite{garcia2} is similar to one
studied here (\ref{eqnse}) since in a 2D billiard
energies of modes are also growing with mode index
$E_m \propto m$. Thus due to such a similarity
between the model \cite{garcia2} and the present
fiber model (\ref{eqnse}) one could expect
to have unlimited norm spreading with time
up to higher and higher modes $m$.
However such an unbounded  scenario has certain difficulties:
indeed, if the total number of populated
linear modes $N \sim \Delta m(t) \sim t^{\nu/2}$
grows with time then due to RJ condensation
a main part of the total norm
should be in the condensate phase being mainly
at the linear ground state.
From the view point of RJS model this norm
should become closer and closer to unity with the
growth of $N$ (see Figure~\ref{fig8}).
Due to presence of stability island with
self-consistent ground state
we see that in the NSE fiber model (\ref{eqnse})
RJ condensate has not more than approximately 90\%
of the total norm (see Figure~\ref{fig12})
so that only about 10\% of norm can propagate to
higher and higher $N$ values.
In the case of Stark ladder model \cite{garcia2}
we have there about 50\% of the norm propagating to
the high energy modes, counting from the initial state
location (and about 50\% of norm moving to the low energy modes).
Thus there can be a certain similarity
between the Stark model \cite{garcia2} and
our fiber NSE model (\ref{eqnse}).
However, in \cite{garcia2} the coupling matrix
elements $V$ are approximately comparable 
for all modes (centered within a localization length $\xi$), independently of m values,
while for the fiber case the dependence
of $V$ on $m$ requires  further analysis.
Indeed, at high $E_m$ (or $m$) values
the linear modes become more and more
oscillating (see Figure~\ref{fig3})
and thus we may expect that
a number of effective components
$\xi$ is growing with $E_m$ 
and thus couplings $V$ are decreasing.
A simple estimate based on the results
for rough billiards \cite{roughbil}
would give $\xi \sim \sqrt{E_m} \sim \sqrt{m}$
and thus $V \sim 1/\xi^{1/2} \sim 1/{E_m}^{1/4}$,
but this expectation requires
separate verification.
If $V $ decreases with $E_m$  then effective
nonlinearity also decreases
and at some high energies $E_m$
an effective nonlinearity
$\beta V$ becomes below a chaos border
that leads us to a bounded scenario
with only finite number of effectively populated modes.
At present we cannot present firm arguments in favor of
bounded scenario of excitation to high energies
or unbounded one.

Also the results presented in \cite{rmtprl}
for RMT with growing energy diagonal term
(see SupMat Figure~S14 in \cite{rmtprl})
show that even in a system with $N=64$ modes
the thermal RJ distribution is reached only at such high times
as $ t \approx 10^8$ ($t \approx 10^6$ is not enough).
Such times cannot be reached in  numerical simulations
of fiber NSE (\ref{eqnse}).
Such high times are also out of reach with  
fiber experiments.

For initial states with vortexes our results in Figures~\ref{fig24},~\ref{fig25},~\ref{fig26}
show that for $\beta_{ch} < \beta < \beta_{sf}$ 
a vortex state is destroyed and we expect that the dynamical RJ thermalization
takes place at long times which however are difficult to reach
in numerical simulations due to high life time of vortex.
For the superfluid phase at $\beta > \beta_{sf} \approx 45$
the question about vortex life time remains open.

\section{Experimental parameters for a quantum chaos fiber}
\label{sec10}

Here we discuss the requested experimental parameters for D-shape fiber
to have sufficiently strong nonlinearity $\beta$ to reach regime of
dynamical thermalization. We start from experimental conditions
already realized in \cite{aplbabin,wabnitz} with a circle fiber cross-section
where the classical dynamics of linear system at $\beta=0$ in (\ref{eqnse})
is integrable. These fiber experiments
were done at high laser beam intensity
and the theoretical description of the complex field envelope
$A(x,y,z,t)$ expressed in $[\sqrt{W}/m]$ is described by the NSE type equation
in which we keep notations of \cite{wabnitz}:
\begin{equation}
  {\partial A \over \partial z} - (i/2k_0) {\nabla^2}_{\bot} A
   = i \gamma {\vert A \vert}^2 A \; .
\label{eqwab}
\end{equation}
Here $z$ is distance along the fiber, corresponding to time in (\ref{eqnse}),
$k_0 = \omega n_{co}/c$ is the light wave vector $k_0 = 2\pi/\lambda$,
$\gamma = \omega n_2/c$ is the fiber nonlinear coefficient
with nonlinear Kerr index $n_2 = 3.2 \times 10^{-20} m^2/W$;
$n_{co} \approx 1.5$ is the refractive index.
Here we omit the term with time derivative $d^2 A/d t^2$
which gives only linear in $A$ contribution in (\ref{eqwab})
and assumed that the refractive index
$n_{co}$ is constant inside the fiber
of core radius $r_c =30 \mu m$.
Thus from (\ref{eqwab}) from the above typical parameters  and a typical wave length
$\lambda = 1000 nm$ we obtain the expression for coefficient $\beta$ in the NSE (\ref{eqnse}):
\begin{equation}
  \beta = 2 n_{co} n_2 (2\pi/\lambda)^2 {r_c}^2 I [W/cm^2] \; ,
\label{eqwabb}
\end{equation}
that gives $\beta \approx 0.32$ for typical parameters of \cite{wabnitz} with
$I = 10 GW/cm^2$, $n_2 = 3.2 \times 10^{-20} m^2/W$,
$n_{co} = 1.5$, $\lambda = 1000 nm$, $r_c = 30 \mu m$.
In (\ref{eqwab}) the length unit is $z_0 = {r_c}^2/ \lambda \approx 0.1 cm$
for the above values $r_c$, $\lambda$. Thus for a fiber length of $1$ meter
we have the dimensionless length of $z/z_0 \approx  1000$
that corresponds to the dimensional time interval $t$ 
of NSE (\ref{eqnse}) being also 1000.
In our numerical simulations of NSE (\ref{eqnse}) we are reaching similar
time scales $t \approx 100$  that is comparable with the fiber length
of $0.4 m$ in \cite{wabnitz}.

We suppose that increasing by a factor 3 the parameters
$r_c \approx 90 \mu m$ and $I \approx 30 GW/cm^2$  in (\ref{eqwabb})
it would be possible to have nonlinear parameter $\beta \approx 10$
at which dynamical thermalization sets in the D-shape
quantum chaos fiber described by the NSE (\ref{eqnse}).

\section{Discussion}
\label{sec11}

Our numerical and analytical studies
of the NSE fiber light fluid dynamics in a chaotic D-shape billiard
show that it is characterized by various unusual and rich phenomena.
Above the chaos border at $|\beta| > \beta_{ch} \sim 10$
the dynamical thermalization leads to the RJ thermal distribution (\ref{eqrj})
with a formation of the RJ condensate in the  ground state of linear system
which fraction has about 80-90\% of total probability.
Certain similarities between RJ and Fro\"ohlich condensates are discussed.
We show that at $\beta > 0$ near the ground state there is a self-consistent
state with a certain island of stable motion around it.

In the regime of chaotic dynamical thermalization
with emergence of RJ condensate we show
that the time dependences of the quantum von Neumann entropy $S_q(t)$
and the classical Bolzmann entropy $S_B(t)$ are non-monotonic in time
being characterized by an  initial increase at short times and later
relaxation from its maximal value to a significantly smaller value
on long times in the steady-state regime with RJ condensate.
Such a result for $S_B(t)$ is
in contrast to the Boltzmann H-theorem \cite{boltzmann1,mayer,landau}
even if no such theorem has been proven for the NSE equation in
a quantum chaos fiber.
We point on a certain divergence problem of $S_B$ definition
in our system (\ref{eqnse}) due to presence of unlimited number of
billiard linear eigenmodes assuming that this can be at the origin  
of this contrast with the H-theorem.

\begin{figure}[H]
	\begin{center}
		\includegraphics[width=0.85\columnwidth]{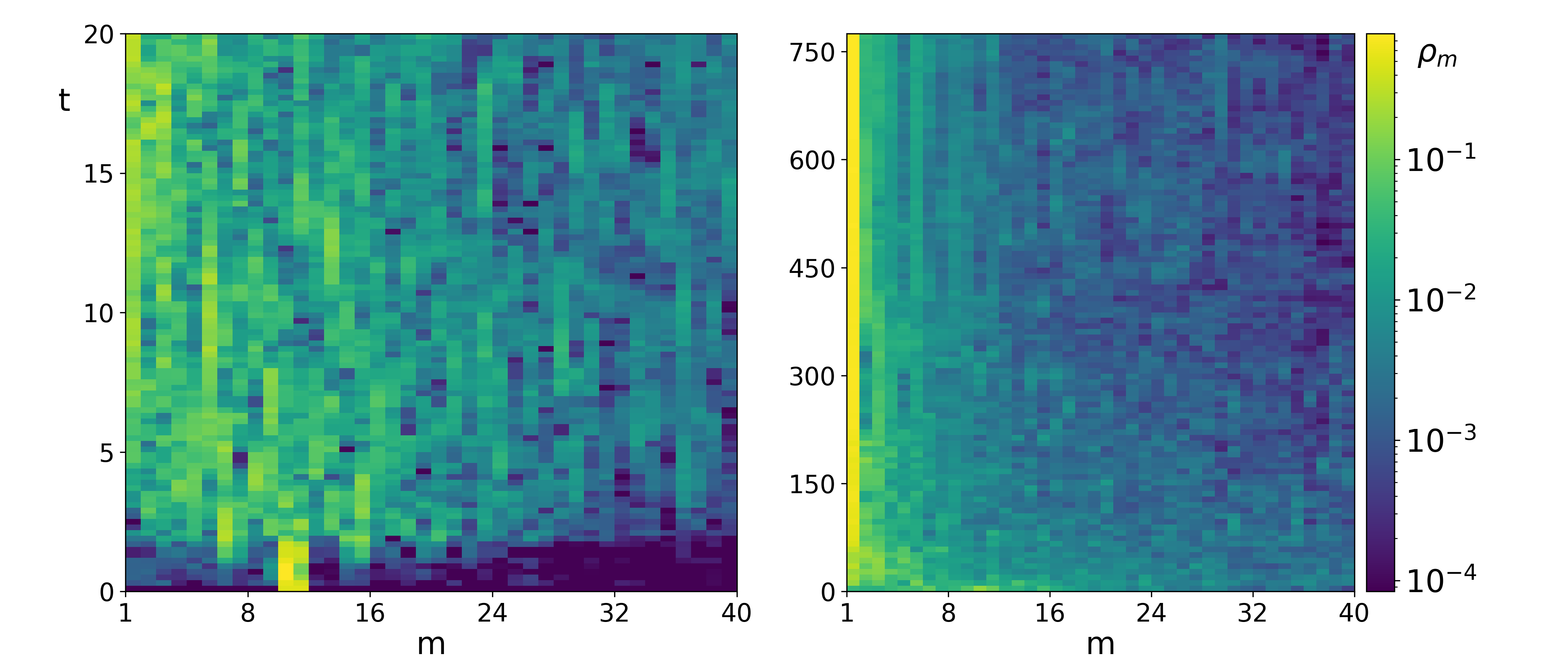}
    \end{center}
	\vglue -0.3cm
	\caption{\label{fig27}
          Time development of dynamical thermalization and RJ condensate formation (yellow band)
          are shown via  probabilitis $\rho_m(t)$ at
          linear eigenmodes $\psi_m$ (color bar)
          at short (left panel) and long (right panel) times;
          here $\beta = -10$, initial state is
          $\psi(t=0) = (\psi_m + \psi_{ma+1})/\sqrt{2}$ at $m=11$
          as in Figures~\ref{fig21},~\ref{fig22}. 
}        
\end{figure}

In absence of nonlinearity
our system belongs to the systems of quantum chaos,
thus having many features as in the Random Matrix Theory.
On these grounds we argue that the dynamical thermalization
induced by nonlinearity represents a generic
features of such thermalization similar
to those discussed for nonlinear perturbation
of Random Matrix theory in \cite{rmtprl}.
We also note that even if the phenomenon of RJ condensation
in nonlinear systems has been discussed for multi-mode fibers
\cite{picozzi1,picozzi2,picozzi3,picozzi2024}
its origin was not associated with the emergence of
dynamical chaos above a certain chaos border
below which the RJ thermalization and RJ condensate are absent
and dynamics is quasi-integrable as in the KAM theory.
We attribute the so-called self-cleaning phenomenon
observed in multi-mode optical fibers \cite{aplbabin,wabnitz,fiber1}
to the formation of RJ condensate induced by
dynamical thermalization and chaos.
A pictorial image of time development of dynamical thermalization and
RJ condensate formation is shown in Figure~\ref{fig27}.

In addition to dynamical thermalization
at certain conditions the focusing fiber NSE (\ref{eqnse}) at $\beta <0$ 
can have the regime with wave collapse which in contrast
to the case of open space  \cite{talanov,kuznetsov}
can happen even at relatively high positive energies
(see Figures~\ref{fig19},~\ref{fig20}).
In such regime there is a nontrivial interplay
between chaotic component and wave collapse.
However, the RJ condensate emerging due to
dynamical thermalization does not lead to the collapse formation.

We also show that for the defocusing fiber NSE (\ref{eqnse})
there are long living excitations in the form
of vortexes which finally are expected
to relax to the RJ thermal state
for nonlinearity $\beta_{ch} < \beta < \beta_{sf}$.
However, for a strong nonlinearity
$\beta > \beta_{sf} \approx 45$  
the vortex life time is strikingly
increased corresponding to the transition to the superfluid phase of light fluid.

We hope that these interesting and generic phenomena described here
can be studied experimentally with
multi-mode fibers similar to those of
\cite{aplbabin,wabnitz,fiber1,fiber2,fiber3,fiber4,fiber5}
(namely with  a quantum chaos fiber cross-section, e.g. D-shape as in Figure~\ref{fig1}).
In fact certain studies of D-shape fiber have been reported recently in \cite{stone}
but without discussion of dynamical thermalization.
Also the quantum fluid experiments with laser beam
propagating in a nonlinear medium formed by atomic vapor, like in \cite{pavloff},
can be used to test interplay of nonlinearity, quantum chaos and dynamical thermalization.

 {\bf Acknowledgments:}
 We thank S.Babin,   D.Kharenko and E.Podivilov (IAE RAS Novosibirsk),
 K.M.Frahm (LPT Toulouse), 
 and N.Pavloff (LPTMS Orsay)
 for  useful discussions.
 
The authors acknowledge support from the grants
 ANR France project OCTAVES (ANR-21-CE47-0007),
NANOX $N^\circ$ ANR-17-EURE-0009 in the framework of 
the Programme Investissements d'Avenir (project MTDINA),
MARS (ANR-20-CE92-0041).

\section*{Appendix A}
\setcounter{equation}{0}
\renewcommand{\theequation}{A\arabic{equation}}
\setcounter{figure}{0}
\setcounter{section}{0}
\renewcommand\thefigure{A\arabic{figure}}
\renewcommand\thesection{A\arabic{section}}
\renewcommand{\figurename}{Appendix Figure}

Here we present additional Figures related to the main part of the article;
as an additional Supplementary Material we provide video of vortex dynamics
in the file videovortex.mp4 (see text and Figure~\ref{fig26} with its description).

\begin{figure}[H]
	\begin{center}
           \includegraphics[width=0.8\columnwidth]{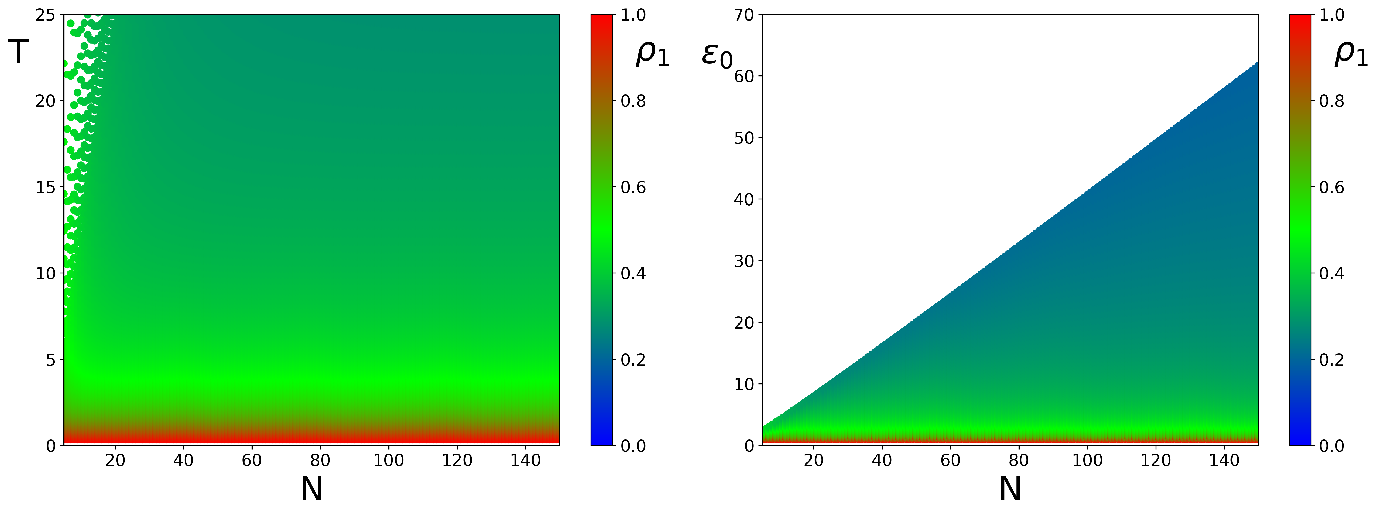}
	\end{center}
	\vglue -0.3cm
	\caption{ \label{figA1} Color map of condensate probability $\rho_1$
          at corresponding system temperature $T$ and number of system eigenmodes $N$ (left panel);
          right panel shows the same $\rho_1$ as in left panel
          but in the plane $(N, \epsilon_0)$. Results are 
           obtained in the frame of BE thermal distribution (\ref{eqbe}), here $\Delta =1$.
	}
\end{figure}

\begin{figure}[H]
	\begin{center}
           \includegraphics[width=0.8\columnwidth]{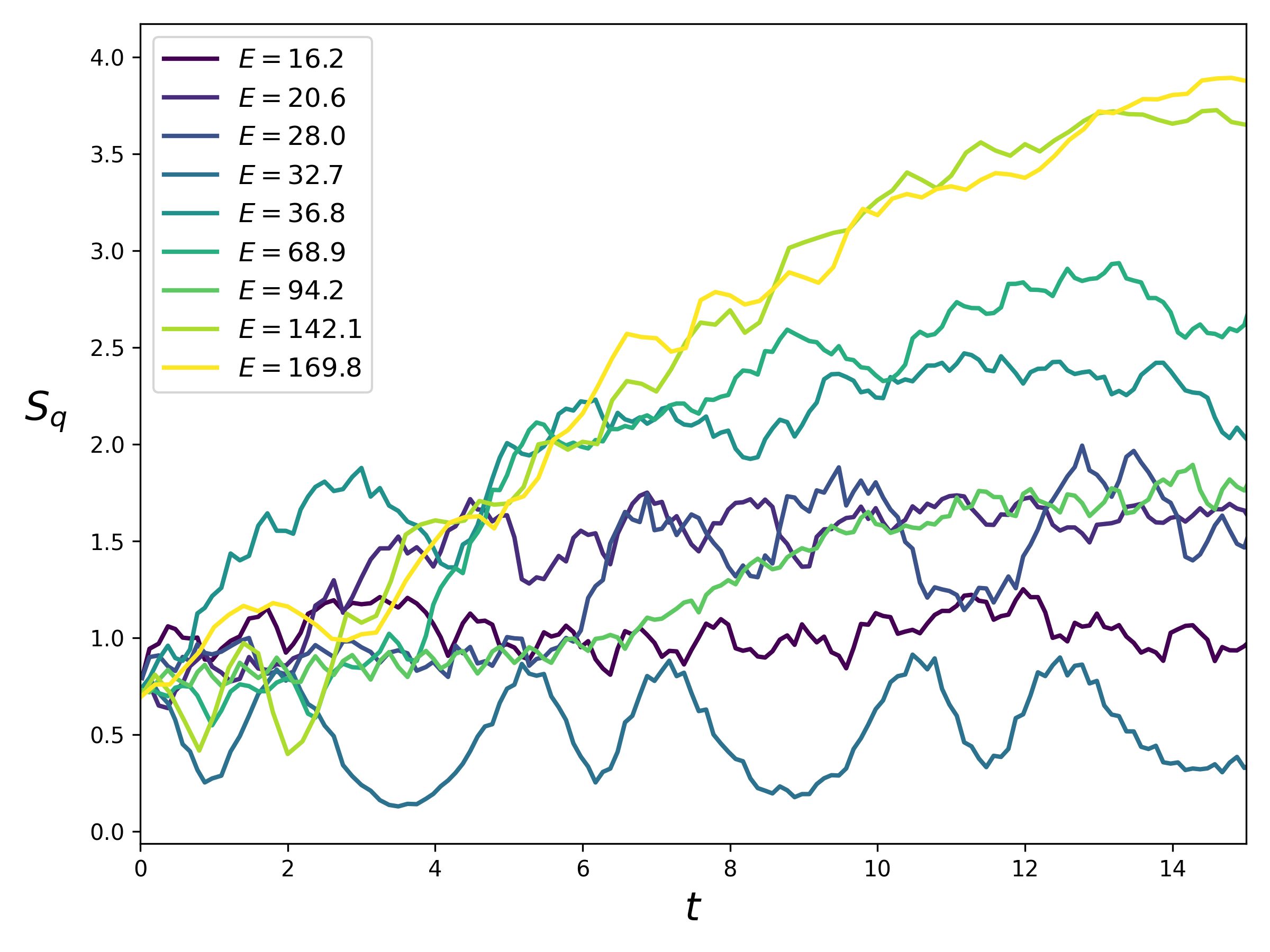}
	\end{center}
	\vglue -0.3cm
	\caption{\label{figA2} Dependence  of quantum von Neumann entropy $S_q(t)$ from Figure~\ref{fig14}
          is shown here for an initial short time interval
          $0 \leq t \leq 15$; averaging is done on a small time window
          $\delta t =0.1$ depicting initial oscillations between
          states $m=1$ and $m=2$ for initial state
          $\psi(t=0) = (\psi_1 + \psi_2)/\sqrt{2}$ at total energy $E=16.2$.
	}
\end{figure}

\begin{figure}[H]
	\begin{center}
           \includegraphics[width=0.8\columnwidth]{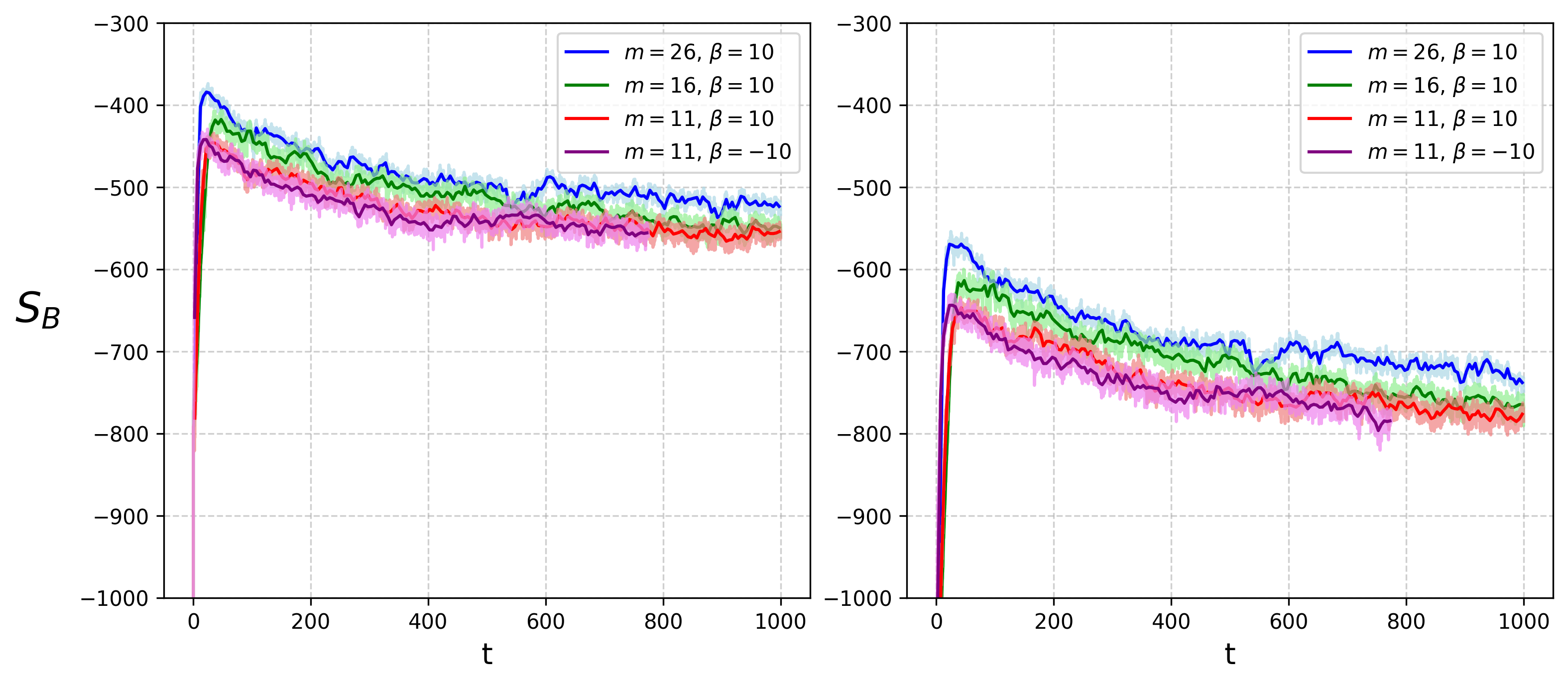}
	\end{center}
	\vglue -0.3cm
	\caption{\label{figA3}
	Time dependence  of the classical Boltzmann entropy $S_B(t) = \sum_m \ln \rho_m(t)$. 
	The entropy is defined as $S_B =  \sum_{m=1}^{m_{max}}  \ln(\rho_m)$,
        time average is done  over a window of $\delta t=5$.
        The curves correspond to initial states  $\psi(t=0) = (\psi_m+\psi_{m+1})/\sqrt{2}$
        with $m$ values given in the panels; system parameters are the same as in Figure~\ref{fig14}.
        The blue, green, and red curves show the cases with $\beta=10$ for $m=26, 16, 11$, respectively. 
        The violet curve represents the case with $m=11$ and $\beta=-10$. 
        The left panel uses a cutoff of $m_{max}=75$, while the right panel uses $m_{max}=150$.
        For each color, the semi-transparent curve shows
        the evolution of the entropy, while the solid, 
        opaque curve represents its mean value, smoothed over a time window of $\delta t=5$.
	}
\end{figure}

\begin{figure}[H]
	\begin{center}
          \includegraphics[width=0.7\columnwidth]{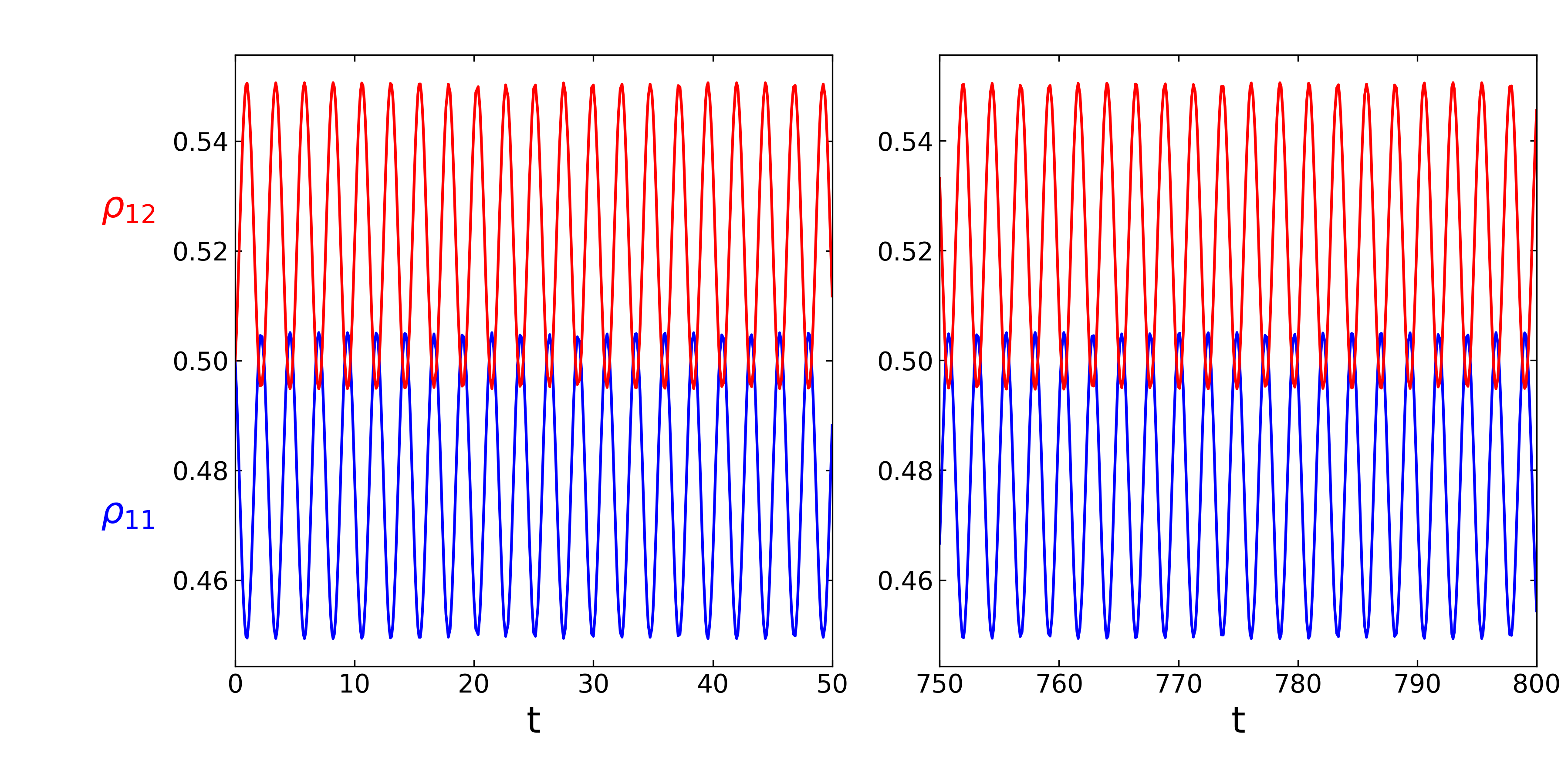}\\
          \includegraphics[width=0.7\columnwidth]{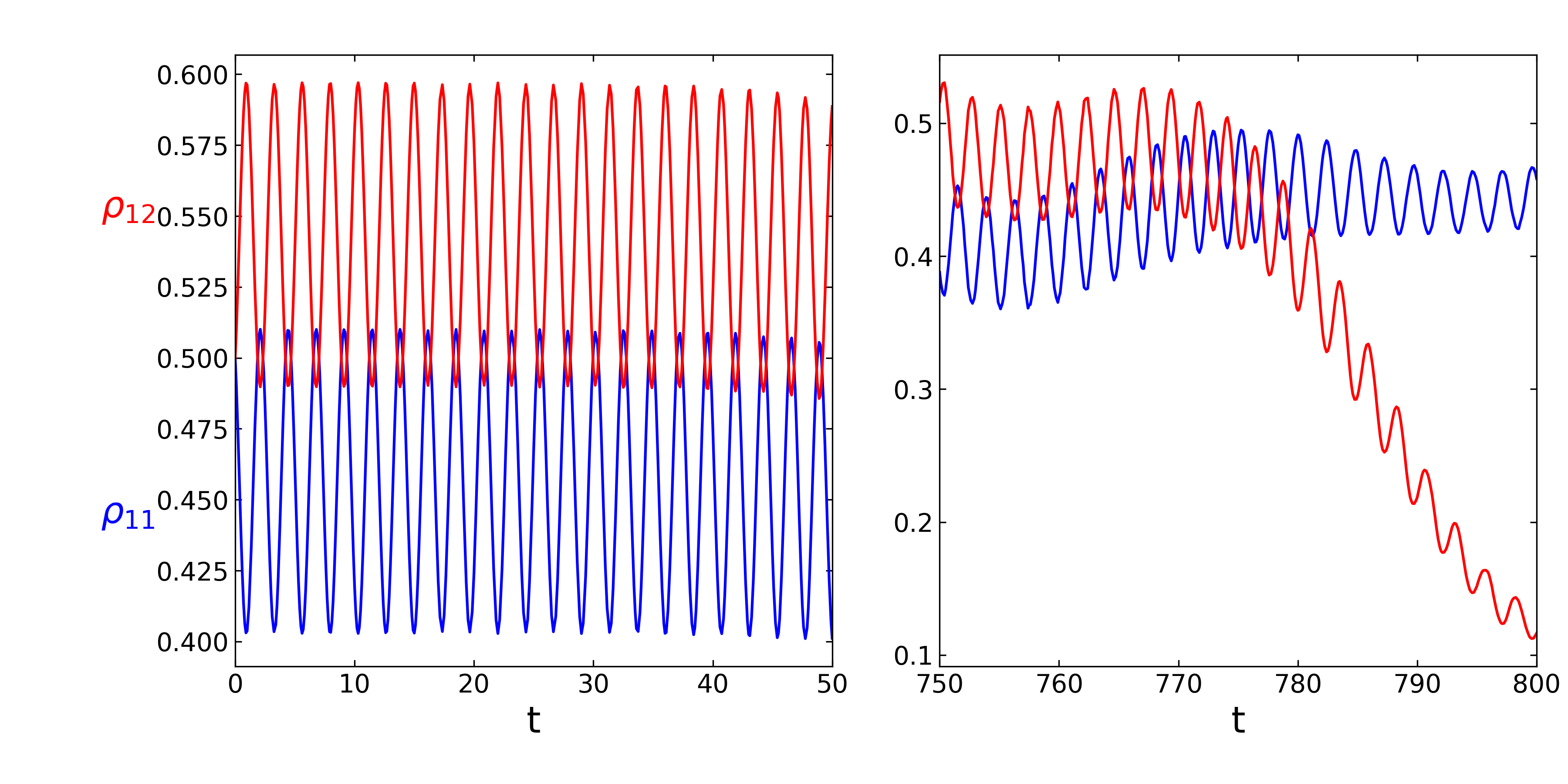}
	\end{center}
	\vglue -0.3cm
	\caption{\label{figA4} Top panels show probability oscillations $\rho_{11}$ (blue),
          $\rho_{12}$ (red) with time, left/tight panels are at short/long times at
          $\beta =0.5$, $E=64.99$.
          Bottom panels: same as in top panels but for $\beta=1$, $E=65.2$.
          Initial state in the NSE (\ref{eqnse}) is
          $\psi(t=0) = (\psi_m + \psi_{m+1})/\sqrt{2}$ at $m=11$. 
	}
\end{figure}

\begin{figure}[H]
	\begin{center}
           \includegraphics[width=0.8\columnwidth]{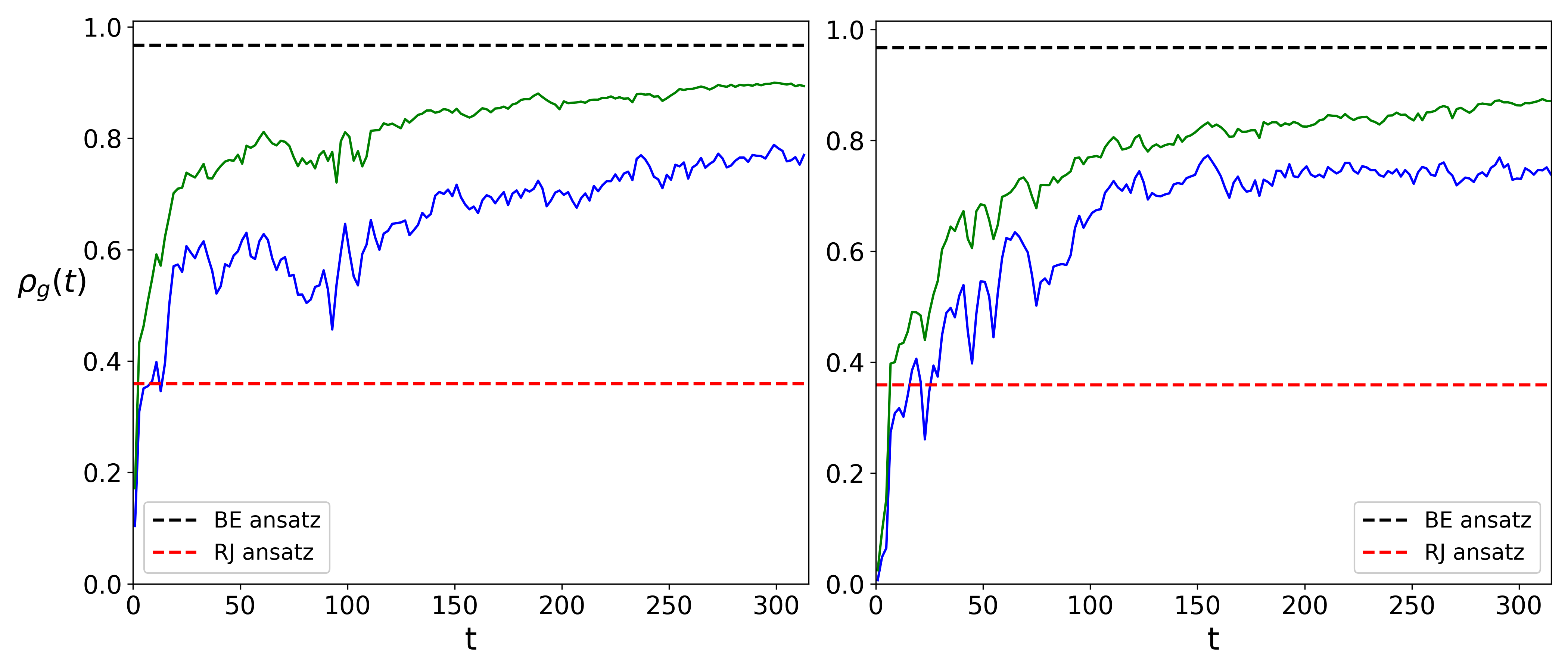}
	\end{center}
	\vglue -0.3cm
	\caption{\label{figA5} Dependence of ground state probability $\rho_g(t)$ on time;
          same parameters as Figure~\ref{fig21} at $\beta=-10$ but the
          initial state is $\psi(t=0) = (\psi_m+\psi_{m+1})/\sqrt{2}$ at
          $m=5$ (left panel) and $m=7$ (right panel); blue/green curves are
          the probability in the linear/instantaneous ground state as in Figure~\ref{fig21}
          with the same dashed lines.
	}
\end{figure}

\begin{figure}[H]
	\begin{center}
          \includegraphics[width=0.7\columnwidth]{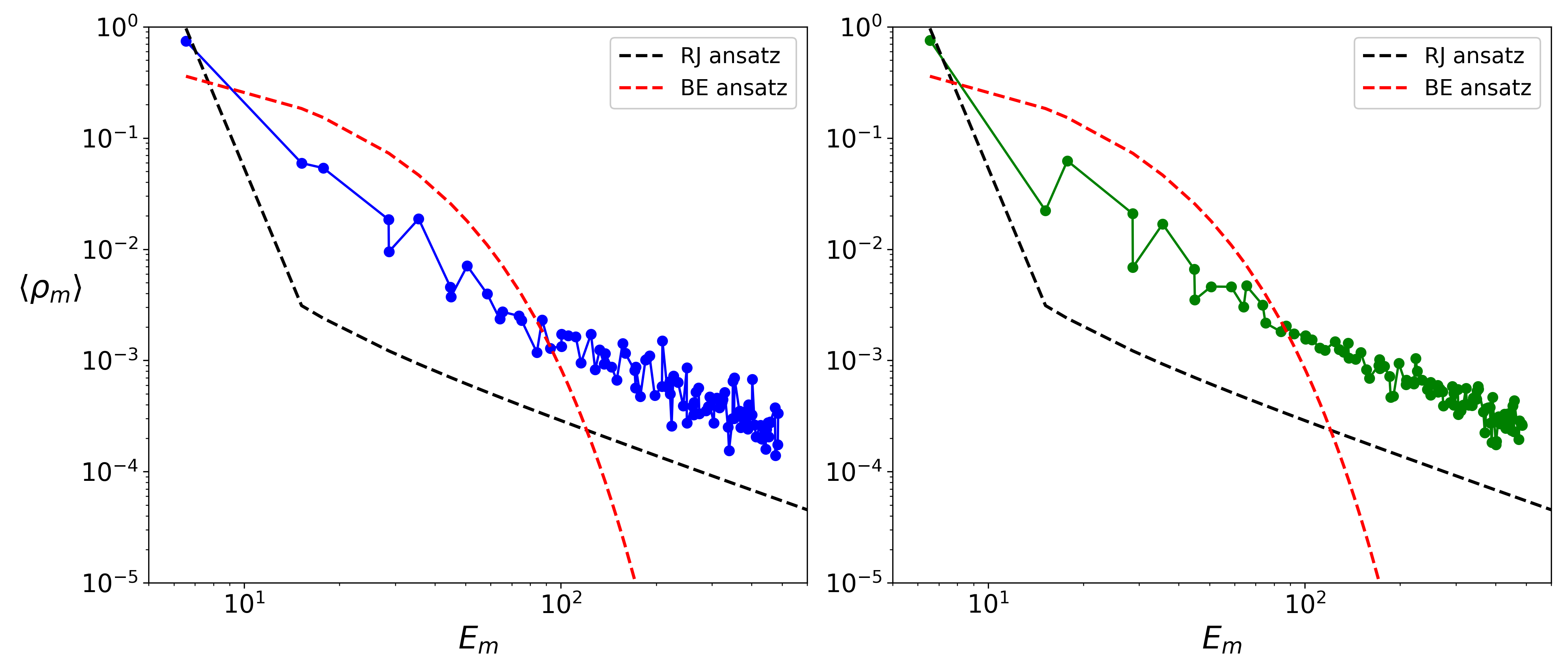}\\
          \includegraphics[width=0.7\columnwidth]{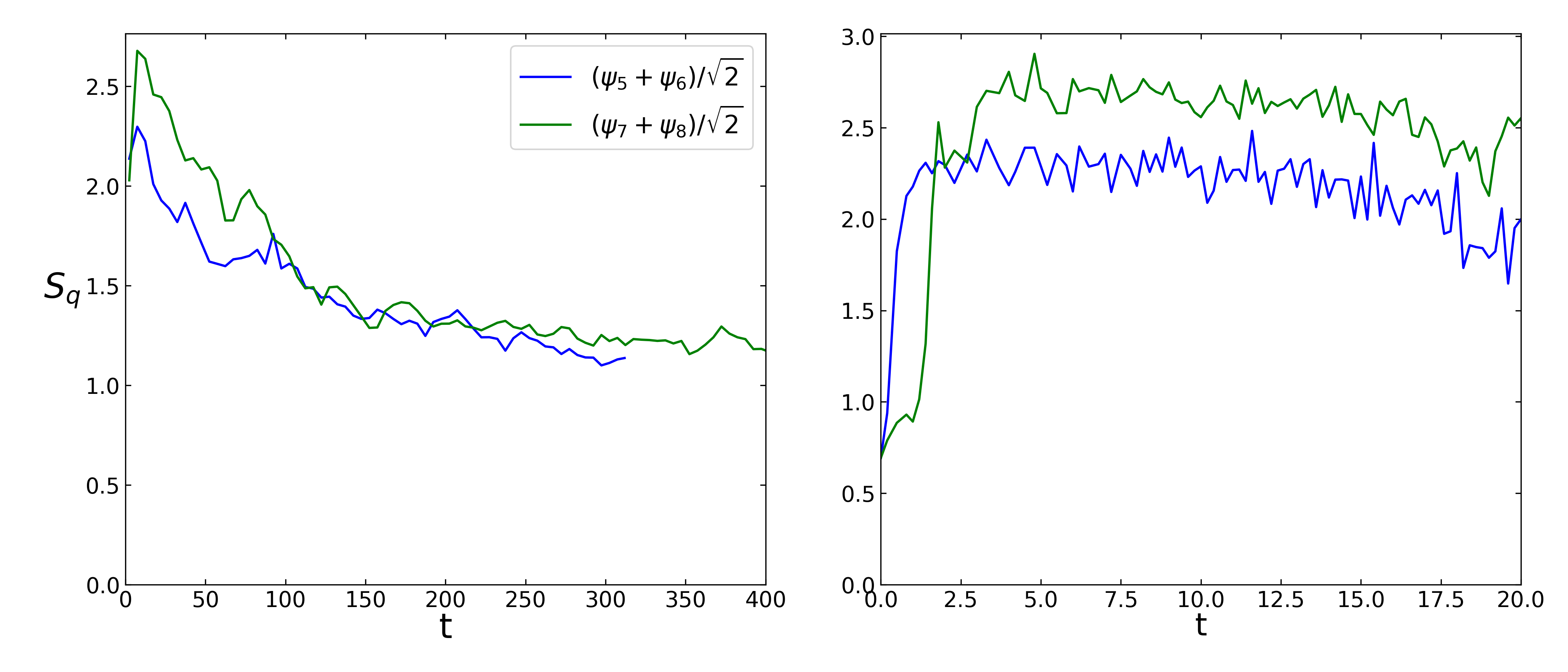}
	\end{center}
	\vglue -0.3cm
	\caption{\label{figA6} Same as Figure~\ref{fig22} at $\beta=-10$
          and initial state $\psi(t=0) = (\psi_m+\psi_{m+1})/\sqrt{2}$ at
          $m=5$ (top left panel, blue color) and $m=7$ (top right panel, green color);
          data are averaged over the range $200 \leq t \leq 400$;
          dashed curves are as in Figure~\ref{fig21}.
          Top panels show averaged probabilities $< \rho_m >$ vs. energies $E_m$
          and corresponding bottom panels show the quantum von Neumann
          entropy dependence $S_q(t)$ on long/short times
          in right/left panels.
	}
\end{figure}

\begin{figure}[H]
	\begin{center}
           \includegraphics[width=0.8\columnwidth]{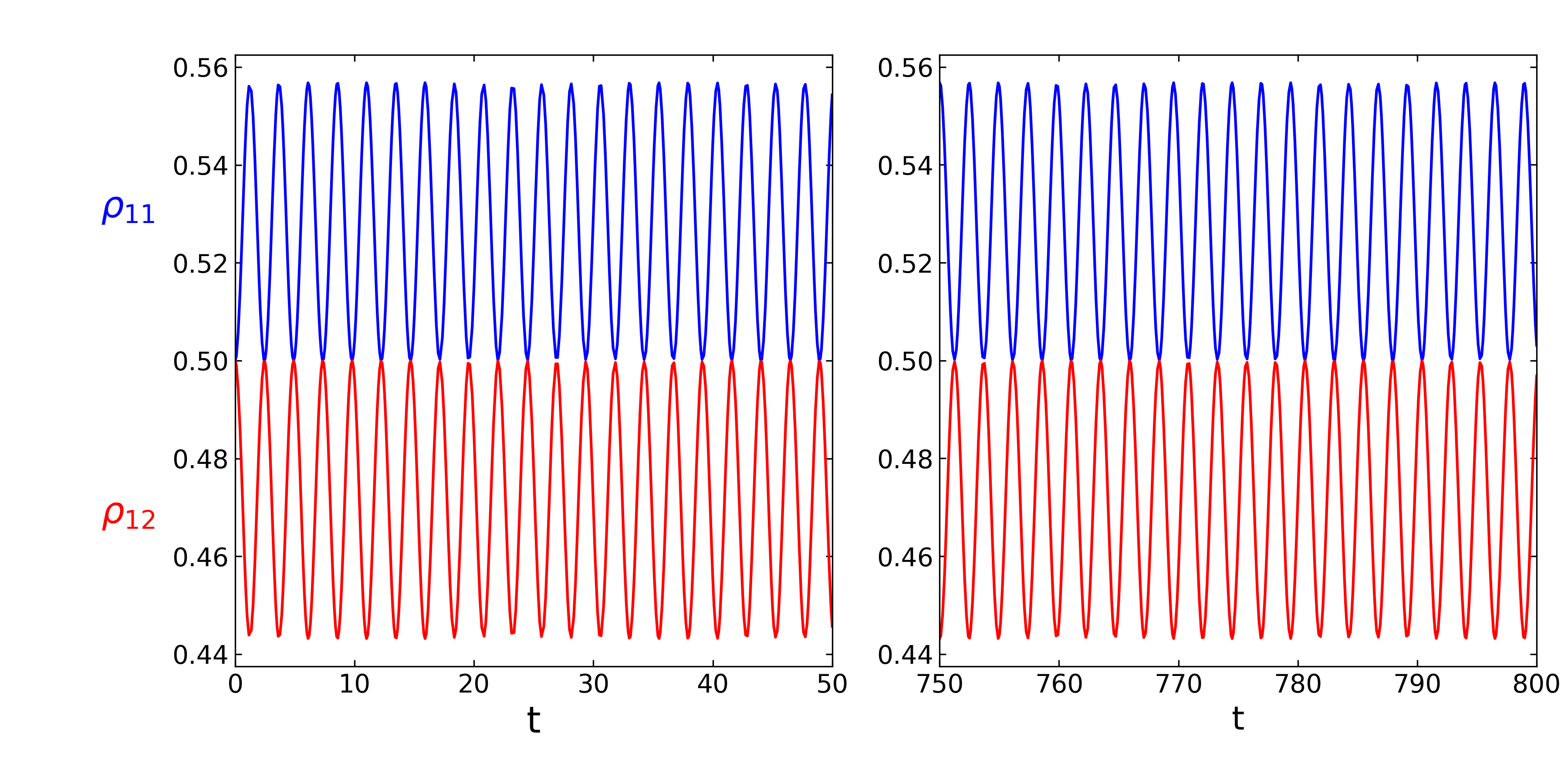}
	\end{center}
	\vglue -0.3cm
	\caption{\label{figA7} Probability oscillations $\rho_{11}$ (blue),
          $\rho_{12}$ (red) with time, left panel for $0\leq t \leq 50$,
          right panel for $750 \leq t \leq 800$ at $\beta =-0.5$.
	}
\end{figure}

\begin{acknowledgements} 
We thank S.Babin,   D.Kharenko and E.Podivilov (IAE RAS Novosibirsk),   K.M.Frahm (LPT Toulouse) and N.Pavloff (LPTMS Orsay) for  useful discussions.

The authors acknowledge support from the grants
 ANR France project OCTAVES (ANR-21-CE47-0007),
NANOX $N^\circ$ ANR-17-EURE-0009 in the framework of 
the Programme Investissements d'Avenir (project MTDINA),
MARS (ANR-20-CE92-0041).
\end{acknowledgements} 



\end{document}